\documentclass[a4paper,11pt]{article}
\pdfoutput=1 

\usepackage{jcappub} 

\usepackage[T1]{fontenc} 
\usepackage{verbatim}
\usepackage[utf8]{inputenc}
\usepackage[american]{babel}
\usepackage{graphicx}
\usepackage{epsfig}
\usepackage{adjustbox}
\usepackage{booktabs}
\usepackage{multirow}
\usepackage{dcolumn}
\usepackage{amsmath}
\usepackage{amsfonts}
\usepackage{mathtools}
\usepackage{amssymb}
\usepackage{empheq}
\usepackage{subcaption} 
\usepackage{caption}
\usepackage{epstopdf}
\usepackage{bm}
\usepackage{bbm}
\usepackage{dsfont}
\usepackage{mathrsfs}
\usepackage{version}
\usepackage{enumerate}
\usepackage{braket}
\usepackage{enumitem}
\usepackage{transparent}
\usepackage{pifont}
\usepackage{colortbl}
\usepackage{rotating}
\usepackage{adjustbox}
\usepackage{cancel}
\usepackage{tabularx}
\usepackage{soul}
\usepackage{pdfpages}
\usepackage{microtype}
\usepackage[table]{xcolor}
\usepackage{siunitx}
\usepackage{float} 
\usepackage[nocompress]{cite} 
\usepackage{marginnote}
\usepackage{relsize}
\usepackage{feynmp-auto}

\usepackage{hyperref}
\hypersetup{colorlinks=true,
linkcolor=navyblue,citecolor=coralred,urlcolor=green(munsell),pdfencoding=auto}

\definecolor{navyblue}{rgb}{0.0, 0.0, 0.5}
\definecolor{bleudefrance}{rgb}{0.19, 0.55, 0.91}
\definecolor{coralred}{rgb}{1.0, 0.25, 0.25}
\definecolor{royalblue}{rgb}{0.25, 0.41, 0.88}
\definecolor{cadmiumgreen}{rgb}{0.0, 0.42, 0.24}
\definecolor{green(munsell)}{rgb}{0.0, 0.66, 0.47}
\definecolor{blue-violet}{rgb}{0.54, 0.17, 0.89}
\definecolor{darkviolet}{rgb}{0.58, 0.0, 0.83}
\definecolor{orange(colorwheel)}{rgb}{1.0, 0.5, 0.0}
\definecolor{internationalorange}{rgb}{1.0, 0.31, 0.0}

\definecolor{magenta(process)}{rgb}{1.0, 0.0, 0.56}

\definecolor{darkspringgreen}{rgb}{0.09, 0.45, 0.27}

\definecolor{royalblue(web)}{rgb}{0.25, 0.41, 0.88}

\definecolor{cadmiumorange}{rgb}{0.93, 0.53, 0.18}

\definecolor{heliotrope}{rgb}{0.87, 0.45, 1.0}

\makeatletter
\renewcommand*{\@textcolor}[3]{%
\protect\leavevmode
\begingroup
\color#1{#2}#3%
\endgroup
}
\makeatother

\newcommand{\myfloatalign}{\centering}

\makeatletter
\newlength{\apb@width}
\newcommand{\autoparbox}[2][c]{\settowidth{\apb@width}{#2}\parbox[#1]{\apb@width}{#2}}
\newcommand{\includegraphicsbox}[2][]{\autoparbox{\includegraphics[#1]{#2}}}
\makeatother

\DeclareCaptionLabelSeparator{colquad}{.\,\,\,}
\renewcommand\[{\left[}

\DeclarePairedDelimiter{\abs}{\lvert}{\rvert}

\makeatletter
\let\save@mathaccent\mathaccent
\newcommand*\if@single[3]{%
\setbox0\hbox{${\mathaccent"0362{#1}}^H$}%
\setbox2\hbox{${\mathaccent"0362{\kern0pt#1}}^H$}%
\ifdim\ht0=\ht2 #3\else #2\fi
}
\newcommand*\rel@kern[1]{\kern#1\dimexpr\macc@kerna}
\newcommand*\widebar[1]{\@ifnextchar^{{\wide@bar{#1}{0}}}{\wide@bar{#1}{1}}}
\newcommand*\wide@bar[2]{\if@single{#1}{\wide@bar@{#1}{#2}{1}}{\wide@bar@{#1}{#2}{2}}}
\newcommand*\wide@bar@[3]{%
\begingroup
\def\mathaccent##1##2{%
\let\mathaccent\save@mathaccent
\if#32 \let\macc@nucleus\first@char \fi
\setbox\z@\hbox{$\macc@style{\macc@nucleus}_{}$}%
\setbox\tw@\hbox{$\macc@style{\macc@nucleus}{}_{}$}%
\dimen@\wd\tw@
\advance\dimen@-\wd\z@
\divide\dimen@ 3
\@tempdima\wd\tw@
\advance\@tempdima-\scriptspace
\divide\@tempdima 10
\advance\dimen@-\@tempdima
\ifdim\dimen@>\z@ \dimen@0pt\fi
\rel@kern{0.6}\kern-\dimen@
\if#31
\overline{\rel@kern{-0.6}\kern\dimen@\macc@nucleus\rel@kern{0.4}\kern\dimen@}%
\advance\dimen@0.4\dimexpr\macc@kerna
\let\final@kern#2%
\ifdim\dimen@<\z@ \let\final@kern1\fi
\if\final@kern1 \kern-\dimen@\fi
\else
\overline{\rel@kern{-0.6}\kern\dimen@#1}%
\fi
}%
\macc@depth\@ne
\let\math@bgroup\@empty \let\math@egroup\macc@set@skewchar
\mathsurround\z@ \frozen@everymath{\mathgroup\macc@group\relax}%
\macc@set@skewchar\relax
\let\mathaccentV\macc@nested@a
\if#31
\macc@nested@a\relax111{#1}%
\else
\def\gobble@till@marker##1\endmarker{}%
\futurelet\first@char\gobble@till@marker#1\endmarker
\ifcat\noexpand\first@char A\else
\def\first@char{}%
\fi
\macc@nested@a\relax111{\first@char}%
\fi
\endgroup
}
\makeatother

\newcommand\ee{\end{equation}}
\newcommand\be{\begin{equation}}
\newcommand\eea{\end{eqnarray}}
\newcommand\bea{\begin{eqnarray}}
\newcommand{\bsp}{\begin{split}}
\newcommand{\esp}{\end{split}}
\newcommand{\bit}{\begin{itemize}[leftmargin=*]}
\newcommand{\eit}{\end{itemize}}
\newcommand{\ben}{\begin{enumerate}[leftmargin=*]}
\newcommand{\een}{\end{enumerate}}

\newcommand{\ie}{
i.e.
}
\newcommand{\eg}{
e.g.
}

\newcommand\eq[1]{Eq.~\eqref{eq:#1}}
\newcommand\eqsI[1]{Eqs.~\eqref{eq:#1}}
\newcommand{\eqsII}[2]{Eqs.~\eqref{eq:#1}, \eqref{eq:#2}}
\newcommand{\eqsIII}[3]{Eqs.~\eqref{eq:#1}, \eqref{eq:#2}, \eqref{eq:#3}}
\newcommand{\eqsIV}[4]{Eqs.~\eqref{eq:#1}, \eqref{eq:#2}, \eqref{eq:#3}, \eqref{eq:#4}}

\newcommand{\iu}{\mathrm{i}}
\newcommand{\eu}{\mathrm{e}}
\newcommand{\dif}{\mathrm{d}}

\renewcommand{\vec}{\bm}

\newcommand\eps{\varepsilon}

\newcommand{\vk}{\vec{k}}

\def\vk{\vec{k}}

\def\<{\left\langle}
\def\>{\right\rangle}

\def\comment#1{}

\title{The EFT Likelihood for Large-Scale Structure}

\author[a]{Giovanni Cabass,}
\author[a]{Fabian Schmidt}

\affiliation[a]{Max-Planck-Institut f\"{u}r Astrophysik, 
Karl-Schwarzschild-Str. 1, 85741 Garching, Germany}

\emailAdd{gcabass@mpa-garching.mpg.de}
\emailAdd{fabians@mpa-garching.mpg.de}

\abstract{\noindent We derive, using functional methods and the bias expansion, 
the conditional likelihood for observing a specific tracer field given an underlying matter field. 
This likelihood is necessary for Bayesian-inference methods. 
If we neglect all stochastic terms apart from the ones appearing in the auto two-point function of tracers, 
we recover the result of {Schmidt et al., 2018} \cite{Schmidt:2018bkr}. 
We then rigorously derive the corrections to this result, such as those coming from a non-Gaussian stochasticity 
(which include the stochastic corrections to the tracer bispectrum) and higher-derivative terms. 
We discuss how these corrections can affect current applications of Bayesian inference. 
We comment on possible extensions to our result, with a particular eye towards 
primordial non-Gaussianity. This work puts on solid theoretical grounds the effective-field-theory-(EFT-)based approach to Bayesian forward modeling.} 

\begin{document}
\maketitle
\flushbottom



\section{Introduction and summary of main results}
\label{sec:introduction}

\noindent Standard approaches to infer cosmology from large-scale structure observations make use of $n$-point correlation functions 
of some tracer of the dark matter distribution, like galaxies. 
These correlation functions can be robustly predicted using the effective field theory of biased tracers in 
large-scale structure, whose main advantage 
is the fact that it allows to keep theoretical uncertainties under control (a generic feature 
of effective field theories). Then, estimators for the correlation functions are 
constructed and constraints on cosmological parameters are derived 
(see e.g.~Section 4.1 of \cite{Desjacques:2016bnm} for a review). 

This method has two drawbacks. First, the higher the $n$ the more difficult it is 
to both predict the correlation function and measure it from data. 
Second, computing the covariance of an $n$-point function, 
which is required to correctly interpret the measurements, typically requires the knowledge of 
the $2n$-point function (or its numerical estimation based on mock catalogs), 
and this makes going to high $n$ cumbersome, if not unfeasible. 
In absence of a clear hint on which $n$ (if any) we have to stop at in order to 
obtain the maximum amount of cosmological information from galaxy clustering, 
and given the rapidly increasing number of bias coefficients at high $n$, we see how this can turn into a very complicated endeavor. 

An elegant way to bypass these problems is by making use of the so-called Bayesian forward modeling (see 
\cite{1995MNRAS.272..885F, 
2010MNRAS.406...60J, 
2010MNRAS.409..355J, 
2013MNRAS.432..894J, 
Wang:2014hia, 
2015MNRAS.446.4250A} 
for applications and \cite{1989ApJ...336L...5B, 
Schmittfull:2017uhh, Seljak:2017rmr, Modi:2019hnu} for related approaches). 

This method aims at {exploiting} directly {the amplitudes and} 
the phases of the tracer field, instead of its correlation functions. 
A rough picture is the following. Given a realization of the initial conditions, we will have a particular distribution of tracers. 
We can then vary the initial conditions (and bias as well as cosmological parameters) until we match to the observed tracer field. 
The initial conditions are not deterministic, though: they have their own probability density functional. 
Therefore, instead of looking for the realization that exactly matches the observed Universe 
(i.e.~looking at the best-fit values), we can also just integrate over all the realizations and 
obtain marginalized errors on the parameters of the background cosmology. This is no different, in spirit, 
from what is done when performing Markov-Chain Monte Carlo sampling to obtain bounds on cosmology from CMB data. 

From this discussion, we see that what we need is a prior on the initial conditions $\delta_{\rm in}$ 
(besides that on cosmological and bias parameters), 
and the conditional probability density functional ${\cal P}[\delta_g|\delta_{\rm in}]$ 
of observing a specific tracer field $\delta_g$ given a realization of the initial conditions. 
Via simple manipulations with nested probabilities, we can rewrite ${\cal P}[\delta_g|\delta_{\rm in}]$ 
in terms of two quantities (we refer to \cite{Schmidt:2018bkr} for more details):
\begin{enumerate}[leftmargin=*]
\item the conditional probability ${\cal P}[\delta|\delta_{\rm in}]$, i.e.~the probability 
of having a total (dark+luminous) matter field $\delta$ given the initial conditions $\delta_{\rm in}$;
\item the conditional probability ${\cal P}[\delta_g|\delta]$ of observing $\delta_g$ given a matter field $\delta$. 
\end{enumerate}

The prior for $\delta_{\rm in}$ is determined by inflation 
(which provides Gaussian initial conditions, apart from a possible small primordial non-Gaussianity), 
together with any evolution up to the onset of nonlinearities. 
${\cal P}[\delta|\delta_{\rm in}]$ is instead given by the forward model for matter and gravity: 
for a deterministic evolution, this is simply a Dirac delta functional 
$\smash{\delta^{(\infty)}_{\rm D}(\delta - \delta_{\rm fwd}[\delta_{\rm in}])}$, 
where $\delta_{\rm fwd}[\delta_{\rm in}]$ is the forward model for matter (such as 
N-body simulations, for example).\footnote{Even if we use an N-body simulation 
to forward-model the matter field, the evolution is never deterministic 
(due to the fact that, by construction, N-body simulations still integrate out all sub-grid modes), 
so the assumption of having a Dirac delta functional is not technically correct. 
We will see below that this is naturally taken into account by the approach used in this paper.} 
\emph{The goal of this paper is to compute ${\cal P}[\delta_g|\delta]$ using the effective field theory of biased tracers.} 

Let us outline our general strategy to compute ${\cal P}[\delta_g|\delta]$. 
First, we can use the properties of conditional probabilities to express this as the ratio 
of the \emph{joint probability} ${\cal P}[\delta_g,\delta]$ and the probability ${\cal P}[\delta]$ for the matter field itself. 
How do we compute these two quantities? We know how to compute correlation functions 
of $\delta_g$ and $\delta$ using the effective field theory of large-scale structure (EFT of LSS)/bias expansion. More generally, 
we know how to compute the full generating functionals $Z[J]$ and $Z[J_g,J]$, 
the first for the correlation functions of $\delta$ alone, and the second for those of $\delta_g$ and $\delta$ together. 
The generating functional for matter correlation functions in the EFT of LSS 
was discussed for the first time (to the authors' knowledge) in 
\cite{Carroll:2013oxa}: here we extend it for the first time to include also biased tracers. 

Once we have the two generating functionals $Z[J]$ and $Z[J_g,J]$, an expression for ${\cal P}[\delta]$ and ${\cal P}[\delta_g,\delta]$ 
can be easily obtained via the inverse functional Fourier transform. 
This allows us to cast the problem of computing these probability density functionals (henceforth ``likelihoods'', 
for simplicity) using methods of functional integration commonly employed in quantum field theory. 
As we will see below, this method successfully recovers the 
result of \cite{Schmidt:2018bkr}, where the authors computed ${\cal P}[\delta_g|\delta]$ 
using the assumption that the noise field $\eps_g$ (that enters in 
the linear bias expansion of $\delta_g$ as $\delta_g = b_1\delta + \eps_g$) is a Gaussian-distributed variable. 

On top of this, the approach developed here gives a simple and rigorous way of computing the corrections to this result which come, 
for example, from the wrong assumption of having Gaussian noise. While Ref.~\cite{Schmidt:2018bkr} roughly estimated 
the magnitude of some of these terms, it remained unclear how to rigorously derive these corrections. 
We will show that the amplitude of the noise field with respect to the amplitude of 
$\delta$ is what essentially controls the size of these corrections, 
effectively making it a third expansion parameter in addition to the expansion in perturbations 
(i.e.~in the smallness of $\delta$ on large scales) and the expansion in derivatives. 
Finally, as we will see at the end of the paper, the method employed here easily allows 
to include the impact of primordial non-Gaussianity. 

Before proceeding to the main body of the paper, we emphasize that functional methods feature prominently also in 
the recent works \cite{Blas:2015qsi,Blas:2016sfa,McDonald:2017ths}. 
In the first two papers the authors develop, roughly speaking, an analytic approach 
to compute correlators of $\delta$ by deriving an evolution equation for 
${\cal P}[\delta]$, the likelihood of the nonlinear matter field. Ref.~\cite{McDonald:2017ths}, 
instead, develops a method to compute the generating functional for matter in Lagrangian space including shell crossing. 

Their approach and the one followed in this paper have some overlap. 
For example, the authors of \cite{Blas:2015qsi,Blas:2016sfa} also conclude that the 
problem of computing correlation functions for the matter field 
can be recast in a way reminiscent of a Euclidean quantum field theory in three dimensions. 
In all three papers the discussion about how to account for 
the impact of primordial non-Gaussianity is also similar to ours. 
We stress, however, that the goal of this work is very different from that of these three papers: 
what we care about is the conditional likelihood ${\cal P}[\delta_g|\delta]$ for the \emph{galaxy} field, 
and not the generating functional or the likelihood for the \emph{matter} field.

\subsection{Summary of main results}

\noindent Given the length of the paper, and its inherently technical nature, 
before proceeding we want to emphasize here its main takeaways: 
\begin{itemize}[leftmargin=*]
\item in our calculation of the conditional likelihood we use the effective field theory 
of large-scale structure, i.e.~an intrinsically perturbative forward model, 
for the evolution of the matter field. This is done for purely technical reasons, 
since we cannot compute the functional integrals analytically otherwise. 
However, \emph{at the end we arrive at a resummation at all orders in perturbations in the evolution of the matter field.} 
More precisely, the details of the evolution of the matter field are ``absorbed'' into the matter likelihood, and they disappear from 
${\cal P}[\delta_g|\delta]$ once we divide the joint likelihood by ${\cal P}[\delta]$ 
(we will actually be able to rigorously prove this only for Gaussian noise and at leading order in derivatives: 
we will only sketch what happens if this assumption is dropped); 
\item thanks to the above result, the field $\delta$ employed in the conditional likelihood 
is the fully nonlinear matter field that we can evolve, 
e.g., with N-body gravity-only simulations. By using a fully nonlinear forward model, then, 
we are able to consider at once \emph{all} the information contained in the nonlinear displacement of the initial matter distribution. 
This information is protected by the equivalence principle, 
that ensures that tracers such as galaxies move on the same trajectories as matter on large scales, 
and it is the only information that we can extract from observations of galaxy clustering 
that is not degraded once we marginalize over the bias coefficients proper to the tracer considered
\mbox{(see also \cite{Schmidt:2018bkr,Elsner:2019rql} for a discussion);} 
\item recasting the computation in terms of an action makes it easy to read off the scaling dimensions 
of the different terms that appear in the conditional likelihood (i.e.~how relevant or irrelevant they 
are on large scales), like those coming from the deterministic bias expansion or from non-Gaussian stochasticities. 
This makes Section~\ref{sec:setting_up_the_integrals}, and Section~\ref{subsec:action_and_RG_scalings} in particular, 
\mbox{arguably the most important section of this paper;} 
\item the results of Section~\ref{subsec:action_and_RG_scalings} are also useful to understand the presence of the amplitude of 
the noise $\eps_g$ as a new expansion parameter, beyond the expansion in powers of the matter field $\delta$ and its derivatives. 
This is also a key new result: it proves that extracting cosmological information from very noisy tracers (through either Bayesian inference or measurements of correlation functions) will be very difficult; 
\item the formalism used in this work makes the inclusion of primordial non-Gaussianity straightforward. 
Besides modifying the prior on $\delta_{\rm in}$, 
PNG affects the form of the bias expansion and consequently of correlation functions. 
For example, the coupling between these two sources of nonlinearities shows up in the 
galaxy bispectrum (see e.g.~\cite{Scoccimarro:2003wn,2011JCAP...04..006B,Assassi:2015fma}). Another example is the fact that PNG induces new terms in the 
deterministic bias expansion due to the coupling between long and short modes, 
the most famous being the ``scale-dependent bias'' $\sim \nabla^{-2}\delta$ in case of local-type PNG. 
These effects are all captured by suitable change of the deterministic relation between 
the galaxy field and $\delta$. In this paper we show that, even after these modifications 
are taken into account, the presence of PNG does not affect the form of the 
conditional likelihood in the limit of Gaussian stochasticities and only the noise in the auto correlation 
function of tracers being non-vanishing. In this limit, that we will argue throughout the paper gives the leading 
result on large scales, the conditional likelihood is still given by the result derived in \cite{Schmidt:2018bkr}. 
\end{itemize}

\subsection{Structure of the paper}

\noindent Our paper is composed of six main sections and six appendices. 
\begin{description}[leftmargin=*]
\item[Section~\ref{sec:notation_and_conventions}] contains the details on the notation and the conventions. 
\item[Section~\ref{sec:setting_up_the_integrals}] shows how we set up the functional 
integrals for ${\cal P}[\delta]$ and ${\cal P}[\delta_g,\delta]$. 
We start with a definition of the generating functionals $Z[J]$ and $Z[J_g,J]$: 
this is done in Sections~\ref{subsec:matter} and \ref{subsec:joint}. 
In these sections, together with Section~\ref{subsec:stochastic_terms_and_tadpoles}, we also 
discuss in more detail how the stochasticities are included in the functional formalism employed in this paper. 
Section~\ref{subsec:action_and_RG_scalings} then shows how the computation of the likelihood 
is very similar to that of a generating functional for a Euclidean quantum field theory in three dimensions. 
This section explains what is the equivalent of the ``action'' for 
this theory, establishes a parallel between the terms appearing in the action and the bias expansion 
for a generic tracer, and (briefly) discusses its Feynman rules. 
Moreover, via simple dimensional analysis it shows how to estimate the relative importance 
on large scales of the different terms appearing in the action. 
\item [Section~\ref{sec:calculation_of_likelihood-gaussian_stochasticities}] 
contains a first computation of the likelihood where, using a terminology 
that reflects the assumptions of \cite{Schmidt:2018bkr}, we assume ``Gaussian stochasticities''. 
{The precise meaning of Gaussian stochasticities in this paper 
is that there is no noise in the bias coefficients, and the difference between 
data and theoretical prediction for the galaxy field is distributed as a Gaussian.} 
In Section~\ref{subsec:P_eps_g} we consider the case where only the noise in the tracer auto two-point function 
is non-vanishing (and our result matches that of \cite{Schmidt:2018bkr} in this limit). 
In Section~\ref{subsec:P_eps_g_eps_m} we include the cross stochasticity between matter and tracer, and we show how the likelihood now 
gains some additional terms that were not considered in \cite{Schmidt:2018bkr}. 
\item [Section~\ref{sec:calculation_of_likelihood-nongaussian_stochasticities}] goes beyond Gaussian stochasticities. 
It discusses the impact of two terms: the stochasticity in the linear local-in-matter-density (LIMD) bias coefficient 
$b_1$ and the bispectrum of the noise $\eps_g$ (this is done in Section~\ref{subsec:bispectrum_and_noise_in_LIMD_bias}). 
Most importantly, in Section~\ref{subsec:NG_scalings} we show that it is the amplitude of $\eps_g$ 
with respect to the amplitude of the matter field $\delta$ that determines the importance on large scales of the corrections 
coming from these terms. This is one of the key results of this work. 
\item [Section~\ref{sec:discussion}] contains a discussion on the calculations and the results of the preceding 
sections. First, we study in more detail the three parameters in which we can expand the likelihood on large scales, i.e.~the 
smallness of derivatives and perturbations together with the expansion in the amplitude of the noise field $\eps_g$. 
This is done in Section~\ref{subsec:expansion_parameters}, while 
Section~\ref{subsec:loops} gives an idea of how to deal with loop corrections in presence of non-Gaussian stochasticities. 
\item [Section~\ref{sec:conclusions}] concludes the paper and presents a summary of future directions. 
The impact of our results on current applications of Bayesian forward modeling 
is discussed in Section~\ref{subsec:bayesian_forward_modeling}, while Section~\ref{subsec:further_developments} 
discusses other future developments, \ie~redshift-space distortions and how to account for primordial non-Gaussianity (PNG). 
Finally, Section~\ref{subsec:comparison_with_fabian} compares our results with those of Schmidt et al., 2018 \cite{Schmidt:2018bkr}. 
\item [{Appendix~\ref{app:scalings}}] {elaborates briefly on how the scalings of Section~\ref{subsec:action_and_RG_scalings}, 
which allow to estimate the relative importance of the different contributions to the likelihood, are derived.} 
\item [Appendices~\ref{app:tree_level}, 
\ref{app:higher_derivative_stochasticities}, \ref{app:higher_order_stochasticities} and \ref{app:saddle_point_formulas}] contain 
most of the technical 
details of the computations of Sections~\ref{sec:calculation_of_likelihood-gaussian_stochasticities} 
and \ref{sec:calculation_of_likelihood-nongaussian_stochasticities}. 
\item [{Appendix~\ref{app:loops}}] {contains some details on the loop corrections discussed in
Section~\ref{subsec:loops} that can be omitted on first reading.} 
\end{description}

\section{Notation and conventions}
\label{sec:notation_and_conventions}

\noindent We mostly follow the notation of \cite{Desjacques:2016bnm} for the quantities that appear both there and in this paper. 
Table~\ref{tab:conventions} contains a list of the new symbols that do not appear in \cite{Desjacques:2016bnm}. 

We will use the word ``galaxy'' or ``tracer'' interchangeably: since the bias expansion applies 
equally to any tracer of large-scale structure, there is no loss of generality in doing so. 
We denote by $P_{\rm in}$ the power spectrum of $\delta_{\rm in}$. The linear matter power spectrum 
$P_{\rm L}$ is then $D^2_1P_{\rm in}$, where $D_1$ is the linear growth {factor}. 
We define the nonlinear scale $k_{\rm NL}$ following the standard
convention, i.e.~by $k^3_{\rm NL}P_{\rm L}(k_{\rm NL})/2\pi^2 = 1$. 
That is, $k_{\rm NL}$ is defined as the scale at which the dimensionless linear power spectrum is equal to $1$. 
The power spectra of the noise fields, such as $\eps_g$, are expanded in powers of $k^2$ (derivatives) as 
\begin{equation} 
\label{eq:noise_expansion}
P_{\eps_g}(k) = P_{\eps_g}^{\{0\}} + P_{\eps_g}^{\{2\}}k^2 + \cdots\,\,.
\end{equation}
Our Fourier convention and short-hand notation are
\begin{subequations}
\label{eq:fourier}
\begin{alignat}{3}
f(\vec{k}) &= \int\dif^3x\,f(\vec{x})\,\eu^{-\iu\vec{k}\cdot\vec{x}} 
&&\equiv \int_{\vec{x}}f(\vec{x})\,\eu^{-\iu\vec{k}\cdot\vec{x}}\,\,, \label{eq:fourier-1} \\
f(\vec{x}) &= \int\frac{\dif^3k}{(2\pi)^3}\,f(\vec{k})\,\eu^{\iu\vec{k}\cdot\vec{x}} 
&&\equiv \int_{\vec{k}}f(\vec{k})\,\eu^{\iu\vec{k}\cdot\vec{x}}\,\,. \label{eq:fourier-2}
\end{alignat}
\end{subequations}
We do not use a different symbol for Fourier-space and real-space quantities, letting the arguments 
($\vec{k}$, $\vec{p}$, etc. for Fourier space; $\vec{x}$, $\vec{y}$, etc. for real space) make the distinction. 
The {$D$}-dimensional Dirac delta function is denoted by {\smash{$\delta^{(D)}_{\rm D}$}}. 
Often we will use $\vec{k}_{12\dots n}$ to denote the sum $\vec{k}_1+\vec{k}_2+\cdots\,\vec{k}_n$. Similarly, we denote by 
$\int_{\vec{k}_1,\dots\vec{k}_n}$ a multiple integration over the momenta $\vec{k}_1,\dots\vec{k}_n$. 
We use a prime to denote that we have stripped 
a correlation function (or, in general, a diagram) of its $(2\pi)^3\delta^{(3)}_{\rm D}(\vec{k}_{12\dots n})$ 
momentum-conserving delta function. 

We use the basis of \cite{Mirbabayi:2014zca} (see also \cite{Senatore:2014eva}) for the bias expansion. 
In this basis the nonlocality in time is reabsorbed order-by-order in perturbation theory using the fact that the 
evolution of the linear matter field is scale-independent 
(see Section 2.5 of \cite{Desjacques:2016bnm} for more details). 
Therefore, we will avoid writing the time dependence of the bias coefficients, 
of the linear growth factor, and of the tracer and matter fields in general: our results 
will hold at any given redshift.\footnote{Notice however that we will assume an 
Einstein-de Sitter cosmology in order to have time-independent kernels for the SPT 
solution for the matter field. This is done purely to simplify the intermediate calculations: 
our main final result, i.e.~the fact that for Gaussian noise with constant power spectrum we are able 
to resum the nonlinear gravitational evolution of matter exactly, is completely independent of this assumption.} 

We will never explicitly need the expression for the various operators in the bias expansion. 
It will be enough to introduce two families of Fourier-space 
kernels, $K_n$ and $K_{g,n}$, to represent the \emph{deterministic evolution} 
(denoted by a subscript ``$\rm fwd$'') of the matter and tracer fields. More precisely, we write 
\begin{subequations}
\label{eq:G_and_M_kernels}
\begin{align}
\delta_{{\rm fwd}}[\delta_{\rm in}](\vec{k}) &= D_1K_1(k)\delta_{\rm in}(\vec{k}) \nonumber \\
&\;\;\;\; + \sum_{n=2}^{+\infty}\int_{\vec{p}_1,\dots\vec{p}_n}(2\pi)^{3}\delta^{(3)}_{\rm D}(\vec{k}-\vec{p}_{1\dots n})\,D_1^n\,
K_n(\vec{k};\vec{p}_1,\dots\vec{p}_n)\,\delta_{\rm in}(\vec{p}_1)\cdots\delta_{\rm in}(\vec{p}_n)\,\,, \label{eq:G_and_M_kernels-1} \\
\delta_{g,{\rm fwd}}[\delta_{\rm in}](\vec{k}) &= D_1K_{g,1}(k)\delta_{\rm in}(\vec{k}) \nonumber \\
&\;\;\;\; + \sum_{n=2}^{+\infty}\int_{\vec{p}_1,\dots\vec{p}_n}(2\pi)^{3}\delta^{(3)}_{\rm D}(\vec{k}-\vec{p}_{1\dots n})\,D_1^n\,
K_{g,n}(\vec{k};\vec{p}_1,\dots\vec{p}_n)\,\delta_{\rm in}(\vec{p}_1)\cdots\delta_{\rm in}(\vec{p}_n)\,\,. \label{eq:G_and_M_kernels-2} 
\end{align}
\end{subequations}

\begin{table}
\myfloatalign
\caption[.]{Summary of symbols used in the paper, and their meaning. Note that we use the subscript ``$g$'' 
for the quantities $\vec{\phi}_g$, $\vec{\mathcal{J}}_g$, 
$S_g[\vec{\phi}_g] $ and ${\cal M}_g^{ab}(\vec{k},\vec{k}')$ that appear in the computation of {the \emph{joint} likelihood.}} 
\label{tab:conventions}
\centering
\medskip
\begin{tabular}{ll}
\toprule
symbol & meaning \\
\midrule
$J_g$ & current associated with $\delta_g$ \\
$J$ & current associated with $\delta$ \\
$\delta_{\rm in}$ & initial matter field \\
$X_g$ & $\iu J_g$ \\
$X$ & $\iu J$ \\ 
\midrule 
$Z[J]$ & generating functional for $\delta$ \\
$Z[J_g,J]$ & generating functional for $\delta_g$ and $\delta$ \\
$W = W^{(2)} + W^{(3)} + \cdots$ & field-dependent part of $\ln Z$ \\
${\cal P}[\delta]$ & matter likelihood \\
${\cal P}[\delta_g,\delta]$ & joint likelihood \\
${\cal P}[\delta_g|\delta]$ & conditional likelihood \\
$\wp = \wp^{(2)} + \wp^{(3)} + \cdots$ & field-dependent part of $\ln{\cal P}$ \\
$\Delta\wp[\delta_g|\delta] = \Delta\wp^{(2)}[\delta_g|\delta] + \Delta\wp^{(3)}[\delta_g|\delta] + \cdots$ & 
corrections to $\ln{\cal P}[\delta_g|\delta]$ of \cite{Schmidt:2018bkr} \\ 
\midrule 
$\delta_{\rm fwd}[\delta_{\rm in}]$ & deterministic forward model for $\delta$ \\
$\delta_{g,{\rm fwd}}[\delta_{\rm in}]$ & deterministic forward model for $\delta_g$ \\
$\delta_{g,{\rm det}}[\delta] = \delta^{(1)}_{g,{\rm det}}[\delta] + \delta^{(2)}_{g,{\rm det}}[\delta] +\cdots$ 
& deterministic bias expansion for $\delta_g$ \\ 
\midrule 
$K_n(\vec{k};\vec{p}_1,\dots\vec{p}_n)$ & kernels for $\delta_{\rm fwd}[\delta_{\rm in}]$ \\
$K_{g,n}(\vec{k};\vec{p}_1,\dots\vec{p}_n)$ & kernels for $\delta_{g,{\rm fwd}}[\delta_{\rm in}]$ \\
$K_{g,{\rm det},n}(\vec{k};\vec{p}_1,\dots\vec{p}_n)$ & kernels for $\delta_{g,{\rm det}}[\delta]$ \\
$b(k) = K_{g,1}(k)/K_1(k)$ & scale-dependent linear bias \\ 
\midrule 
$\vec{\phi}$ & $(X,\delta_{\rm in})$ \\
$\vec{\phi}_g$ & $(X_g,X,\delta_{\rm in})$ \\
$\vec{\mathcal{J}}$ & $(\iu\delta,0)$ \\
$\vec{\mathcal{J}}_g$ & $(\iu\delta_g,\iu\delta,0)$ \\
$S[\vec{\phi}] = S^{(2)} + S_{\rm int} = S^{(2)} + S^{(3)}_{\rm int} + \cdots$ & action for ${\cal P}[\delta]$ \\
$S_g[\vec{\phi}_g] = S_g^{(2)} + S_{g,{\rm int}} = S_g^{(2)} + S_{g,{\rm int}}^{(3)} + \cdots$ & 
action for ${\cal P}[\delta_g,\delta]$ \\
${\cal M}^{ab}(\vec{k},\vec{k}')$ & $S^{(2)}[\vec{\phi}]
=\frac{1}{2}\int_{\vec{k},\vec{k}'}\phi^a(\vec{k}){\cal M}^{ab}(\vec{k},\vec{k}')\phi^b(\vec{k}')$ \\
${\cal M}_g^{ab}(\vec{k},\vec{k}')$ & $S^{(2)}_g[\vec{\phi}_g]
=\frac{1}{2}\int_{\vec{k},\vec{k}'}\phi_g^a(\vec{k}){\cal M}^{ab}_g(\vec{k},\vec{k}')\phi_g^b(\vec{k}')$ \\ 
\midrule 
$n_\delta$ & $k^3_{\rm NL}P_{\rm L}(k) = 2\pi^2(k/k_{\rm NL})^{n_\delta}$ \\
$k_{\eps_g}$ & ${{\sqrt{P^{\{0\}}_{\eps_g}/P_{\rm L}(k_{\eps_g})}=1}}$ \\ 
\bottomrule
\end{tabular}
\end{table}

\noindent These definitions become clearer with some examples: 
\begin{itemize}[leftmargin=*]
\item the SPT solution for matter at second order in perturbations is obtained by setting $K_1=1$ in \eq{G_and_M_kernels-1}, 
and $K_2 = F_2$, where $F_2$ is the familiar perturbation theory kernel 
(see, e.g., Appendix B of \cite{Desjacques:2016bnm} for a review);
\item consider a biased tracer defined by a second-order bias relation $\delta_g = b_1\delta + b_2\delta^2/2$, {where 
the matter field is evolved using SPT.} Then, $K_{g,1} = b_1$, while $K_{g,2}$ is \mbox{given by} 
\begin{equation}
\label{eq:K_2_example-A}
\frac{b_2}{2} + b_1F_2(\vec{p}_1,\vec{p}_2)\,\,;
\end{equation}
\item finally, consider a higher-derivative operator $\delta_g = b_{\nabla^2\delta}\nabla^2\delta$, 
again stopping at second order in perturbations. We then have\footnote{In the first lines of 
\eqsI{G_and_M_kernels} we have allowed for a generic dependence of $K_1$ and $K_{g,1}$ on $k$. 
Importantly, in absence of any preferred direction $K_1$ and $K_{g,1}$ can only be functions of $k^2$.}
\begin{equation}
\label{eq:K_2_example-B}
K_{g,1} = -b_{\nabla^2\delta}k^2\,\,, \quad K_{g,2} = -b_{\nabla^2\delta}k^2F_2(\vec{p}_1,\vec{p}_2)\,\,.
\end{equation}
\end{itemize}
From \eq{K_2_example-B} we can understand why we allowed for a dependence of $K_n$ and $K_{g,n}$ on $\vec{k}$. 
Indeed, the higher-derivative corrections to SPT kick in at the same order as 
$\delta_g\supset b_{\nabla^2\delta}\nabla^2\delta$: together with the $c^2_{\rm s}$ counterterm, 
that contributes to the matter power spectrum as 
$\sim c^2_{\rm s}(k^2/k^2_{\rm NL})P_{\rm L}(k)$, we can allow for higher-derivative corrections to the kernel $F_2$ as well, 
which contribute in the same way as \eq{K_2_example-B}. These higher-derivative terms are controlled by the nonlinear scale $k_{\rm NL}$, 
while we expect the nonlocality scale controlling $b_{\nabla^2\delta}$ to be at least of order of the halo Lagrangian radius $R(M_h)$. 
In this paper we will not explicitly need to know the form of these corrections to SPT, so we will not discuss them further. 

Two final definitions that will become useful in the remainder of the paper are that of the scale-dependent linear bias $b(k)$, \ie~ 
\begin{equation}
\label{eq:scale_dependent_linear_bias}
b(k)\equiv\frac{K_{g,1}(k)}{K_1(k)} = b_1-b_{\nabla^2\delta}k^2+\cdots\,\,,
\end{equation}
and that of the kernels $K_{g,{\rm det},n}$ for the deterministic bias expansion. These kernels are defined by the relation
\begin{equation}
\label{eq:deterministic_bias_expansion-A}
\delta_{g,{\rm fwd}}[\delta_{\rm in}] = \delta_{g,{\rm det}}\big[\delta_{{\rm fwd}}[\delta_{\rm in}]\big]\,\,.
\end{equation}
For example, we have that at second order 
\begin{equation}
\label{eq:deterministic_bias_expansion-B}
K_{g,{\rm det},2}(\vec{k};\vec{p}_1,\vec{p}_2) = K_{g,2}(\vec{k};\vec{p}_1,\vec{p}_2)-b(k)K_{2}(\vec{k};\vec{p}_1,\vec{p}_2)\,\,.
\end{equation}
A word of caution. A key object will be $\delta_{g,{{\rm det}}}[\delta]$. 
By this, we mean that we take a realization of the fully nonlinear matter field $\delta$ 
and we construct the deterministic bias expansion out of this field. 
For example, when going to second order in the bias expansion but at leading order in derivatives, in real space we construct
\begin{equation} 
\label{eq:tidal_field_squared} 
\delta_{g,{{\rm det}}}[\delta] = b_1\delta + \frac{b_2}{2}\delta^2 + b_{K^2} K^2[\delta]\,\,, 
\end{equation}
where $K^2 = K_{ij} K^{ij}$ and the tidal field $K_{ij}[\delta]$ is $(\partial_i\partial_j/\nabla^2 - \delta_{ij}/3)\delta$. 
$\delta$ is not necessarily given by the functional $\delta_{{\rm fwd}}[\delta_{\rm in}]$ of 
the initial conditions $\delta_{\rm in}$ as in \eq{G_and_M_kernels-1}. 

Finally, since this work relies heavily on functional methods, we lay out our conventions for functional derivatives and integrals. 
Functional integration over a field $\varphi$ is denoted by $\int{\cal D}\varphi$. 
Dirac delta functionals are denoted by $\delta^{(\infty)}_{\rm D}(\varphi-\chi)$, and are defined by 
\begin{equation}
\label{eq:dirac_delta_functional-A}
\int{\cal D}\chi\,\delta^{(\infty)}_{\rm D}(\varphi-\chi)\,{\cal F}[\chi] = {\cal F}[\varphi]\,\,,
\end{equation}
for some functional ${\cal F}[\varphi]$. Our convention for the functional derivatives in Fourier space is 
\begin{equation}
\label{eq:functional_derivatives-A}
\frac{\partial\varphi(\vec{k})}{\partial\varphi(\vec{k}')} = (2\pi)^3\delta^{(3)}_{\rm D}(\vec{k}+\vec{k}')\,\,.
\end{equation}
Notice that we use $\partial$, and not $\delta$, to denote a functional derivative, 
in order to avoid confusion with the matter field. Equation~\eqref{eq:functional_derivatives-A} tells us that 
\begin{equation}
\label{eq:functional_derivatives-B}
\frac{\partial}{\partial\varphi(\vec{k}')}\int_{\vec{k}}\varphi(\vec{k})\chi(-\vec{k}) 
= \frac{\partial}{\partial\varphi(\vec{k}')}\int_{\vec{k}}\varphi(-\vec{k})\chi(\vec{k}) = 
\int_{\vec{k}}(2\pi)^3\delta^{(3)}_{\rm D}(\vec{k}+\vec{k}')\,\chi(-\vec{k}) = \chi(\vec{k}')\,\,.
\end{equation}

We will often consider functional integrals and derivatives over multiple fields. 
To lighten the notation in some of the calculations 
(mostly those of Appendix~\ref{app:saddle_point_formulas}), 
we will use ``multi-indices'' $i,j,k,\dots$ that represent both momenta and ``internal'' indices. 
For example, we will write $\phi^i = \big(\varphi_1(\vec{k}),\varphi_2(\vec{k})\big)$. 
The Einstein summation convention will be employed, with no difference between upper and lower indices. 

The functional techniques used in this paper can be found in textbooks on quantum field theory. 
We have followed mainly the Chapter ``Quantum Field Theory: Functional Methods'' of \cite{ZinnJustin:2002ru}, 
while the discussion about the ``shell-by-shell integration'' 
and the scaling dimensions (Sections~\ref{subsec:stochastic_terms_and_tadpoles} and \ref{subsec:action_and_RG_scalings}), 
and the brief discussion of the Polchinski equation (Appendix~\ref{app:loops}), follow respectively 
\cite{Peskin:1995ev} and \cite{SkinnerLectures}. 

Finally, we will see in Section~\ref{subsec:action_and_RG_scalings} that the computation 
of the likelihood is similar to a field theory of three interacting fields in three spatial dimensions. 
The three fields will basically be the initial field $\delta_{\rm in}$, and the two currents $J_g$ and $J$ associated with 
the galaxy field and the nonlinear matter field, respectively. 
In Tab.~\ref{tab:feynman} we collect the symbols used for internal and external lines in Feynman diagrams.

\section{Setting up the functional integrals}
\label{sec:setting_up_the_integrals}

\noindent In this section we discuss how we can set up the integral for the matter and joint likelihoods. 
Let us assume that we have an expression 
for the generating functionals $Z[J]$ and $Z[J_g,J]$ (defined below in \eq{matter_likelihood-B} 
{and \eq{joint_likelihood-B}, respectively}). 
Then, using the integral representation of the Dirac delta functional, \ie\footnote{The constant ${\cal N}_{\delta^{(\infty)}}$ 
can always be reabsorbed in the normalization of the likelihoods. Therefore its precise value is irrelevant.} 
\begin{equation}
\label{eq:dirac_delta_functional-B}
\delta^{(\infty)}_{\rm D}(\varphi-\chi) = {\cal N}_{\delta^{(\infty)}}
\int{\cal D}X\,\eu^{\iu\int_{\vec{k}}X(\vec{k})(\varphi(-\vec{k})-\chi(-\vec{k}))}\,\,,
\end{equation}
and its analog for a ``doublet'' of fields $(\varphi_1,\varphi_2)$, 
we can obtain an expression for the likelihoods ${\cal P}[\delta]$ and ${\cal P}[\delta_g,\delta]$. More precisely, 
the matter likelihood is given by 
\begin{equation}
\label{eq:matter_likelihood-A}
{\cal P}[\delta] = {\cal N}_{\delta^{(\infty)}}\int{\cal D}X\,\eu^{\iu\int_{\vec{k}}X(\vec{k})\delta(-\vec{k})}\,Z[-\iu X]\,\,,
\end{equation}
while the joint likelihood is
\begin{equation}
\label{eq:joint_likelihood-A}
{\cal P}[\delta_g,\delta] = {\cal N}^2_{\delta^{(\infty)}}\int{\cal D}X_g\,{\cal D}X\,
\eu^{\iu\int_{\vec{k}}X_g(\vec{k})\delta_g(-\vec{k})}\,\eu^{\iu\int_{\vec{k}}X(\vec{k})\delta(-\vec{k})}\,Z[-\iu X_g,-\iu X]\,\,.
\end{equation}

In the next sections, we will derive expressions for both $Z[J]$ and $Z[J_g,J]$ 
as functional integrals over $\delta_{\rm in}$. Therefore, 
we will reduce the problem of computing the likelihood to the computation of two functional integrals: 
one over $X$, $\delta_{\rm in}$, and one over $X_g$, $X$, $\delta_{\rm in}$.

\subsection{Integral for the matter likelihood}
\label{subsec:matter}

\noindent {Generalizing the approach of} \cite{Carroll:2013oxa}, we can write the generating functional 
for the correlation functions of the matter field as the integral
\begin{equation}
\label{eq:matter_likelihood-B}
Z[J] = \int{\cal D}\delta_{\rm in}\,
\exp\left\{\int_{\vec{k}} \left(\frac{1}{2} P_{\eps_m}(k)J(\vec{k})J(-\vec{k})
+ J(\vec{k})\delta_{\rm fwd}[\delta_{\rm in}](-\vec{k}) \right)\right\}\,{\cal P}[\delta_{\rm in}]\,\,,
\end{equation}
where ${\cal P}[\delta_{\rm in}]$ is the likelihood of the initial density field,
{and $P_{\eps_m}(k)$ is the power spectrum of the noise in the matter density field; we will return to this new ingredient below.}
In the following, we assume that the likelihood for $\delta_{\rm in}$ is Gaussian (see 
Section~\ref{subsec:further_developments} for a discussion on how to go beyond this assumption), i.e.~we take 
\begin{equation}
\label{eq:initial_likelihood}
{\cal P}[\delta_{\rm in}] = \eu^{-\frac{1}{2}\int_{\vec{k}}\frac{\delta_{\rm in}(\vec{k})\delta_{\rm in}(-\vec{k})}{P_{\rm in}(k)}}\,\,.
\end{equation}
Notice that we have not included the normalization of this Gaussian. The reason is that such normalization is irrelevant if we 
just care about the contribution to the logarithm of ${\cal P}[\delta]$ 
that is dependent on the field $\delta$. The same goes for ${\cal P}[\delta_g,\delta]$. 
For the same reason, from now on we will drop the factors ${\cal N}_{\delta^{(\infty)}}$, ${\cal N}^2_{\delta^{(\infty)}}$ 
in \eqsII{matter_likelihood-A}{joint_likelihood-A}. {We will deal with them in full generality in Section~\ref{subsec:P_eps_g}.} 

Let us discuss \eq{matter_likelihood-B} in more detail: 
\begin{itemize}[leftmargin=*]
\item if we take functional derivatives of \eq{matter_likelihood-B} with 
respect to the current $J$ at $J=0$, the terms in the {exponent in curly brackets} 
that contain one power of the current $J$ and one or more powers of 
the initial field $\delta_{\rm in}$ reproduce exactly the known rules of perturbation 
theory for the computation of $n$-point functions of the nonlinear matter field \cite{Carroll:2013oxa};
\item in addition, we have a term with two powers of the current $J$. This term gives us the noise in the matter power spectrum. 
More precisely, if we drop all nonlinear and higher-derivative terms in $\delta_{\rm fwd}[\delta_{\rm in}](-\vec{k})$ 
(which allows us to carry out the resulting Gaussian integral exactly), we see that the power spectrum of the matter field is equal to 
\begin{equation}
\label{eq:linear_matter_power_spectrum}
\begin{split}
\braket{\delta(\vec{k})\delta(\vec{k}')}' 
&= \Bigg(\frac{\partial^2\ln(Z[J]/Z[0])}{\partial J(\vec{k})\partial J(\vec{k}')}\bigg|_{J=0}\Bigg)' 
= D^2_1 P_{\rm in}(k) + P_{\eps_m}(k) = P_{\rm L}(k) + P_{\eps_m}(k)\,\,.
\end{split}
\end{equation}
\end{itemize}
Let us stress that the noise term $\ln Z[J]\supset JJ$ (the ``$\supset$'' symbol meaning 
that $\ln Z[J]$ contains terms with two powers of $J$ in its functional Taylor series) 
would be present even if we had considered fully deterministic initial conditions. 
It arises from integrating out short-scale modes in order to arrive at 
a hydrodynamical description of the matter field on large scales \cite{Baumann:2010tm}. 
Because of matter and momentum conservation, the noise power spectrum scales as $P_{\eps_m}(k)\sim k^4$ on large scales. 

The stochasticity in the power spectrum is not the only one that is generated, actually: 
the EFT coefficients describing the corrections to SPT such as the speed of sound, for example, 
also gain some noise. These new terms are captured in our functional formalism by 
additional contributions to the {exponent in the curly brackets} of \eq{matter_likelihood-B}. 
For example, the leading stochasticity in $c^2_{\rm s}$ is captured by a term of order $J J \delta_{\rm in}$. Moreover, 
just as we have a stochastic contribution to the two-point function, 
we can have stochastic contributions to all $n$-point functions. These are captured 
by terms $\ln Z[J]\supset JJ{J}\cdots$, and in the terminology of the bias expansion of \cite{Desjacques:2016bnm} 
they correspond to higher-order correlation functions like $\braket{\eps_m\eps_m\eps_m}$.

\subsection{Integral for the joint likelihood}
\label{subsec:joint}

\noindent It is straightforward to extend the expressions of the previous section to the case of the joint likelihood. 
The expression for $Z[J_g,J]$ is 
\begin{equation}
\label{eq:joint_likelihood-B}
\begin{split}
Z[J_g,J] = \int{\cal D}\delta_{\rm in}\,&\eu^{\frac{1}{2}\int_{\vec{k}}P_{\eps_g}(k)J_g(\vec{k})J_g(-\vec{k})\, 
+\, \int_{\vec{k}}P_{\eps_g\eps_m}(k)J_g(\vec{k})J(-\vec{k})\, 
+\, \frac{1}{2}\int_{\vec{k}}P_{\eps_m}(k)J(\vec{k})J(-\vec{k})} \\ 
&\times\eu^{\int_{\vec{k}}J_g(\vec{k})\delta_{g,{\rm fwd}}[\delta_{\rm in}](-\vec{k})\, 
+\, \int_{\vec{k}}J(\vec{k})\delta_{\rm fwd}[\delta_{\rm in}](-\vec{k})}\,{\cal P}[\delta_{\rm in}]\,\,.
\end{split}
\end{equation}
Here, the noise terms are ({a}) the stochasticity for galaxies $P_{\eps_g}(k)\sim k^0$, 
({b}) the cross stochasticity between galaxies and matter $P_{\eps_g\eps_m}(k)\sim k^2$, 
and ({c}) the matter stochasticity $P_{\eps_m}(k)\sim k^4$, where the powers of $k$ 
given correspond to the leading term in the limit $k\to 0$. 
If we again drop nonlinear terms in $\delta_{g,{\rm fwd}}[\delta_{\rm in}]$ and $\delta_{\rm fwd}[\delta_{\rm in}]$ 
we recover the {expected} power spectra and cross spectrum, \ie~
\begin{subequations}
\label{eq:joint_spectra_with_noise}
\begin{alignat}{3}
\braket{\delta_g(\vec{k})\delta_g(\vec{k}')}' &= 
\Bigg(\frac{\partial^2\ln(Z[J_g,J]/Z[0,0])}{\partial J_g(\vec{k})\partial J_g(\vec{k}')}\bigg|_{J_g=0,J=0}\Bigg)' &&= 
b_1^2P_{\rm L}(k) + P_{\eps_g}(k)\,\,, \label{eq:joint_spectra_with_noise-1} \\
\braket{\delta_g(\vec{k})\delta(\vec{k}')}' &= 
\Bigg(\frac{\partial^2\ln(Z[J_g,J]/Z[0,0])}{\partial J_g(\vec{k})\partial J(\vec{k}')}\bigg|_{J_g=0,J=0}\Bigg)' &&= 
b_1P_{\rm L}(k) + P_{\eps_g\eps_m}(k)\,\,, \label{eq:joint_spectra_with_noise-2} \\
\braket{\delta(\vec{k})\delta(\vec{k}')}' &= 
\Bigg(\frac{\partial^2\ln(Z[J_g,J]/Z[0,0])}{\partial J(\vec{k})\partial J(\vec{k}')}\bigg|_{J_g=0,J=0}\Bigg)' &&= 
P_{\rm L}(k) + P_{\eps_m}(k)\,\,. \label{eq:joint_spectra_with_noise-3}
\end{alignat}
\end{subequations}

As in the case of the matter likelihood, we do not have only terms of the form 
$\ln Z[J_g,J]\supset J_gJ_g,\ J_gJ,\ JJ$, but also stochasticities in 
higher-order $n$-point functions. A three-point function $\braket{\eps_g\eps_g\eps_g}$ is captured by $\ln Z[J_g,J]\supset J_gJ_gJ_g$, 
$\braket{\eps_g\eps_g\eps_m}$ by $\ln Z[J_g,J]\supset J_gJ_gJ$, and so on. 
Then, we also have stochasticities in the bias coefficients. 
For example, the stochasticity in $b_1$ is captured by a term 
of order $J_g J_g \delta_{\rm in}$, 
while the stochasticities in $b_2$ or $b_{K^2}$ by terms 
of order $J_g J_g \delta_{\rm in}\delta_{\rm in}$. 
These will be discussed in more detail in Section~\ref{subsec:bispectrum_and_noise_in_LIMD_bias}. 
For a summary of these correspondences in the computation of ${\cal P}[\delta]$ 
and ${\cal P}[\delta_g,\delta]$, see Tab.~\ref{tab:summary_Z}.

\subsection{More about stochastic terms and tadpoles}
\label{subsec:stochastic_terms_and_tadpoles}

\noindent The stochastic contributions discussed in the previous section are generated 
when we coarse-grain the matter field to obtain a large-scale description of galaxy clustering, 
even if the initial conditions were deterministic (see e.g.~Section~2.10.3 of \cite{Desjacques:2016bnm} for a review). 
We will now show explicitly that they are generated also by loops of $\delta_{\rm in}$, 
which require us to include $P_{\eps_g}$, $P_{\eps_g\eps_m}$ and $P_{\eps_m}$ in \eqsII{matter_likelihood-B}{joint_likelihood-B} 
in order to remove the UV dependence of these loops. Moreover, we will show explicitly 
that these power spectra are analytic in $k^2$, with their leading-order terms being 
$P_{\eps_g}\sim k^0$, $P_{\eps_g\eps_m}\sim k^2$ and $P_{\eps_m}\sim k^4$. 
For simplicity, in the following we will focus only on \eq{joint_likelihood-B}: 
everything we will say translates straightforwardly to \eq{matter_likelihood-B} as well. 
We also emphasize that a discussion about the necessity of these counterterms with an analytic structure in $k^2$ has 
been discussed already in the literature. We refer, for example, to \cite{McDonald:2006mx,2014JCAP...08..056A} for more details 
(see e.g. Section~2.2 of \cite{2014JCAP...08..056A}): 
here we just want to show how this terms arise in the generating functional $Z[J_g,J]$. 

We use a ``Wilson-like'' approach of shell-by-shell integration, similarly to what is done, for example, in Chapter 12 of 
\cite{Peskin:1995ev} (see its Section 12.1). This will also allow us to introduce some 
concepts that will be used in the next section. The procedure is as follows: 
\begin{itemize}[leftmargin=*] 
\item the integral in ${\cal D}\delta_{\rm in}$ must be regularized. 
We do so by introducing a hard cutoff in Fourier space, \ie~we set all modes $\delta_{\rm in}(\vec{k})$ 
with $\abs{\vec{k}} > \Lambda$ to zero. The resulting field is called $\delta_{\rm in,\Lambda}$, 
and the measure ${\cal D}\delta_{\rm in}$ is 
consequently denoted by $[{\cal D}\delta_{\rm in}]_\Lambda$;
\item then, we split the integration in two: given a parameter $0\leq b < 1$, first we consider the modes 
$b\Lambda\leq k < \Lambda$, and then those with $k< b\Lambda$. 
The field $\delta_{\rm in,\Lambda}$ is then given by the sum of $\delta_{{\rm in},b\Lambda}$ and 
what we call, following \cite{Peskin:1995ev}, $\hat{\delta}_{\rm in}$, 
i.e.~the contribution to the field from modes in the shell $b\Lambda\leq k < \Lambda$. 
The measure $[{\cal D}\delta_{\rm in}]_\Lambda$ factorizes into $[{\cal D}\delta_{\rm in}]_{b\Lambda}$ 
times ${\cal D}\hat{\delta}_{\rm in}$;
\item finally, to avoid unnecessary clutter, we drop the subscripts $b\Lambda$ 
and denote {$\delta_{{\rm in},b\Lambda} \to \delta_{\rm in}$}. 
\end{itemize}

For this calculation, it is sufficient to consider currents $J_g$ and $J$ that only consist of long-wavelength modes, 
\ie~we can put $\hat{J}_g=0$ and $\hat{J}=0$. Clearly, we will need to reevaluate this assumption later, 
since when we compute the likelihood we are actually integrating over the currents as well. 
This is done in Section~\ref{subsec:loops}. Then, 
for vanishing $P_{\eps_g}$, $P_{\eps_g\eps_m}$ and $P_{\eps_m}$, \eq{joint_likelihood-B} becomes 
\begin{equation}
\label{eq:shell_by_shell}
\begin{split}
Z[J_g,J] = \int{\cal D}\delta_{\rm in}\,{\cal D}\hat{\delta}_{\rm in}\,&
\eu^{\int_{\vec{k}}J_g(\vec{k})\delta_{g,{\rm fwd}}[\delta_{\rm in}+\hat{\delta}_{\rm in}](-\vec{k})\, 
+\, \int_{\vec{k}}J(\vec{k})\delta_{\rm fwd}[\delta_{\rm in}+\hat{\delta}_{\rm in}](-\vec{k})} \\
&\times\eu^{-\frac{1}{2}\int_{\vec{k}}\frac{\delta_{\rm in}(\vec{k})\delta_{\rm in}(-\vec{k})}{P_{\rm in}(k)}}\,
\eu^{-\frac{1}{2}\int_{\vec{k}}\frac{\hat{\delta}_{\rm in}(\vec{k})\hat{\delta}_{\rm in}(-\vec{k})}{P_{\rm in}(k)}}\,\,,
\end{split}
\end{equation}
where we have used the fact that $\delta_{\rm in}(\vk)\hat{\delta}_{\rm in}(-\vk) = 0$ for all $\vk$, 
since their respective support does not overlap.
Notice that, differently from what usually happens in quantum field theory, 
the currents are not coupled linearly to $\delta_{\rm in}$. Therefore, in the above equations we cannot substitute 
$\delta_{g,\rm fwd}[\delta_{\rm in}+\hat{\delta}_{\rm in}]$ and $\delta_{\rm fwd}[\delta_{\rm in}+\hat{\delta}_{\rm in}]$ with 
$\delta_{g,\rm fwd}[\delta_{\rm in}]$ and $\delta_{\rm fwd}[\delta_{\rm in}]$.\footnote{Of course, 
we can do this substitution in the terms 
that are linear in the initial matter field.}

\begin{table}
\myfloatalign
\caption[.]{Internal and external lines of Feynman diagrams for $Z[J]$ and $Z[J_g,J]$. 
Notice that only the initial density field can enter in loops 
when we compute the generating functionals, since both $Z[J]$ and $Z[J_g,J]$ are given by a path integral over $\delta_{\rm in}$. 
Conversely, only the two currents can be external lines. This will change when we move to the computation of the likelihoods. 
Thick lines will denote hard modes in the loops (the function $\Theta_b(k)$ 
restricting $k$ to lie within the shell $b\Lambda\leq k < \Lambda$). 
Finally, we stress that the conventions for our diagrams are slightly different from those 
commonly employed to compute polyspectra in perturbation theory, 
since external lines carry the currents $J_g$ and $J$. 
This is simply because we are interested in computing the generating functionals, whose derivatives give the 
$n$-point correlation functions. \mbox{For example, compare \eq{matter_stochasticity_loop-A} 
with Eqs.~(B.18), (B.25) of \cite{Desjacques:2016bnm}.}} 
\label{tab:feynman}
\centering
\medskip
\begin{tabular}{ll}
\toprule
symbol & meaning \\
\midrule
\raisebox{-0.0cm}{\includegraphicsbox[scale=0.1,trim={1.5cm 8cm 1.5cm 8cm},clip]{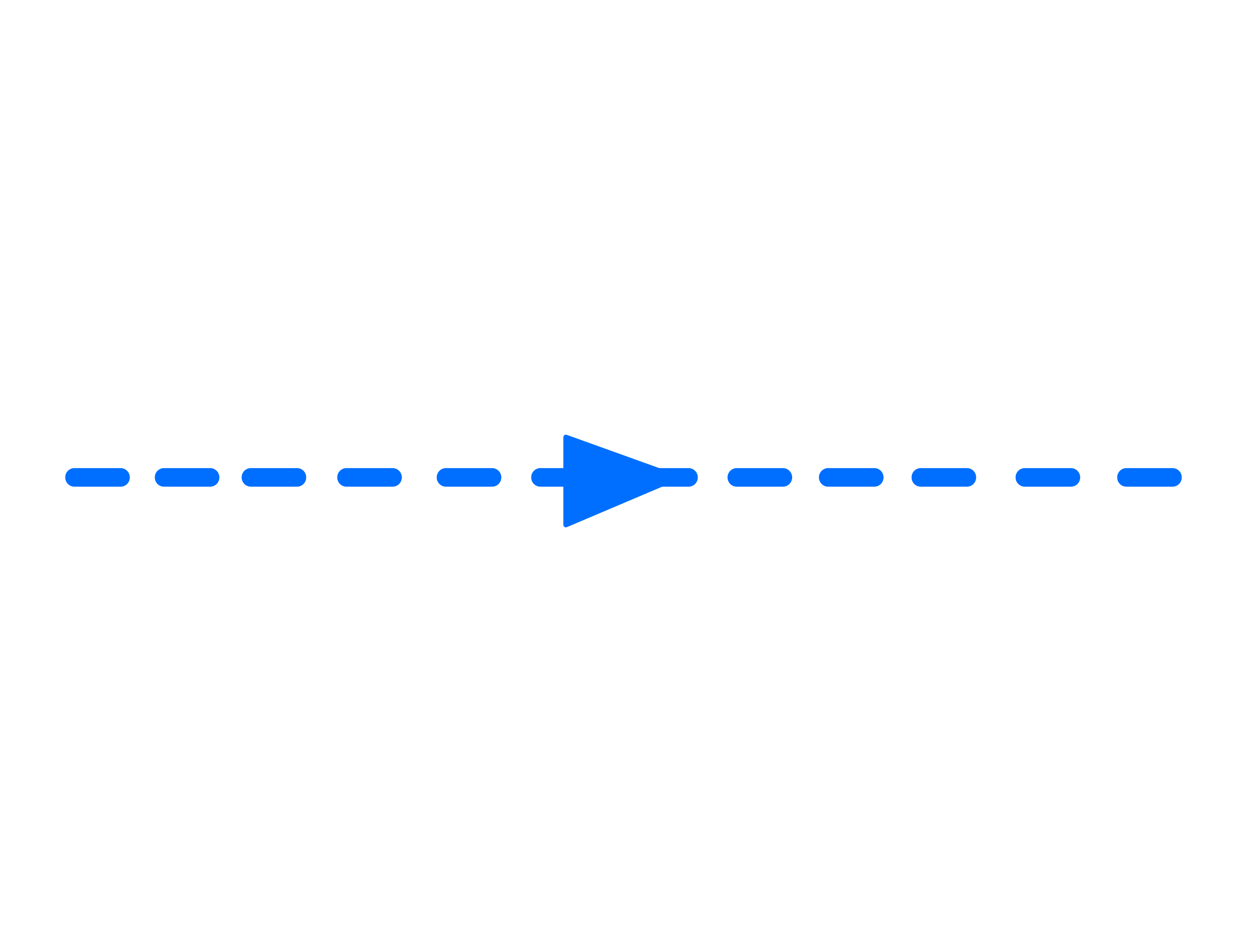}} & $J_g(\vec{k})$ \\
\raisebox{-0.0cm}{\includegraphicsbox[scale=0.1,trim={1.5cm 8cm 1.5cm 8cm},clip]{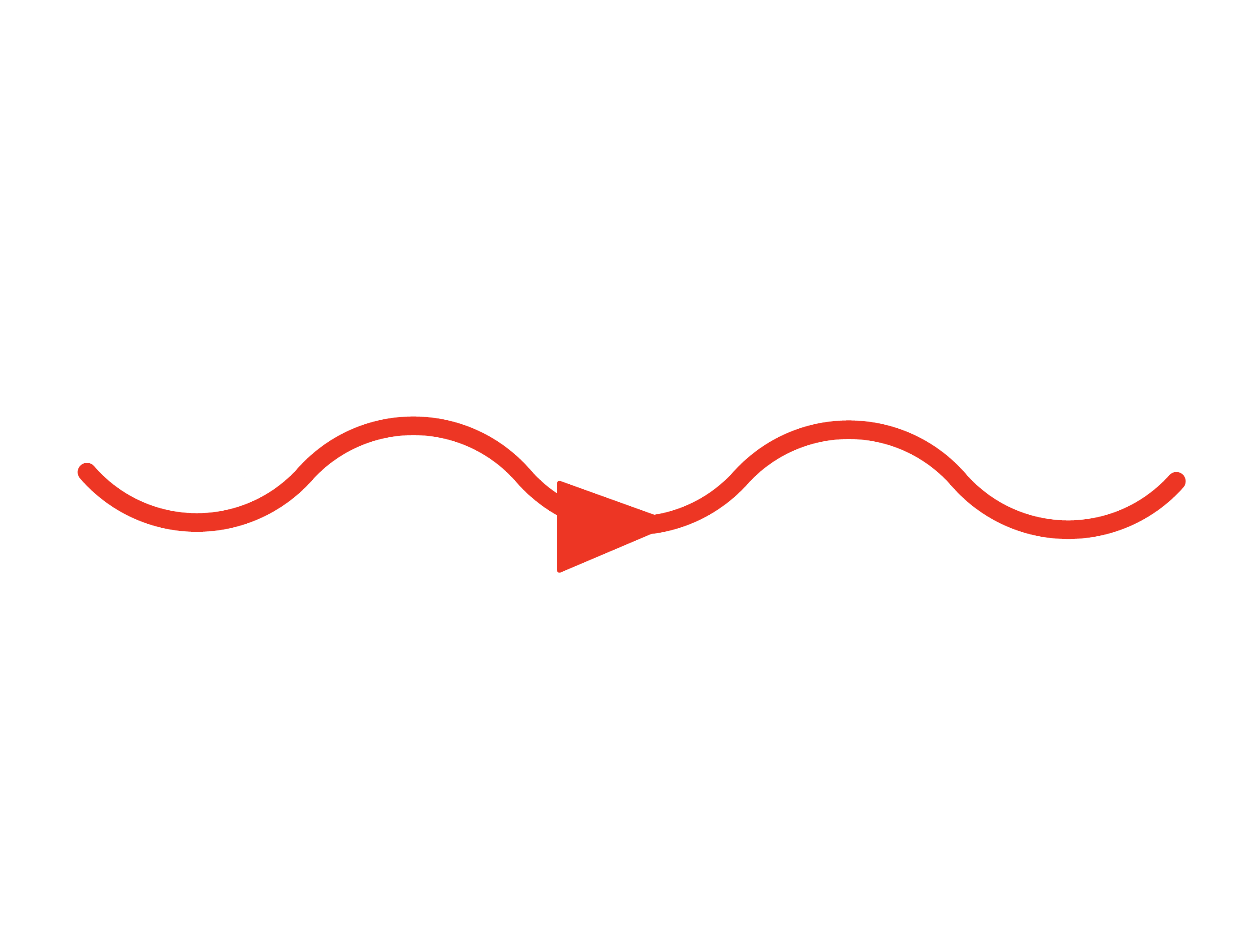}} & $J(\vec{k})$ \\
\raisebox{-0.0cm}{\includegraphicsbox[scale=0.1,trim={1.5cm 8cm 1.5cm 8cm},clip]{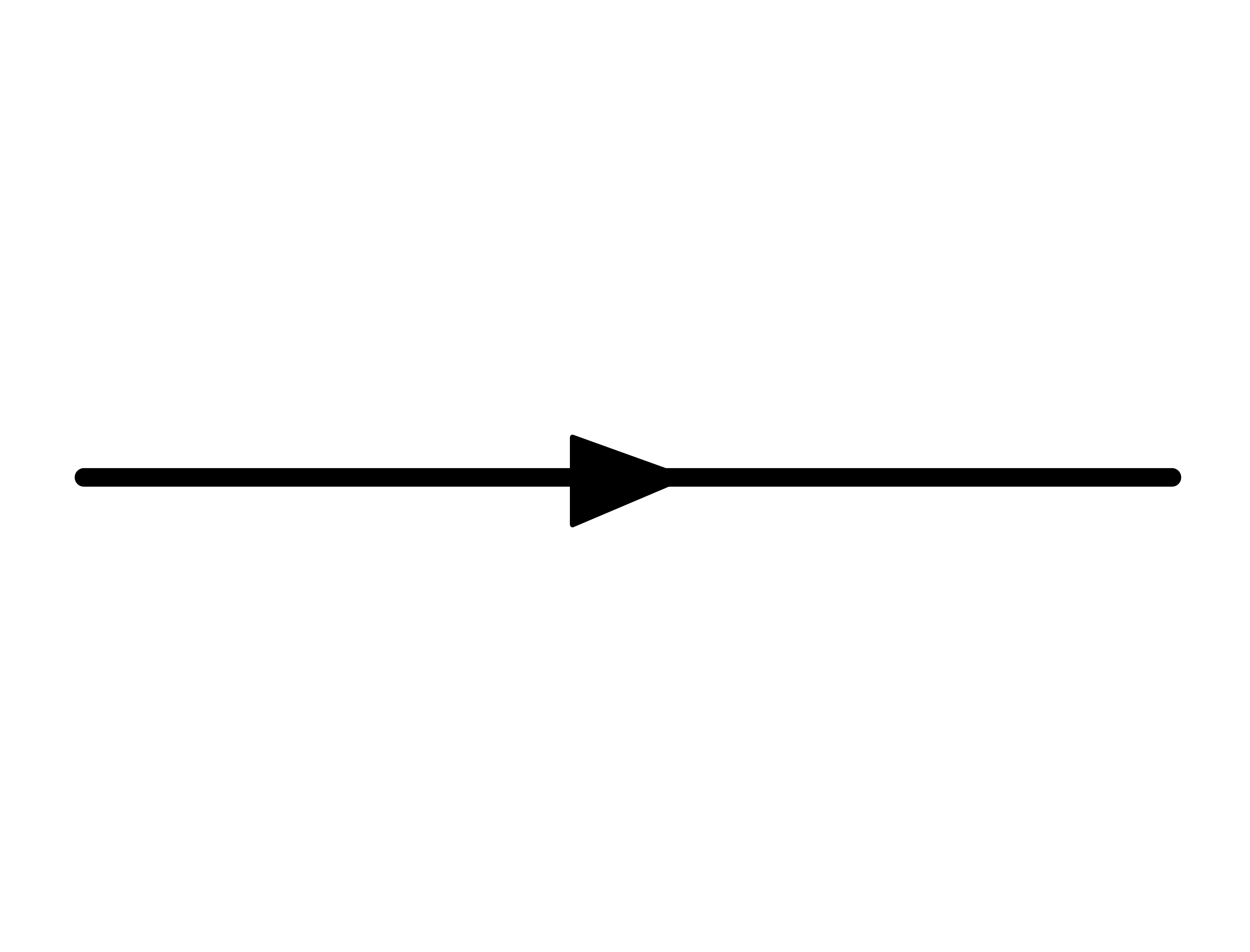}} & $\delta_{\rm in}(\vec{k})$ \\ 
\raisebox{-0.0cm}{\includegraphicsbox[scale=0.1,trim={1.5cm 8cm 1.5cm 8cm},clip]{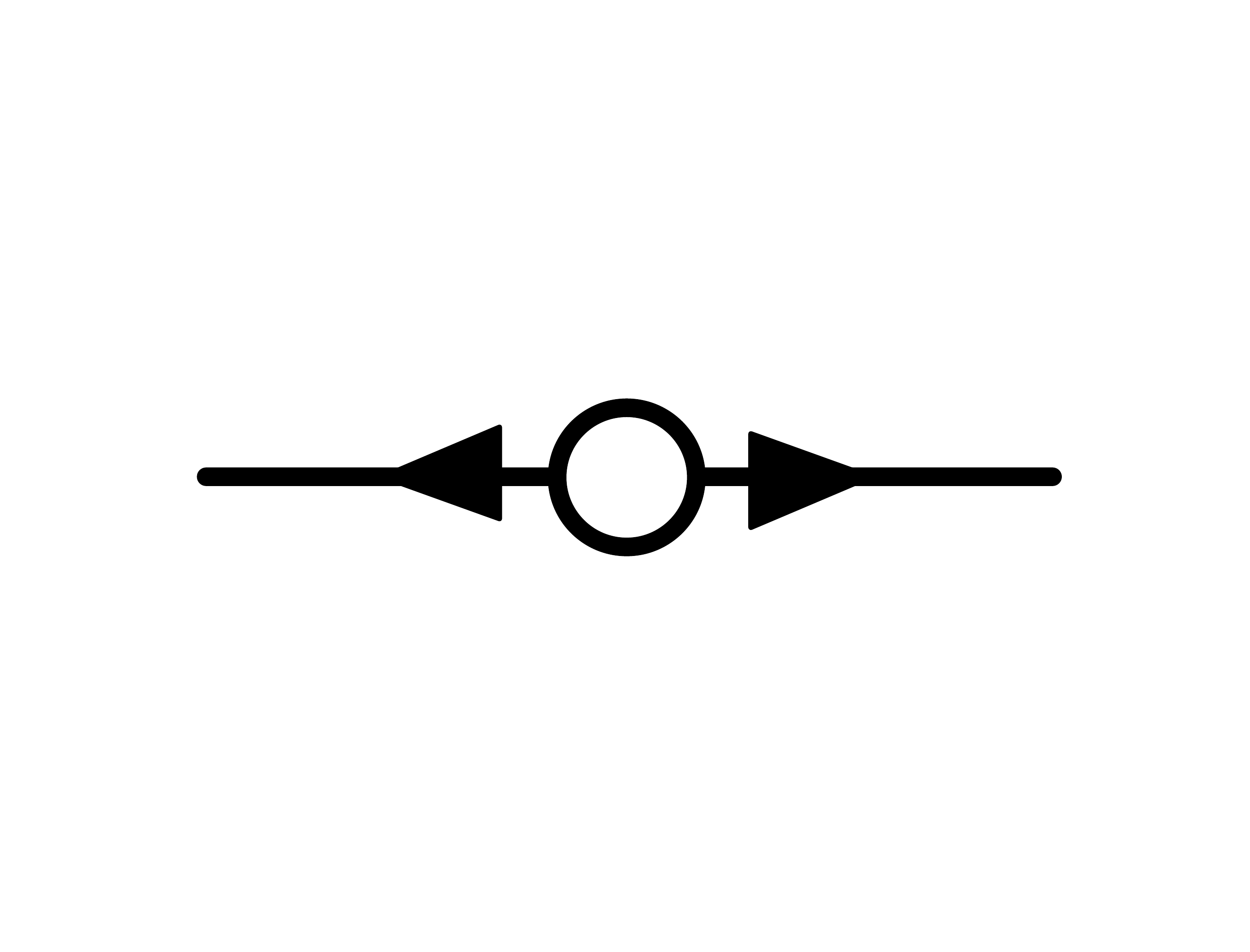}} 
& $\braket{\delta_{\rm in}(\vec{k})\delta_{\rm in}(\vec{k}')}' = P_{\rm in}(k)$ \\
\midrule
\raisebox{-0.0cm}{\includegraphicsbox[scale=0.1,trim={1.5cm 8cm 1.5cm 8cm},clip]{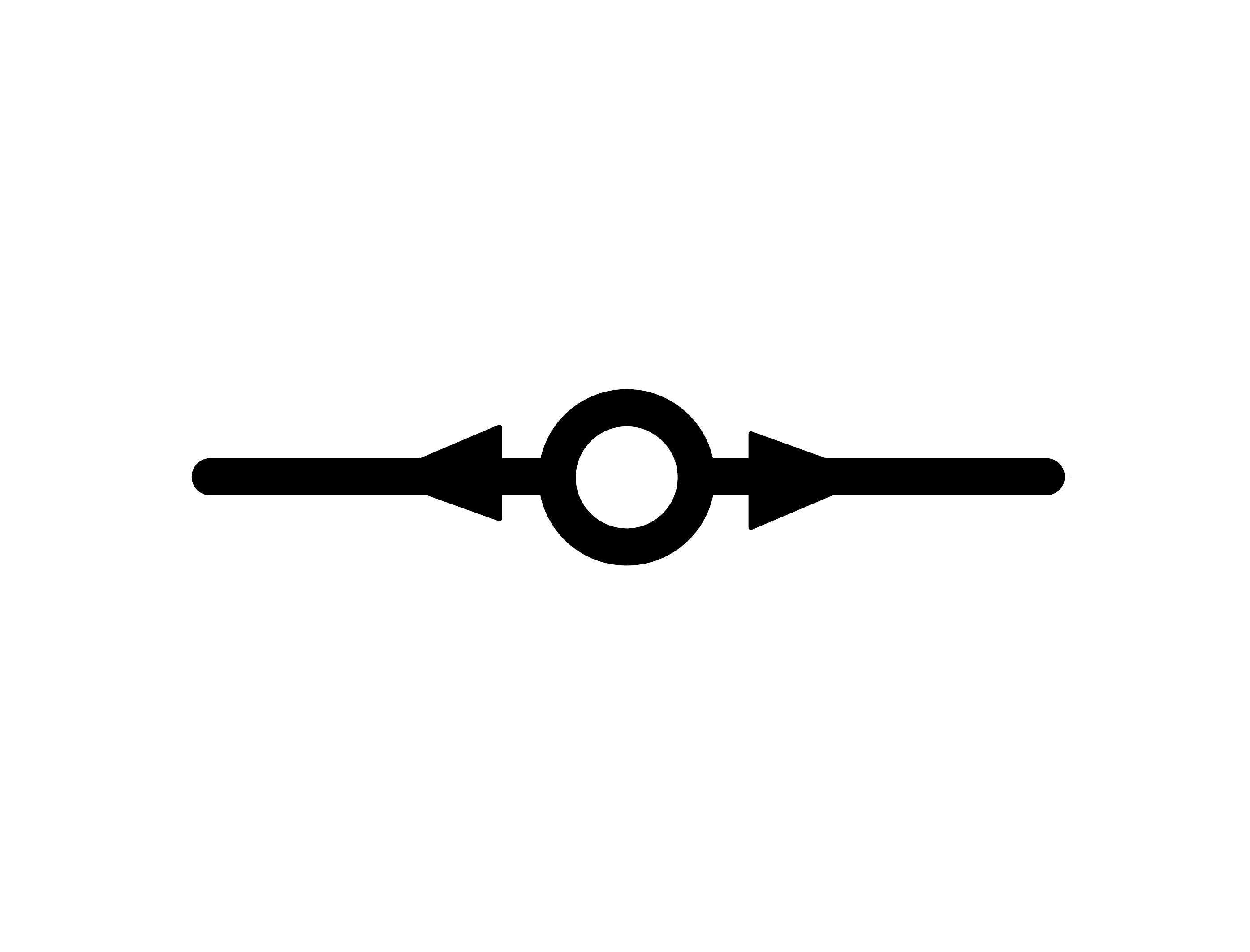}} 
& $\braket{\hat{\delta}_{\rm in}(\vec{k})\hat{\delta}_{\rm in}(\vec{k}')}' = P_{\rm in}(k)\Theta_b(k)$ \\
\bottomrule
\end{tabular}
\end{table}

Expanding the exponentials in the first line of \eq{shell_by_shell}, 
we can now perturbatively do the integrals over $\hat{\delta}_{\rm in}$ 
(see Tab.~\ref{tab:feynman} for the list of symbols used in Feynman diagrams). 
For simplicity, we stop at second order in perturbations and zeroth order in derivatives in \eqsI{G_and_M_kernels}. 
Then, $\delta_{\rm fwd}[\delta_{\rm in}]$ is given by 
\begin{equation}
\label{eq:delta_fwd_recall}
\delta_{{\rm fwd}}[\delta_{\rm in}](\vec{k}) = D_1\delta_{\rm in}(\vec{k}) + 
\int_{\vec{p}_1,\vec{p}_2}(2\pi)^{3}\delta^{(3)}_{\rm D}(\vec{k}-\vec{p}_{12})\,D_1^2\,
F_2(\vec{p}_1,\vec{p}_2)\,\delta_{\rm in}(\vec{p}_1)\delta_{\rm in}(\vec{p}_2)\,\,,
\end{equation}
{where} we used that $K_1=1$ and $K_2 = F_2$ at leading order in derivatives. Let us then consider the diagram 
\begin{equation}
\label{eq:matter_stochasticity_loop-A}
\begin{split}
&\left(\raisebox{-0.0cm}{\includegraphicsbox[scale=0.25,trim={1.5cm 6cm 1.5cm 6cm},clip]{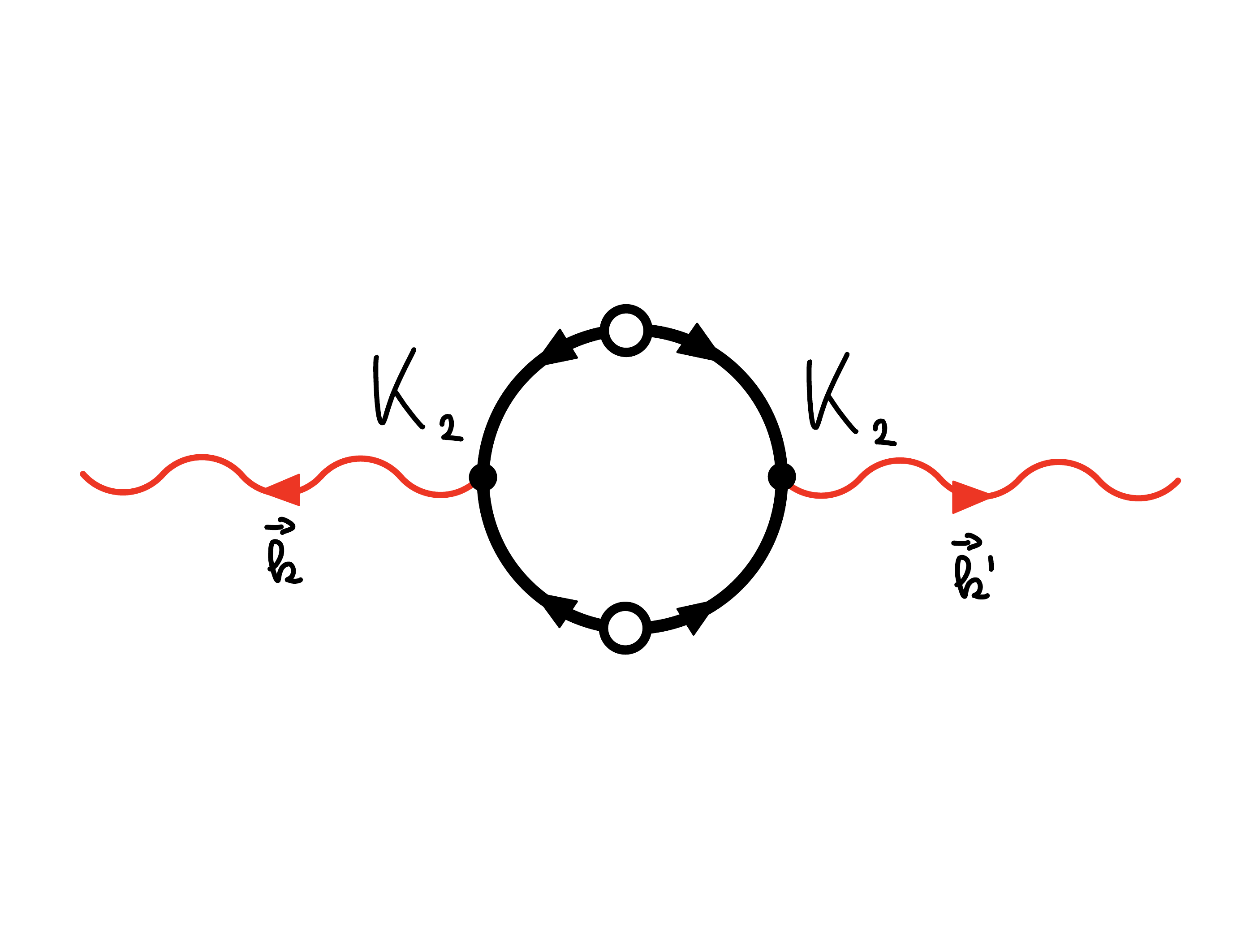}}\right)' = \\
&\;\;\;\;J(\vec{k})J(\vec{k}')\int_{\vec{p}}F_2(\vec{p}_-,-\vec{p}_+)F_2(-\vec{p}_-,\vec{p}_- - \vec{k}')
\Theta_{{b}}(\abs{\vec{p}_-})P^2_{\rm L}(\abs{\vec{p}_-})\,\,,
\end{split}
\end{equation}
where $\vec{p}_-$ and $\vec{p}_+$ are defined as 
\begin{equation}
\label{eq:matter_stochasticity_loop-B}
\vec{p}_{\pm} = \vec{p} \pm \frac{\vec{k}}{2}
\end{equation}
and the Heaviside theta-like function $\Theta_b = \Theta^2_b$ restricts $\vec{p}_-$ to lie within 
the shell $b\Lambda\leq\abs{\vec{p}_-}<\Lambda$.\footnote{Also notice 
that the arrows reproduce the momentum conservation described by the vertices of \eqsI{G_and_M_kernels}.} 
This diagram appears in both $Z[J]$ and $Z[J_g,J]$: {it is UV{-}sensitive and forces us to introduce} 
a stochastic contribution to the power spectrum of matter $P_{\eps_m}\sim k^4$ in the $k/p\to 0$ limit 
(thanks to the double-softness of the $F_2$ kernel \cite{Abolhasani:2015mra}, 
reflecting matter and momentum conservation {for the short modes}). 

A similar diagram shows up in the generating functional $Z[J_g,J]$. More precisely, the diagram 
\begin{equation}
\label{eq:galaxy_stochasticity_loop}
\begin{split}
&\left(\raisebox{-0.0cm}{\includegraphicsbox[scale=0.25,trim={1.5cm 6cm 1.5cm 6cm},clip]{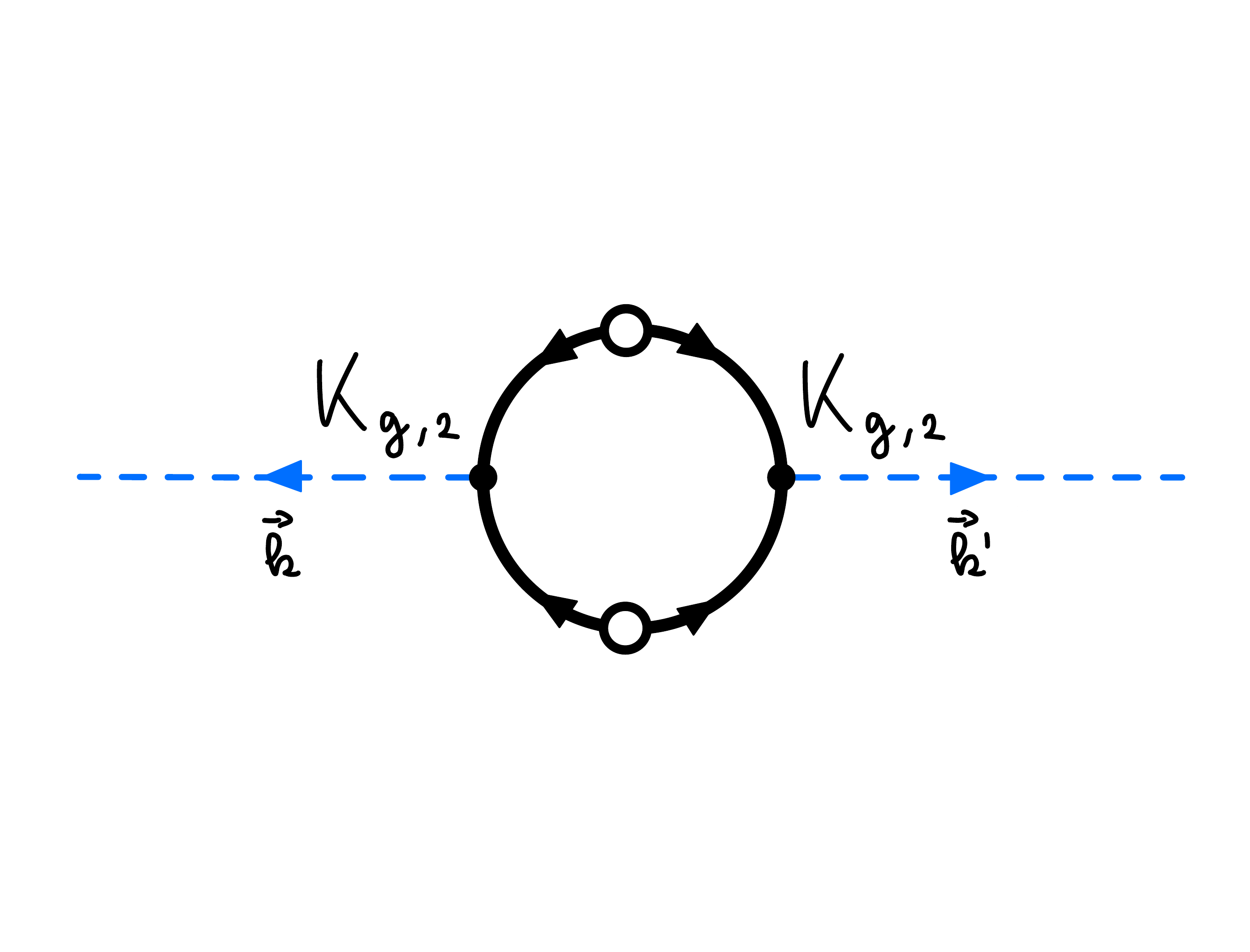}}\right)' = \\
&\;\;\;\;
J_g(\vec{k})J_g(\vec{k}')\int_{\vec{p}}K_{g,2}(\vec{p}_-,-\vec{p}_+) K_{g,2}(-\vec{p}_-,\vec{p}_- - \vec{k}')
\Theta_{{b}}(\abs{\vec{p}_-})P^2_{\rm L}(\abs{\vec{p}_-})
\end{split}
\end{equation}
gives rise to a stochastic correction to the galaxy power spectrum. The difference with \eq{matter_stochasticity_loop-A} 
is that the kernel $K_{g,2}$ is not double-soft. To see this, 
just consider the contribution $\delta_g\supset b_2\delta^2/2$ in the deterministic bias expansion: 
from \eq{K_2_example-A} we have that $K_{g,2}$ is simply equal to $b_2/2$ in the squeezed limit $k/p\to0$. 
Therefore, we find {that a constribution $P_{\eps_g}\sim k^0$ is needed to absorb the contribution from this UV-sensitive diagram}. 

The mixed loop $\sim K_2 K_{g,2}$ generates the cross stochasticity $P_{\eps_g\eps_m}$. 
More precisely, its diagram is the same as that of \eq{galaxy_stochasticity_loop}, where 
one external $J_g$ leg is replaced by $J$, and one of the $K_{g,2}$ kernels is replaced by $K_2$ (equal to 
$F_2$ at leading order in derivatives). By expanding in $k/p$ we see that to absorb the UV-sensitivity 
of this loop we need a counterterm $\smash{\int_{\vec{k}}P_{\eps_g\eps_m}(k)J_g(\vec{k})J(-\vec{k})}$ in 
$\ln Z[J_g,J]$, where the expansion of $\smash{P_{\eps_g\eps_m}(k)}$ in powers of $\smash{k^{2}}$ starts at $\smash{{\cal O}(k^2)}$. 
Moreover, loops will also generate many other terms that we did not include in 
\eqsII{matter_likelihood-B}{joint_likelihood-B}. Among these, there are the non-Gaussian corrections 
to the stochasticities (encoded in terms with more than two 
powers of $J_g$ {or} $J$), and also the stochastic corrections to the bias coefficients 
(encoded, for example, in terms with two powers of the currents and one of the initial matter field), 
as mentioned in the previous section. 
These will be discussed in more \mbox{detail in Sections~\ref{sec:calculation_of_likelihood-nongaussian_stochasticities}, 
\ref{sec:discussion} and \ref{subsec:bayesian_forward_modeling}.} 

Loops of the initial matter field will generate also tadpoles, besides the stochastic terms 
(see also \cite{2014JCAP...08..056A} for a discussion). For example, let us focus on $Z[J_g,J]$ and consider the diagram 
\begin{equation}
\label{eq:galaxy_tadpole_loop}
\begin{split}
\left(\raisebox{-0.0cm}{\includegraphicsbox[scale=0.25,trim={5.5cm 6cm 5.5cm 6cm},clip]{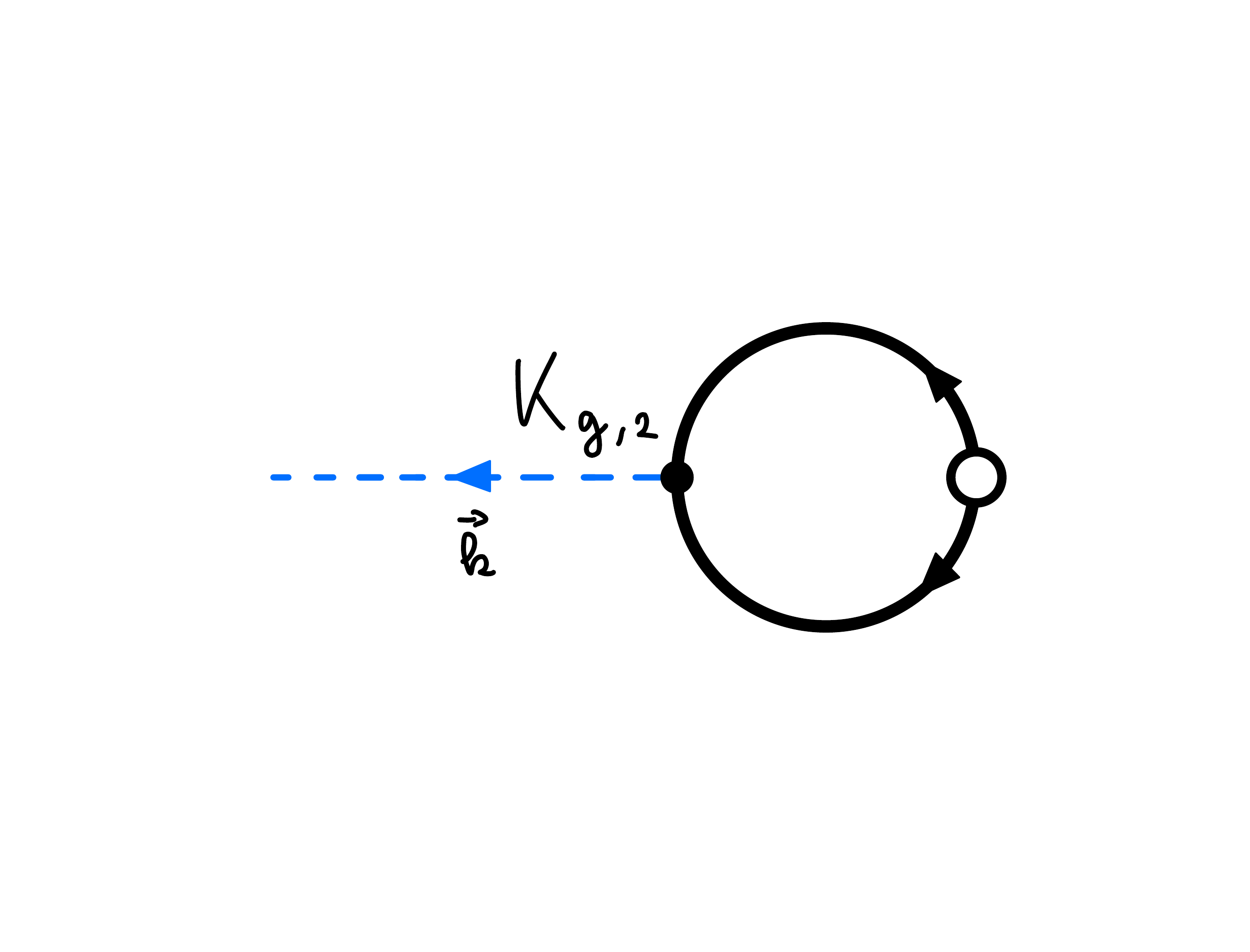}}\right)' &= 
J_g(\vec{k})\int_{\vec{p}}K_{g,2}(\vec{p},-\vec{p})\Theta_{{b}}(p)P_{\rm L}(p)\,\,.
\end{split}
\end{equation}
This diagram generates a linear term in the logarithm of $Z[J_g,J]$, i.e.~a tadpole. 
What happens if we consider the matter-only case? The diagram is the same, 
with the only difference that the kernel controlling it is $F_2$ instead of $K_{g,2}$. 
{Then, because of matter and momentum conservation (which implies $F_2(\vec{p},-\vec{p})=0\,\,\forall\,\vec{p}$), 
we do not generate any tadpole term in $Z[J]$.} 
One might wonder if some term linear in $J$ can be generated in $Z[J_g,J]$. 
This cannot happen, again as a consequence of matter and momentum 
conservation for the matter kernels (we have checked this only at one-loop order, 
but we expect it to be true at all loops).\footnote{As we will see in more detail in Section~\ref{subsec:loops}, 
once we integrate over the currents to arrive at the likelihood it is possible for loops 
{of $X_g = \iu J_g$ or $X = \iu J$} to generate 
a term with one external $X$ leg in the exponent of the integrand for ${\cal P}[\delta_g,\delta]$, cf.~\eq{joint_likelihood-A}. 
However, matter and momentum conservation guarantees that the coefficient 
of this term is always zero. The same happens for {loops of $X = \iu J$} in ${\cal P}[\delta]$.} 

How do we get rid of a tadpole for $\delta_g$? The answer is clear once we look at 
\eq{joint_likelihood-A}: we can always redefine the field $\delta_g$ to reabsorb 
any term linear in $X_g {= \iu J_g}$. More precisely, assume that we have a term 
\begin{equation}
\label{eq:tadpole_removal-A}
\ln Z[J_g,J]\supset\int_{\vec{k}}J_g(\vec{k})v(-\vec{k})\,\,,
\end{equation}
where $v(\vec{k})\propto\delta^{(3)}(\vec{k})$ because of translational and rotational symmetry. 
Then, in \eq{joint_likelihood-A} we get a contribution $-\iu\int_{\vec{k}}X_g(\vec{k})v(-\vec{k})$ in the exponential. 
Redefining $\delta_g\to\delta_g+v$ we can get rid of this contribution. 

Finally, since no tadpole for matter is generated, neither in \eq{matter_likelihood-A} nor in \eq{joint_likelihood-A}, 
we conclude that loop corrections will not spoil the equality between the two matter fields $\delta$ appearing in these two likelihoods.

\subsection{Actions and dimensional analysis}
\label{subsec:action_and_RG_scalings}

\noindent The manipulations of the previous sections suggest that we should be able to reduce 
the problem of computing the likelihood ${\cal P}[\delta_g,\delta]$ to that of a {Euclidean} 
field theory in three dimensions. 
Let us see how this works. We start from the matter likelihood: without loss of generality, we can write it as 
\begin{equation}
\label{eq:action_matter-A}
{\cal P}[\delta] = \int{\cal D}X\,{\cal D}\delta_{\rm in}\,
\eu^{\int_{\vec{k}}\vec{\phi}(\vec{k})\cdot\vec{\mathcal{J}}(-\vec{k})-S[\vec{\phi}]}\,\,,
\end{equation}
where $\vec{\phi}$ and $\vec{\mathcal{J}}$ are given by 
\begin{equation}
\label{eq:action_matter-B}
\vec{\phi} = (X,\delta_{\rm in})\,\,,\quad\vec{\mathcal{J}} = (\iu\delta,0)\,\,.
\end{equation}
From this, we see that indeed the calculation of ${\cal P}[\delta]$ 
amounts to computing the correlation functions for the three-dimensional field theory described by 
the action $S[\vec{\phi}]$, with only the field $X$ appearing in the external legs. 
What is the form of the action? We can obtain it from \eqsII{matter_likelihood-A}{matter_likelihood-B}. 
First, we write it as a quadratic part plus an interaction part, \ie~{(we use Einstein's summation 
convention on the ``internal'' indices $a,b$, without distinction between upper and lower indices)} 
\begin{subequations}
\label{eq:action_matter-C}
\begin{align}
S[\vec{\phi}] &= S^{(2)}[\vec{\phi}] + S_{\rm int}[\vec{\phi}]\,\,, 
\label{eq:action_matter-C-1} \\
S^{(2)}[\vec{\phi}] &= \frac{1}{2}\int_{\vec{k},\vec{k}'}
\phi^a(\vec{k}){\cal M}^{ab}(\vec{k},\vec{k}')\phi^b(\vec{k}')\,\,, \label{eq:action_matter-C-2} \\
S_{\rm int}[\vec{\phi}] &= S^{(3)}_{\rm int}[\vec{\phi}]+\cdots\,\,, \label{eq:action_matter-C-3} 
\end{align}
\end{subequations}
with the matrix ${\cal M}$ given by (recalling that $K_1=1$ at zeroth order in derivatives) 
\begin{equation}
\label{eq:kinetic_matrix_for_matter}
{\cal M}(\vec{k},\vec{k}') = (2\pi)^3\delta^{(3)}(\vec{k}+\vec{k}')
\begin{pmatrix}
P_{\eps_m}(k) & \iu K_1(k)D_1 \\
\iu K_1(k)D_1 & P^{-1}_{\rm in}(k)
\end{pmatrix}\,\,.
\end{equation}
Then, we can discuss what is the form of $S_{\rm int}$. To do this, we need to recall what are the terms entering 
in the exponential of \eq{matter_likelihood-B} and put $J=-\iu X$. For example, 
{a term $S_{\rm int}\supset X\delta_{\rm in}\delta_{\rm in}$ 
gives us the deterministic evolution at second order in perturbations,} a term $S_{\rm int}\supset XXX$ corresponds 
to a three-point function $\braket{\eps_m\eps_m\eps_m}$, 
a term $S_{\rm int}\supset XX\delta_{\rm in}$ to the stochasticity in $c^2_{\rm s}$, and so on.
This is summarized in Tab.~\ref{tab:summary_Z}. 

We can then move to the joint likelihood. We write it as 
\begin{equation}
\label{eq:action_joint-A}
{\cal P}[\delta_g,\delta] = \int{\cal D}X_g\,{\cal D}X\,{\cal D}\delta_{\rm in}\,
\eu^{\int_{\vec{k}}\vec{\phi}_g(\vec{k})\cdot\vec{\mathcal{J}}_g(-\vec{k})-S_g[\vec{\phi}_g]}\,\,.
\end{equation}
The fields $\vec{\phi}_g$ and $\vec{\mathcal{J}}_g$, and the action $S_g[\vec{\phi}_g]$, are now given by 
(we warn the reader that we use a subscript ``$g$'' to denote the field ``multiplet'' and the action for the \emph{joint} likelihood) 
\begin{subequations}
\label{eq:action_joint-B-temp}
\begin{align}
\vec{\phi}_g &= (X_g,X,\delta_{\rm in})\,\,, \label{eq:action_joint-B-temp-1} \\ 
\vec{\mathcal{J}}_g &= (\iu\delta_g,\iu\delta,0)\,\,, \label{eq:action_joint-B-temp-2} 
\end{align}
\end{subequations}
and 
\begin{subequations}
\label{eq:action_joint-B}
\begin{align}
S_g[\vec{\phi}_g] &= S_g^{(2)}[\vec{\phi}_g] + S_{g,\rm int}[\vec{\phi}_g]\,\,, \label{eq:action_joint-B-1} \\
S_g^{(2)}[\vec{\phi}_g] &= 
\frac{1}{2}\int_{\vec{k},\vec{k}'}\phi^a_g(\vec{k}){\cal M}^{ab}_g(\vec{k},\vec{k}')\phi^b_g(\vec{k}')\,\,, 
\label{eq:action_joint-B-2} \\
S_{g,{\rm int}}[\vec{\phi}_g] &= S_{g,{\rm int}}^{(3)}[\vec{\phi}_g]+\cdots \,\,, \label{eq:action_joint-B-3}
\end{align}
\end{subequations}
where ${\cal M}_g$ is equal to 
\begin{equation}
\label{eq:action_joint-C}
{\cal M}_g(\vec{k},\vec{k}') = (2\pi)^3\delta^{(3)}(\vec{k}+\vec{k}')
\begin{pmatrix}
P_{\eps_g}(k) & P_{\eps_g\eps_m}(k) & \iu K_{g,1}(k) D_1 \\
P_{\eps_g\eps_m}(k) & P_{\eps_m}(k) & \iu K_1(k)D_1 \\
\iu K_{g,1}(k) D_1 & \iu K_1(k)D_1 & P^{-1}_{\rm in}(k)
\end{pmatrix}\,\,,
\end{equation}
where $K_{g,1} = b_1$ and $K_1=1$ at zeroth order in derivatives. 
As in the matter-only case, we can discuss the form of $S_{g,{\rm int}}$. 
{The deterministic evolution of the galaxy field at second and higher order in perturbations 
is captured by $S_{g,{\rm int}}\supset X_g\delta_{\rm in}\delta_{\rm in}\cdots$.} 
Then, similarly to what happens to $S_{\rm int}$ (since the action for the matter likelihood is part of 
the action for the joint likelihood), we can have terms involving powers of $\delta_{\rm in}$ 
together with more than one power of $X$ {and terms $S_{g,{\rm int}}\supset XXX\cdots$}: 
these represent {the stochastic corrections to the EFT coefficients 
for the total matter field and the non-Gaussianity of $\eps_m$}. 
{Moreover, we can now} have terms involving 
{more than one power of} $X_g$: terms of the form $S_{g,{\rm int}}\supset X_gX_gX_g\cdots$ 
represent the non-Gaussianity of $\eps_g$; mixed terms such as $S_{g,{\rm int}}\supset X_g X_g X$ which represents 
a three-point function $\braket{\eps_g\eps_g\eps_m}$; terms of the form 
$S_{g,{\rm int}}\supset X_gX_g\cdots\delta_{\rm in}\cdots$ 
represent the stochasticity in the bias coefficients, and so on. We summarize this in Tab.~\ref{tab:summary_Z}.

\begin{table}
\myfloatalign
\caption[.]{Summary of the terms in the expansion of $S_{g,{\rm int}}$ 
in powers of $X_g$, $X$ and $\delta_{\rm in}$, and what they correspond to. 
Since the action for the matter likelihood is contained in the action for the 
joint likelihood, there is no need to separately list the terms that make up $S_{\rm int}$. 
The terms are organized according to their relevance in the infrared, the most relevant being on top. 
The scalings of the different operators are derived below (see also Appendix~\ref{app:scalings}). 
The operators with the lowest number of fields in the second-to-last line are $X_gX_gX$ and $X_gXX$.} 
\label{tab:summary_Z}
\centering
\medskip
\begin{tabular}{ll}
\toprule
$S_{g,{\rm int}}\supset{}$ & corresponds to \\
\midrule
$X\delta_{\rm in}\delta_{\rm in}\cdots$ & nonlinear deterministic evolution for $\delta$ \\ 
$X_g\delta_{\rm in}\delta_{\rm in}\cdots$ & nonlinear deterministic evolution for $\delta_g$ \\ 
$X_gX_g\cdots \delta_{\rm in}\cdots$ & stochasticities in bias coefficients for $\delta_g$ \\
$XX\cdots \delta_{\rm in}\cdots$ & stochasticities in EFT coefficients for $\delta$ \\
$X_gX_gX_g\cdots$ & higher-order $n$-point functions of $\eps_g$ \\
{$X_g\cdots X\cdots$} & mixed ($\eps_g${--}$\eps_m$) higher-order stochasticities \\
$XXX\cdots$ & higher-order $n$-point functions of $\eps_m$ \\
\bottomrule
\end{tabular}
\end{table}

Now that we have recast the problem in the language of field theory, 
it pays to look at the scaling dimensions of the various fields, in order to asses the relative importance of the various terms 
{in the actions}. 
That is, we do some dimensional analysis and study the relevance (or irrelevance) of the various ``operators'' in the infrared. 
As it will become clear below, and also once we compute the likelihood, this makes this a key section, 
since it allows us to estimate immediately the importance of the various terms {in Tab.~\ref{tab:summary_Z}} 
before doing any calculation. 

In order to gain insight about the scaling of the various {operators}, we assume that 
the linear matter power spectrum can be approximated by a power law, i.e.~
\begin{equation}
\label{eq:RG-A}
P_{\rm L}(k) = \frac{2\pi^2}{k^3_{\rm NL}}\bigg(\frac{k}{k_{\rm NL}}\bigg)^{n_\delta}\,\,,\quad 
n_\delta = \frac{\dif\ln P_{\rm L}(k)}{\dif\ln k}\bigg|_{k=k_{\rm NL}} = \num{-1.7}\,\,.
\end{equation}
We also drop all higher-derivative terms, so that $K_{g,1} = b_1$ and $K_1=1$. 
Then, the power spectrum $P_{\rm in}$ of the initial matter field scales with $k$ in the same way. 
Now, an important point to realize is that in the EFT of LSS/bias expansion 
we are never in the situation that the cross stochasticity and the matter 
stochasticity dominate the quadratic action. They are higher-derivative terms, and must be treated perturbatively. 
For this reason, it is not really interesting to look at the scaling of $X$ 
{(but see Appendix~\ref{app:scalings} for a more detailed discussion)}, but only {at} those of $X_g$ and $\delta_{\rm in}$, 
\emph{as determined by \eq{action_joint-C} with $P_{\eps_g\eps_m}$ and $P_{\eps_m}$ both vanishing.} 
Moreover, $P_{\eps_g}$ is a constant in $k$ only on large scales: it has an expansion in powers of $k^2$ like the other stochasticities. 
Since we must treat also this expansion perturbatively, 
for the purposes of this discussion we can just consider $P_{\eps_g}$ to be a constant in $k$ 
(we can think of $X_g$ as a very massive field, more massive than any scale that we can probe: 
{we elaborate on this analogy in Appendix~\ref{app:scalings}}). 

The scalings of the fields $X_g$ and $\delta_{\rm in}$ can then be derived in 
the same way as what is usually done in quantum field theory 
(see, e.g., Section 12.1 of \cite{Peskin:1995ev}). Let us assume that we have only long-wavelength modes 
$k < b\Lambda$ in \eq{action_joint-A}, and put the ``external currents'' $\delta_g$ and $\delta$ to zero. 
\emph{As we lower the cutoff \emph{($b\to0^+$)}, 
the scaling of the various operators tells us how much they are relevant (or irrelevant) on large scales.} 
Rescaling the momentum by $\vec{k} = b\vec{k}'$, 
the functional integration measure changes by an unimportant $b$-dependent factor. 
The integration measure $\dif^3 k$ changes into $b^3\dif^3 k'$. 
The power spectrum $P_{\rm in}$ scales as $P_{\rm in}(k) = b^{n_\delta}P_{\rm in}(k')$, while $P_{\eps_g}$ remains invariant. Then, 
the terms in the action that are quadratic in $X_g$ and $\delta_{\rm in}$ remain invariant if we redefine 
\begin{subequations}
\label{eq:RG-B}
\begin{align}
X_g(b\vec{k}') &= b^{-\frac{3}{2}}X'_g(\vec{k}')\,\,, \label{eq:RG-B-1} \\
\delta_{\rm in}(b\vec{k}') &= b^{\frac{n_\delta-3}{2}}\delta'_{\rm in}(\vec{k}')\,\,. \label{eq:RG-B-2} 
\end{align}
\end{subequations}
These are the Fourier-space scaling dimensions. The real-space ones can be obtained straightforwardly from these 
(by accounting for the $\dif^3k$ relating a field to its Fourier components): they are given by 
\begin{subequations}
\label{eq:RG-C}
\begin{align}
X_g(\vec{x}'/b) &= b^{\frac{3}{2}}X'_g(\vec{x}')\,\,, \label{eq:RG-C-1} \\
\delta_{\rm in}(\vec{x}'/b) &= b^{\frac{3+n_\delta}{2}}\delta'_{\rm in}(\vec{x}')\,\,. \label{eq:RG-C-2} 
\end{align}
\end{subequations}
\eq{RG-C-2} should be familiar: for $n_\delta = -2$, we recover the scaling dimension of 
a scalar field with canonical kinetic term in three spacetime dimensions. 

{What about the derivative expansion? As expected, higher-derivative terms are irrelevant in the infrared: indeed, 
we have $\vec{\nabla}\sim b$ so that adding more and more derivatives makes a term less and less important as $b\to 0^+$.} 

{Now that we have the scalings of \eqsI{RG-C} we can estimate the relative importance in the infrared 
of the different operators collected in Tab.~\ref{tab:summary_Z}. First, we can check that \eq{RG-C-2} reproduces 
the expected scalings for the deterministic bias expansion, i.e.~it gives us the familiar expansion 
parameters in which we expand our computation of $n$-point correlation functions. In order to do so, 
we can compare the operators that give the deterministic evolution for $\delta_g$ at order $n+1$ and at order $n$. 
We also use the fact that on large scales the evolution for matter is given by the $F_n$ kernels of SPT, 
which are invariant under rescaling of momenta. 
This holds also for the kernels $K_{g,{\rm det},n}$ of the deterministic bias expansion, at leading order in derivatives. 
From this we see that we can indeed use the real-space scalings to discuss the relative importance of these two contributions, 
without loss of generality. Moreover, since the scaling of the volume element $\dif^3x$ is common to all terms we can forget about it. 
Then, using the first line of Tab.~\ref{tab:summary_Z}, we have} 
\begin{equation}
\label{eq:deterministic_confirmation}
\frac{\text{bias expansion at $(n+1)$th order}}{\text{bias expansion at $n$th order}} = 
\frac{X_g\overbrace{\delta_{\rm in}\delta_{\rm in}\cdots\delta_{\rm in}}^{\text{$n+1$ times}}}
{X_g\underbrace{\delta_{\rm in}\delta_{\rm in}\cdots\delta_{\rm in}}_{\text{$n$ times}}} 
\sim\frac{b^{\frac{3}{2}}\,b^{\frac{(n+1)(3+n_\delta)}{2}}}{b^{\frac{3}{2}}\,b^{\frac{n(3+n_\delta)}{2}}} 
= b^{\frac{3+n_\delta}{2}}\,\,,
\end{equation}
{which is what we expect: 
see, e.g., Section~(4.1.4) of \cite{Desjacques:2016bnm}.}\footnote{{More precisely, 
we refer to their Eq.~(4.27). Notice that the scaling there is $b^{3+n_\delta}$, 
and not $b^{(3+n_\delta)/2}$, since the 
next-to-leading{-}order correction to any $n$-point function 
comes with exactly two powers of $\delta_{\rm in}$. Our discussion is more general, 
and not restricted to a particular $n$-point function.}} 
{The relative importance of the terms describing the deterministic evolution of the matter fields 
can be worked out in the same way (we just have to use the second line of Tab.~\ref{tab:summary_Z}), 
and it is still controlled by \eq{RG-C-2}.} 

{We can then move to study the terms involving higher powers of $X_g$.} 
From \eqsI{RG-C} we see that the relative importance in the infrared of $X_g$ with respect to $\delta_{\rm in}$ scales as 
\begin{equation}
\label{eq:RG-D}
\frac{X_g}{\delta_{\rm in}}\sim b^{\frac{3}{2}-\frac{3}{2}-\frac{n_\delta}{2}} = b^{-\frac{n_\delta}{2}} = b^{0.9}\,\,. 
\end{equation}
In hindsight, this could have been anticipated simply by recalling that the typical size 
of a perturbation of momentum $k$ of a field $\varphi(\vec{x})$ 
is controlled by $\sqrt{k^3 P_\varphi(k)}$, where $P_\varphi$ is the power spectrum of the field. 
Then, the ratio $\sqrt{k^3 P_{\eps_g}(k)}/\sqrt{k^3 P_{\rm in}(k)}$ scales as $k^{-n_\delta/2}$, 
given \eq{RG-A} and a $k$-independent noise. 
Whatever the route we use to arrive at this result, \emph{it tells us that terms with more powers of $X_g$ than $\delta_{\rm in}$ 
\emph{(such as the non-Gaussian corrections to the stochasticity)} will be much less relevant on large scales than the 
terms representing the deterministic evolution} (given by more powers of $\delta_{\rm in}$). 

This will be confirmed and expanded upon by the full calculation of 
Sections~\ref{sec:calculation_of_likelihood-gaussian_stochasticities} 
and \ref{sec:calculation_of_likelihood-nongaussian_stochasticities}, 
together with the discussion of Section~\ref{subsec:expansion_parameters}. 
However, we can already try to answer the following question: 
at which order in perturbations do the terms due to the nonlinear deterministic evolution 
become of comparable relevance as those due to a three-point function $\braket{\eps_g\eps_g\eps_g}$? 

First, since the three-point function $\braket{\eps_g\eps_g\eps_g}$ is a constant on large scales, the corresponding term 
in $S_{g,{\rm int}}$ is a local interaction, i.e.~ 
\begin{equation}
\label{eq:RG-E}
\frac{\partial^3S^{(3)}_{g,{\rm int}}[\vec{\phi}_g]}{\partial X_g(\vec{k}_1)\partial X_g(\vec{k}_2)\partial X_g(\vec{k}_3)}
\propto\delta^{(3)}(\vec{k}_1+\vec{k}_2+\vec{k}_3)\,\,.
\end{equation}
Therefore we only have to worry about the scaling of $X_g$. 
Using the invariance of $K_{g,n}$ under rescaling of the momenta (at leading order in derivatives) 
we see that, as we did in \eq{deterministic_confirmation}, 
we can safely use the real-space scalings (and forget about $\dif^3x$) to discuss 
the relative importance of a three-point function $\braket{\eps_g\eps_g\eps_g}$ 
and higher-order deterministic terms. 

Recalling that higher-order deterministic terms correspond to $S_{g,{\rm int}}\supset X_g\delta_{\rm in}(\delta_{\rm in}\cdots)$ 
(see Tab.~\ref{tab:summary_Z}), and using the fact that {(as we have just discussed)} 
we can consider having local interactions without loss of generality, we have
\begin{equation}
\label{eq:RG-F}
\frac{\braket{\eps_g\eps_g\eps_g}}{\text{bias expansion at $n$th order}} = 
\frac{X_gX_gX_g}{X_g\underbrace{\delta_{\rm in}\delta_{\rm in}\cdots\delta_{\rm in}}_{\text{$n$ times}}} 
\sim\frac{b^{\frac{3}{2}}\,b^{\frac{3}{2}}\,b^{\frac{3}{2}}}{b^{\frac{3}{2}}\,b^{\frac{n(3+n_\delta)}{2}}} 
= b^{3-\frac{n(3+n_\delta)}{2}}\,\,.
\end{equation}
For $n_\delta=\num{-1.7}$, we see that the non-Gaussianity of $\eps_g$ coming from $\braket{\eps_g\eps_g\eps_g}$ 
becomes more relevant than the deterministic bias expansion at $n=5$. 

What about the stochasticity in the bias coefficient $b_1$? 
As we will discuss in more detail in the next sections, this corresponds, schematically, to 
a shift $\smash{b_1\to b_1 + P_{\eps_g\eps_{g,\delta}}^{\{0\}}J(\vec{x})}$ in real space (at leading order in derivatives). 
Here $\smash{P_{\eps_g\eps_{g,\delta}}}$ is the cross spectrum between $\eps_g$ and 
the field $\eps_{g,\delta}$, where $\eps_{g,\delta}$ is defined in real space by (see Section~2.8 of \cite{Desjacques:2016bnm})
\begin{equation}
\label{eq:recall_eps_delta}
\delta_g(\vec{x}) = b_1\delta(\vec{x}) + \eps_g(\vec{x}) + \eps_{g,\delta}(\vec{x})\delta(\vec{x})\,\,.
\end{equation}
This gives rise to a local interaction $S_{g,{\rm int}}\supset X_gX_g\delta_{\rm in}$, 
so that it is again sufficient to consider the scalings of the fields. The equivalent of \eq{RG-F} is 
\begin{equation}
\label{eq:RG-G}
\frac{\text{stochasticity in $b_1$}}{\text{bias expansion at $n$th order}} = 
\frac{X_gX_g\delta_{\rm in}}{X_g\underbrace{\delta_{\rm in}\delta_{\rm in}\cdots\delta_{\rm in}}_{\text{$n$ times}}} 
\sim\frac{b^{\frac{3}{2}}\,b^{\frac{3}{2}}\,b^{\frac{3+n_\delta}{2}}}{b^{\frac{3}{2}}\,b^{\frac{n(3+n_\delta)}{2}}} = 
b^{\frac{3}{2}-\frac{(n-1)(3+n_\delta)}{2}}\,\,.
\end{equation}
From this equation we see that, for $n_\delta=\num{-1.7}$, the stochasticity in $b_1$ 
becomes more relevant than the deterministic bias expansion at $n={3}$
(see also Fig.~\ref{fig:scalings} on p.~\pageref{fig:scalings}). 

These two results, \eqsII{RG-F}{RG-G}, already show that the impact of stochasticities 
beyond those in the auto- and cross-correlations of galaxies and matter is 
not very important on large scales, unless we compare it with deterministic terms 
at very high order, e.g.~terms like $\delta_g\supset b_4\delta^4/4!$ at fourth order. 
We will encounter more of these scaling analyses in the following sections, 
in which we {carry out} the actual computation of the likelihood. 

Before proceeding we emphasize that, while in this section we focused 
on the impact that higher-order stochasticities have on the likelihood, 
our conclusions on their importance relative to the deterministic contributions 
apply equally well if one is interested {just in correlation functions.}

\section{Gaussian stochasticities}
\label{sec:calculation_of_likelihood-gaussian_stochasticities}

\noindent We are now in position to compute the conditional likelihood ${\cal P}[\delta_g|\delta]$. In this section we make the 
assumption of Gaussian stochasticities, \ie~we put all correlation functions of the noise 
fields to zero except for the auto and cross two-point functions 
of the galaxy and matter fields. First, we assume that only $P_{\eps_g}$ is not zero (Section~\ref{subsec:P_eps_g}), 
and then we add $P_{\eps_g\eps_m}$ (Section~\ref{subsec:P_eps_g_eps_m}). 
In Section~\ref{subsec:P_eps_g} we also briefly comment on the impact of higher-derivative terms in the deterministic bias expansion.

\subsection{With \texorpdfstring{$P_{\eps_g}$}{P\_\{\textbackslash epsilon\_g\}} only}
\label{subsec:P_eps_g}

\noindent We have to compute the path integrals for the joint and matter likelihood. 
{This is in general a complicated task, and progress is usually made by 
employing a saddle-point expansion and working order-by-order in loops.} 

{We do this calculation (stopping at tree level) in Appendix~\ref{app:tree_level}. 
However, in the case of only $P_{\eps_g}$ being non-vanishing we 
can actually compute the conditional likelihood exactly, and reproduce the 
result of \cite{Schmidt:2018bkr}.}

\subsubsection*{Calculation at all orders in loops}

\noindent Given the actions of \eqsII{action_matter-C}{action_joint-B} with ${P_{\eps_g\eps_m}=0}$ and ${P_{\eps_m}=0}$, 
we can do the path integral exactly. Indeed, we have that\footnote{All 
the manipulations that follow make sense only if we first restrict the functional integrals at a finite cutoff, 
and then send the cutoff to infinity at 
the end of the calculation (see \eg~\cite{SkinnerLectures} for details). 
In any case, we emphasize that in practical applications any integral over $k$ is cut off at a finite momentum, 
as detailed in \cite{Schmidt:2018bkr} (see also Section~\ref{subsec:expansion_parameters}).} 
\begin{equation}
\label{eq:loops_subsubsection_1-A}
\begin{split}
{\cal P}[\delta_g,\delta] &= 
{\cal N}_{\delta^{(\infty)}}^2\int{\cal D}X_g\,{\cal D}X\,{\cal D}\delta_{\rm in}\,
\eu^{\int_{\vec{k}}\vec{\phi}_g(\vec{k})\cdot\vec{\mathcal{J}}_g(-\vec{k})-S_g[\vec{\phi}_g]} \\
&= {\cal N}_{\delta^{(\infty)}}^2\int{\cal D}X_g\,{\cal D}X\,{\cal D}\delta_{\rm in}\,
\eu^{\iu\int_{\vec{k}}X_g(\vec{k})\delta_g(-\vec{k})}\,\eu^{\iu\int_{\vec{k}}X(\vec{k})\delta(-\vec{k})} \\ 
&\hphantom{= {\cal N}_{\delta^{(\infty)}}^2\int{\cal D}X_g\,{\cal D}X\,{\cal D}\delta_{\rm in} } 
\times\eu^{-\frac{1}{2}\int_{\vec{k}}\frac{\delta_{\rm in}(\vec{k})\delta_{\rm in}(-\vec{k})}{P_{\rm in}(k)}}\,
\eu^{-\frac{1}{2}\int_{\vec{k}}P_{\eps_g}(k){X_{g}(\vec{k}) X_{g}(-\vec{k})}} \\ 
&\hphantom{= {\cal N}_{\delta^{(\infty)}}^2\int{\cal D}X_g\,{\cal D}X\,{\cal D}\delta_{\rm in} } 
\times\eu^{-\iu\int_{\vec{k}}X_g(\vec{k})\delta_{g,{\rm fwd}}[\delta_{\rm in}](-\vec{k})}\, 
\eu^{-\iu\int_{\vec{k}}X(\vec{k})\delta_{\rm fwd}[\delta_{\rm in}](-\vec{k})}\,\,,
\end{split}
\end{equation}
where we have put the matter stochasticity and the cross stochasticity to zero, 
and we have reintroduced the factor of ${\cal N}_{\delta^{(\infty)}}^2$ coming from the 
Dirac delta functionals, cf.~\eqsII{matter_likelihood-A}{joint_likelihood-A}. 
The integral over $X_g$ can be done exactly by completing the square: we have that 
\begin{equation}
\label{eq:loops_subsubsection_1-B}
{-\frac{1}{2}}\int_{\vec{k}}P_{\eps_g}(k){X_{g}(\vec{k}) X_{g}(-\vec{k})} 
+ \iu\int_{\vec{k}}X_g(\vec{k})\big(\delta_g(-\vec{k})-\delta_{g,{\rm fwd}}[\delta_{\rm in}](-\vec{k})\big) 
\end{equation}
is equal to 
\begin{equation}
\label{eq:loops_subsubsection_1-C} 
\begin{split}
&{-\frac{1}{2}}\int_{\vec{k}}P_{\eps_g}(k)\bigg(X_{g}(\vec{k}) 
- \frac{\iu\big(\delta_g(\vec{k})-\delta_{g,{\rm fwd}}[\delta_{\rm in}](\vec{k})\big)}
{P_{\eps_g}(k)}\bigg)\times(\vec{k}\to-\vec{k}) \\
&-\frac{1}{2}\int_{\vec{k}}\frac{\abs{\delta_g(\vec{k})-\delta_{g,{\rm fwd}}[\delta_{\rm in}](\vec{k})}^2}{P_{\eps_g}(k)}\,\,,
\end{split}
\end{equation}
where we have again used the fact that all the fields we are considering are real to simplify the second term. 
Then, shifting the integration variable, we see that the integral over $X_g$ is simply equal to 
\begin{equation}
\label{eq:loops_subsubsection_1-D}
{\cal N}_{\delta^{(\infty)}}^{-1}\prod_{\vec{k}}\sqrt{\frac{1}{2\pi P_{\eps_g}(k)}}\,\,.
\end{equation}
Putting apart this field-independent factor, we remain with the integrals over $X$ and $\delta_{\rm in}$. 
The integral over $X$ is straightforward: it is simply equal to 
\begin{equation}
\label{eq:loops_subsubsection_1-E}
{\cal N}_{\delta^{(\infty)}}^{-1}\delta^{(\infty)}_{\rm D}\big(\delta-\delta_{\rm fwd}[\delta_{\rm in}]\big)\,\,,
\end{equation}
from which we also see that all the factors of ${\cal N}_{\delta^{(\infty)}}$ simplify. 
Finally, we carry out the integration over $\delta_{\rm in}$. 
The Dirac delta functional simply puts $\delta_{\rm in} = \delta^{-1}_{\rm fwd}[\delta]$ in \eq{loops_subsubsection_1-C}. 
Using the relation of \eq{deterministic_bias_expansion-A}, \ie~
\begin{equation}
\label{eq:loops_subsubsection_1-F}
\delta_{g,{\rm fwd}}[\delta_{\rm in}] = \delta_{g,{\rm det}}\big[\delta_{{\rm fwd}}[\delta_{\rm in}]\big]\,\,,
\end{equation}
and following these exact same steps for the matter likelihood of 
\eqsIII{action_matter-A}{action_matter-B}{action_matter-C}, we see that 
\begin{equation}
\label{eq:loops_subsubsection_1-G}
{\cal P}[\delta_g,\delta] = {\cal P}[\delta]\,\Bigg(\prod_{\vec{k}}\sqrt{\frac{1}{2\pi P_{\eps_g}(k)}}\Bigg)\,
\exp\left({-\frac{1}{2}}
\int_{\vec{k}} \frac{\abs{\delta_g(\vec{k})-\delta_{g,{\rm det}}[\delta](\vec{k})}^2}{P_{\eps_g}(k)} \right)\,\,.
\end{equation}
Notice that $\delta_{g,{\rm det}}[\delta]$ is the deterministic bias relation constructed 
from the evolved density field $\delta$ that is given as argument to the joint probability. 

Defining the logarithm of the likelihoods (matter, joint and conditional) as 
$\wp[\delta]$, $\wp[\delta_g,\delta]$ and $\wp[\delta_g|\delta]$, 
the \emph{field-dependent part} of the logarithm of \eq{loops_subsubsection_1-G} is 
\begin{equation}
\label{eq:cubic_conditional_likelihood-E-temp}
\wp[\delta_g,\delta]\equiv\ln {\cal P}[\delta_g,\delta] = 
\wp[\delta]-\frac{1}{2}\int_{\vec{k}}\frac{\abs{\delta_g(\vec{k})-\delta_{g,{\rm det}}[\delta](\vec{k})}^2}{P_{\eps_g}(k)}\,\,,
\end{equation}
{i.e.}
\begin{equation}
\label{eq:cubic_conditional_likelihood-E}
\wp[\delta_g|\delta]\equiv\ln {\cal P}[\delta_g|\delta] = 
-\frac{1}{2}\int_{\vec{k}}\frac{\abs{\delta_g(\vec{k})-\delta_{g,{\rm det}}[\delta](\vec{k})}^2}{P_{\eps_g}(k)}\,\,.
\end{equation}
That is, we recover the result of \cite{Schmidt:2018bkr} at all orders in the deterministic bias expansion. 
Moreover, we also obtain the overall \emph{normalization} of the conditional likelihood, 
which matches the one derived in \cite{Schmidt:2018bkr} (obtaining this normalization is not trivial 
if we follow a perturbative approach such as that of Appendix~\ref{app:tree_level}). 
This normalization makes sense physically since in the limit of zero stochasticity 
we expect the conditional likelihood to be a Dirac delta functional 
of $\delta_g-\delta_{g,{\rm det}}[\delta]$. Indeed, 
using the functional generalization of $\lim_{\sigma^2\to 0}(2\pi\sigma^2)^{-1/2}\exp(-x^2/2\sigma^2) = \delta(x)$, we find 
\begin{equation}
\label{eq:loops_subsubsection_1-I}
\lim_{P_{\eps_g}\to 0}{\cal P}[\delta_g|\delta] = \delta^{(\infty)}_{\rm D}\big(\delta_g-\delta_{g,{\rm det}}[\delta]\big)\,\,.
\end{equation}

The fact that ${\cal P}[\delta_g|\delta]$ is a Gaussian in $\delta_g-\delta_{g,{\rm det}}[\delta]$ follows from 
the assumption of having only the field $\eps_g$ as source of noise 
(\ie~having $\delta_g-\delta_{g,{\rm det}}[\delta]=\eps_g$), 
and that this field is Gaussian. Indeed, we have seen in Section~\ref{subsec:action_and_RG_scalings} that the 
non-Gaussianity of $\eps_g$ is associated with terms that are higher order in $X_g$ and are suppressed on large scales. 
Nevertheless, as one includes terms of successively higher order in the 
deterministic bias expansion, this expression for the conditional 
probability ceases to become more accurate since non-Gaussian corrections 
as well as those due to the matter stochasticity become as relevant 
as the deterministic terms included. We will quantify this below.

\subsubsection*{Higher-derivative terms} 

\noindent Let us briefly discuss the higher-derivative terms in the deterministic 
bias expansion for the galaxy field, in the deterministic evolution of the matter field, and 
in the {power spectrum of the galaxy noise $\eps_g$.} 

It is straightforward to see that the higher-derivative contributions to the deterministic evolution 
are automatically included at all orders in perturbations: at no point does the calculation leading to 
\eq{loops_subsubsection_1-G} assume a particular form for the kernels 
$K_{n}$ {for $\delta_{\rm fwd}[\delta_{\rm in}]$} {or} $K_{g,n}$ {for $\delta_{g,{\rm fwd}}[\delta_{\rm in}]$}, 
whatever the $n$. Indeed, the final result is dependent only on $\delta_{g,{\rm det}}[\delta]$ 
as defined by \eqsII{G_and_M_kernels}{deterministic_bias_expansion-A}. 

The same is true for the power spectrum $P_{\eps_g}(k)$: the result of \eq{loops_subsubsection_1-G} is independent on 
its particular form, and then holds at all orders in its expansion in powers of $k^2$. 

It is clear that the (ir)relevance of higher-derivative terms in the infrared is controlled by exactly the same scaling arguments 
we have introduced in Section~\ref{subsec:action_and_RG_scalings}. {Hence, going to very high order in $k^2$ in 
the expansion of $P_{\eps_g}(k)$ could be useless unless 
higher-derivative terms in $\delta_{g,{\rm det}}[\delta]$ of the same (or close) scaling dimension are also included.} 
We leave a more detailed discussion to Section~\ref{subsec:expansion_parameters}.

\subsection{Adding \texorpdfstring{$P_{\eps_g\eps_m}$}{P\_\{\textbackslash epsilon\_g\textbackslash epsilon\_m\}}}
\label{subsec:P_eps_g_eps_m}

\noindent We now see what happens if we allow for the cross stochasticity between galaxies and matter, $P_{\eps_g\eps_m}$. 
Given that we are still not considering vertices that involve more than one $X_g$ or $X$ {field}, we 
expect that a calculation at all orders in loops, 
along the lines of what we have done in \eqsI{loops_subsubsection_1-A} to \eqref{eq:loops_subsubsection_1-I}, should be possible. 

In this paper, instead, we will only include $P_{\eps_g\eps_m}$ at leading order in the saddle-point approximation (tree level). 
The fact that we are doing the calculation perturbatively 
is also why we have not included the matter stochasticity $P_{\eps_m}$: 
the stochasticities are added order-by-order in an expansion in $k^2$ 
(as we discussed in Section~\ref{subsec:action_and_RG_scalings}), and $P_{\eps_m}$ 
starts at a higher order in $k^2$ with respect to $P_{\eps_g\eps_m}$. 
We will discuss the extension to all orders in $P_{\eps_g\eps_m}$ at the end of this section. 

Let us see how this works out. In our tree-level calculation 
(whose details are contained in Appendix~\ref{app:higher_derivative_stochasticities}), 
we stop at cubic order in the fields. 
{If we define the expansion of the field-dependent part of 
$\wp[\delta_g|\delta]$ in powers of the galaxy and matter fields as} 
\begin{equation}
\label{eq:expansion_conditional_likelihood}
\wp[\delta_g|\delta] = \wp^{(2)}[\delta_g|\delta] + \wp^{(3)}[\delta_g|\delta] + \cdots\,\,,
\end{equation}
{this means that we compute $\smash{\wp^{(2)}[\delta_g|\delta]}$ and $\smash{\wp^{(3)}[\delta_g|\delta]}$.} 
At this order, we reproduce the result of \cite{Schmidt:2018bkr} 
(which contains the stochasticities $\smash{P_{\eps_g}}$ and $\smash{P_{\eps_g\eps_m}}$ at all orders in $\smash{k^2}$), 
and we also obtain three new terms that were absent in {that paper} 
(one in $\smash{\wp^{(2)}}$ and two in $\smash{\wp^{(3)}}$). More precisely, we find 
\begin{equation}
\label{eq:gm_subsec-A}
\begin{split}
\wp[\delta_g|\delta] = 
{-\frac{1}{2}}\int_{\vec{k}}\frac{\abs{\delta_g(\vec{k})-\delta_{g,{\rm det}}[\delta]
(\vec{k})}^2}{P_{\eps_g}(k) - 2b_1P_{\eps_g\eps_m}(k)} + \Delta\wp[\delta_g|\delta]\,\,.
\end{split}
\end{equation}
The first term on the right-hand side is, apart from an irrelevant minus sign, 
the result found by \cite{Schmidt:2018bkr} once the galaxy-matter cross stochasticity is included. 
By $\Delta\wp[\delta_g|\delta]=\Delta\wp^{(2)}[\delta_g|\delta]+\Delta\wp^{(3)}[\delta_g|\delta]+\cdots$ 
we denote the corrections to the result of \cite{Schmidt:2018bkr}: 
\begin{itemize}[leftmargin=*]
\item at quadratic order, we find 
\begin{equation}
\label{eq:gm_subsec-B}
\Delta\wp^{(2)}[\delta_g|\delta] = \int_{\vec{k}}\frac{P_{\eps_g\eps_m}(k)
\big(\delta_g(\vec{k})-\delta_{g,{\rm det}}^{(1)}[\delta](\vec{k})\big)\delta(-\vec{k})}{P_{\eps_g}(k)P_{\rm L}(k)}\,\,,
\end{equation}
{where we defined} 
\begin{equation}
\label{eq:cubic_conditional_likelihood-D-help}
\delta_{g,{\rm det}}[\delta] = \delta^{(1)}_{g,{\rm det}}[\delta] + \delta^{(2)}_{g,{\rm det}}[\delta] +\cdots 
\end{equation}
{as the expansion of $\delta_{g,{\rm det}}[\delta]$ in powers of $\delta$}, {so that} 
$\smash{\delta_{g,{\rm det}}^{(1)}[\delta](\vec{k}) = b_1\delta(\vec{k})}$ at leading order in the expansion in $k^2$; 
\item at cubic order we have two new terms. The first is 
\begin{equation}
\label{eq:gm_subsec-C}
\Delta\wp^{(3)}[\delta_g|\delta]\supset {-{}}\int_{\vec{k}}
\frac{P_{\eps_g\eps_m}(k)}{P_{\eps_g}(k)}\frac{\delta(\vec{k})\delta^{(2)}_{g,{\rm det}}[\delta](-\vec{k})}{P_{\rm L}(k)}\,\,.
\end{equation}
The second is more complicated: it is given by 
\begin{equation}
\label{eq:gm_subsec-D}
\begin{split}
\Delta\wp^{(3)}[\delta_g|\delta] \supset {-{}}\int_{\vec{k}}\frac{\delta_g(\vec{k})-\delta_{g,{\rm det}}^{(1)}(\vec{k})}{P_{\eps_g}(k)} 
\int_{\vec{p}_1,\vec{p}_2}\bigg[&(2\pi)^3\delta^{(3)}(-\vec{k}-\vec{p}_{12})\,K_{g,{\rm det},2}(-\vec{k};\vec{p}_1,\vec{p}_2) \\
&\times\bigg(\frac{P_{\eps_g\eps_m}(p_2)}{P_{\eps_g}(p_2)}
\delta(\vec{p}_1)\big(\delta_g(\vec{p}_2) - \delta_{g,{\rm det}}^{(1)}(\vec{p}_2)\big)\bigg) \\
&+ (\vec{p}_1\to\vec{p}_2)\bigg]\,\,,
\end{split}
\end{equation}
where we recognize {once more} 
the kernel for the {deterministic} bias expansion {up to second order} in the {nonlinear} matter field, \ie~ 
\begin{equation}
\label{eq:gm_subsec-E}
K_{g,{\rm det},2}(\vec{k};\vec{p}_1,\vec{p}_2) \equiv K_{g,2}(\vec{k};\vec{p}_1,\vec{p}_2)-b(k)K_{2}(\vec{k};\vec{p}_1,\vec{p}_2)\,\,.
\end{equation}
We can gain insight on this term by rewriting it in real space. Let us stop at leading order in derivatives, \ie~
\begin{equation}
\label{eq:order_of_stochasticities}
P_{\eps_g}(k)=P_{\eps_g}^{\{0\}}\,\,,\quad P_{\eps_g\eps_m}(k)=P_{\eps_g\eps_m}^{\{2\}}k^2\,\,,\quad b(k)=b_1\,\,,
\end{equation}
and take the kernel $K_{g,{\rm det},2}$ as the one for the second-order LIMD 
contribution $\delta_g\supset b_2\delta^2/2$ (at second order in perturbations the 
tidal field squared also appears, cf.~\eq{tidal_field_squared}: since its scaling dimension is 
the same as $\delta^2$, even if we omit it there will be no loss of generality when we discuss 
the relative importance of these corrections). Then, \eq{gm_subsec-D} becomes 
\begin{equation}
\label{eq:gm_subsec-F}
\Delta\wp^{(3)}[\delta_g|\delta]\supset\frac{P^{\{2\}}_{\eps_g\eps_m}}{P^{\{0\}}_{\eps_g}}\frac{b_2}{P^{\{0\}}_{\eps_g}}
\int\dif^3x\,\delta(\vec{x})\big(\delta_g(\vec{x})-b_1\delta(\vec{x})\big)\nabla^2\big(\delta_g(\vec{x})-b_1\delta(\vec{x})\big)\,\,,
\end{equation}
where $\smash{P^{\{2\}}_{\eps_g\eps_m}/P^{\{0\}}_{\eps_g}}$ has the dimensions of length squared. 
\end{itemize}

Notice that in \eqsIII{gm_subsec-B}{gm_subsec-C}{gm_subsec-D} we have kept 
the full scale dependence of the stochasticities, and $\delta_{g,{\rm det}}$ 
is also kept fully general (\ie~all higher-derivative terms are included). 
The reason is that, as detailed in Appendices~\ref{app:tree_level}, 
\ref{app:higher_derivative_stochasticities} and \ref{app:saddle_point_formulas}, 
the tree-level expressions of \eqsI{W_to_cubic_order} for the matter and joint likelihoods 
do not require us to stop at any given order in derivatives. 
Of course, it does not make sense to include terms suppressed by arbitrarily high powers of $k^2$, since we are anyway missing 
the matter stochasticity $P_{\eps_m}(k)\sim k^4$. {Moreover, we see that:} 
\begin{itemize}[leftmargin=*]
\item {in \eq{gm_subsec-A} we have the galaxy-matter stochasticity at the denominator. 
Even if we only keep the leading $\smash{P^{\{2\}}_{\eps_g\eps_m}k^2}$ term in its expansion, it 
still does not make sense to use that expression for practical applications unless we include also 
the matter stochasticity, since it contains also terms of order $k^4$ from $(1-x^2)^{-1}\sim 1+x^2+x^4+\cdots$;}
\item the higher-derivative terms in the deterministic evolution of matter 
can also be included straightforwardly. 
The only ones to play a role are the scale-dependent corrections to the growth factor $D_1$ coming from, for example, 
$c^2_{\rm s}$-like counterterms. This is because of the presence of 
the linear matter power spectrum $P_{\rm L}(k)$ in \eqsII{gm_subsec-B}{gm_subsec-C}. 
Since none of the calculations of Appendix~\ref{app:higher_derivative_stochasticities} rely on 
the assumption of $D_1$ being scale-independent, it is possible to replace $D_1$ with $D_1 K_1(k)$ everywhere. 
Including higher-derivative terms makes it equal to $D^2_1K^2_1(k)P_{\rm in}(k)$ instead of just $D^2_1P_{\rm in}(k)$; 
\item the presence of the terms involving the linear power spectrum can be understood by recalling how 
we derived \eq{loops_subsubsection_1-G}. A fundamental step in that derivation was recognizing that the integral over 
the field $X$ gave a Dirac delta functional for the gravity-only forward model, cf.~\eq{loops_subsubsection_1-E}. The 
presence of the noise $\eps_m$ effectively gives a spread to this Dirac delta functional. 
This spread can be effectively accounted for, in the large-scale limit $k\to 0$, via functional derivatives of 
the probability distribution of $\delta_{\rm in}$ (using the functional generalization of the Laplace method). 
\end{itemize}

Now is a good point to discuss the relative importance, on large scales, 
of the new terms in \eq{gm_subsec-A}. We compare the terms quadratic and cubic in the fields separately. 
Similarly to what we did in Section~\ref{subsec:action_and_RG_scalings}, 
we will work with real-space scalings. Since all we care about are the relative scalings, 
it does not matter whether we work in real or Fourier space 
(we prefer to work in real space, in general, since it is simpler to make contact 
with the well-known scalings for a canonical scalar field theory in three spatial dimensions, 
when the linear matter power spectrum is a power law 
and we take $n_\delta = -2$ in \eq{RG-A}: see also below \eqsI{RG-C}).

\subsubsection*{Quadratic order in the fields}

\noindent At quadratic order in the galaxy and matter fields, and on large scales, we can expand the denominator 
in the integrand of the first term on the right-hand side of \eq{gm_subsec-A}. Then, we find that $\wp[\delta_g|\delta]$ is 
made up of three terms. The first is simply \eq{cubic_conditional_likelihood-E} at second order in the fields, \ie~
\begin{equation}
\label{eq:gm_subsec-quadratic-A}
\wp^{(2)}[\delta_g|\delta] \supset -\frac{1}{2}\int_{\vec{k}}\frac{\abs{\delta_g(\vec{k}) - b(k)\delta(\vec{k})}^2}{P_{\eps_g}(k)} = 
-\frac{1}{2}\int_{\vec{k}}\frac{\abs{\delta_g(\vec{k}) - b_1\delta(\vec{k})}^2}{P_{\eps_g}^{\{0\}}}\,\,,
\end{equation}
where we have taken $b(k)=b_1$ and $\smash{P_{\eps_g}(k) = P_{\eps_g}^{\{0\}}}$ on large scales 
(in the galaxy noise we have dropped the second-order contribution $\smash{P_{\eps_g}(k) \supset P_{\eps_g}^{\{2\}}k^2}$ 
since its impact is exactly the same as that of $P_{\eps_g\eps_m}(k)\sim k^2$). The two other terms are 
\begin{equation}
\label{eq:gm_subsec-quadratic-B}
\begin{split}
\wp^{(2)}[\delta_g|\delta] \supset {-\frac{P^{\{2\}}_{\eps_g\eps_m}}{P^{\{0\}}_{\eps_g}}
\frac{b_1}{P^{\{0\}}_{\eps_g}}}\int_{\vec{k}}k^2\abs{\delta_g(\vec{k})-b_1\delta(\vec{k})}^2 
+ \frac{P^{\{2\}}_{\eps_g\eps_m}}{P^{\{0\}}_{\eps_g}}\int_{\vec{k}}
\frac{k^2\big(\delta_g(\vec{k})-b_1\delta(\vec{k})\big)\delta(-\vec{k})}{P_{\rm L}(k)}\,\,.
\end{split}
\end{equation}

Then, we take $P_{\rm L}(k)\sim k^{n_\delta}$, as we did in \eq{RG-A}. Let us now consider the fields $\delta$ and $\delta_g-b_1\delta$. 
The real-space scaling of the first, for a power-law power spectrum, is simply given by \eq{RG-C-2}, \ie~ 
\begin{equation}
\label{eq:gm_subsec-quadratic-C}
\delta(\vec{x})\sim b^{\frac{3+n_\delta}{2}}\,\,.
\end{equation} 
What about the second? The difference between $\delta_g$ and $b_1\delta$, 
at linear order, is exactly controlled by the noise for the galaxy field, 
which in Section~\ref{subsec:action_and_RG_scalings} we have identified with $X_g$ 
for all practical purposes. The real-space scaling of $X_g(\vec{x})$, 
and then of $\delta_g(\vec{x})-b_1\delta(\vec{x})$, is given by \eq{RG-C-1}, i.e.\footnote{This 
is clear also by looking at the leading quadratic likelihood, {\eq{gm_subsec-quadratic-A}}.} 
\begin{equation}
\label{eq:gm_subsec-quadratic-D}
\delta_g(\vec{x})-b_1\delta(\vec{x})\sim b^{\frac{3}{2}}\,\,.
\end{equation} 
Hence, if we compare the real-space scaling of the three terms in \eqsII{gm_subsec-quadratic-A}{gm_subsec-quadratic-B}, we have
\begin{subequations}
\label{eq:gm_subsec-quadratic-E}
\begin{align}
\text{\eq{gm_subsec-quadratic-A}}&\sim b^{\frac{3}{2}}\,b^{\frac{3}{2}}\,\,, \label{eq:gm_subsec-quadratic-E-1} \\
\text{$1{\rm st}$ term of \eq{gm_subsec-quadratic-B}}&\sim b^2\,b^{\frac{3}{2}}\,b^{\frac{3}{2}}\,\,, 
\label{eq:gm_subsec-quadratic-E-2} \\
\text{$2{\rm nd}$ term of \eq{gm_subsec-quadratic-B}}&
\sim b^2\,b^{\frac{3}{2}}\,b^{\frac{3+n_\delta}{2}}\,b^{-n_\delta}\,\,, \label{eq:gm_subsec-quadratic-E-3}
\end{align}
\end{subequations}
where we have used that $k^2$ scales as $b^2$ and $P_{\rm L}(k)$ scales as $b^{n_\delta}$ and, 
as discussed in detail at the end of Section~\ref{subsec:action_and_RG_scalings}, 
we can forget about the scaling of the volume element $\dif^3x$ since it is common to all terms. 
From this we see that, as expected, the two terms in \eq{gm_subsec-quadratic-B} are 
less relevant in the infrared than the leading term of \eq{gm_subsec-quadratic-A}. We also see that 
\begin{equation}
\label{eq:gm_subsec-quadratic-F}
\frac{\text{$1{\rm st}$ term of \eq{gm_subsec-quadratic-B}}}{\text{$2{\rm nd}$ term of \eq{gm_subsec-quadratic-B}}}\sim 
\frac{b^2\,b^{\frac{3}{2}}\,b^{\frac{3}{2}}}{b^2\,b^{\frac{3}{2}}\,b^{\frac{3+n_\delta}{2}}\,b^{-n_\delta}} = b^{\frac{n_\delta}{2}}\,\,,
\end{equation}
that is the first term of \eq{gm_subsec-quadratic-B} is the more relevant between the two. 
We leave a more detailed discussion to Section~\ref{subsec:expansion_parameters}: for now, let us see what happens at cubic order.

\subsubsection*{Cubic order in the fields}

\noindent {Again expanding the denominator of \eq{gm_subsec-A} and stopping at ${\cal O}(k^2)$,} 
{a}t cubic order we have to compare four terms. 
The first {is} simply the expansion at third order {in the fields} of \eq{cubic_conditional_likelihood-E}, which is given by 
\begin{equation}
\label{eq:gm_subsec-cubic-A}
\frac{1}{P^{\{0\}}_{\eps_g}}\int_{\vec{k}}{\big(\delta_g(\vec{k})-b_1\delta(\vec{k})\big)\,
\delta^{(2)}_{g,{\rm det}}[\delta](-\vec{k})}\,\,.
\end{equation}
{Then, at cubic order in the fields we have the contribution} 
\begin{equation}
\label{eq:gm_subsec-cubic-B}
{\frac{P^{\{2\}}_{\eps_g\eps_m}}{P^{\{0\}}_{\eps_g}}\frac{2b_1}{P^{\{0\}}_{\eps_g}}}
\int_{\vec{k}}k^2\big(\delta_g(\vec{k})-b_1\delta(\vec{k})\big)\,
\delta^{(2)}_{g,{\rm det}}[\delta](-\vec{k})
\end{equation}
{(notice that we have stopped at leading order in derivatives, cf.~\eq{order_of_stochasticities}, in both terms).} 
We can understand the relative importance of these two terms on large scales more easily 
if {until the end of this section} we take the second-order deterministic galaxy field, $\smash{\delta^{(2)}_{g,{\rm det}}}$, 
to be given by the second-order LIMD contribution $\delta_g\supset b_2\delta^2/2$. 
The other two terms are those of \eqsII{gm_subsec-C}{gm_subsec-D}, 
also expanded at leading order in derivatives. They are equal to 
\begin{equation}
\label{eq:gm_subsec-cubic-C}
{-\frac{P^{\{2\}}_{\eps_g\eps_m}}{P_{\eps_g}^{\{0\}}}}\int_{\vec{k}}
\frac{k^2\delta(\vec{k})\delta^{(2)}_{g,{\rm det}}[\delta](-\vec{k})}{P_{\rm L}(k)}
\end{equation}
and to \eq{gm_subsec-F}, \ie~(as above, the scaling is simpler to see in real space) 
\begin{equation}
\label{eq:gm_subsec-cubic-D}
\frac{P^{\{2\}}_{\eps_g\eps_m}}{P^{\{0\}}_{\eps_g}}\frac{b_2}{P^{\{0\}}_{\eps_g}}
\int\dif^3x\,\delta(\vec{x})\big(\delta_g(\vec{x})-b_1\delta(\vec{x})\big)\nabla^2\big(\delta_g(\vec{x})-b_1\delta(\vec{x})\big)\,\,.
\end{equation} 
Then, with the scalings of \eqsII{gm_subsec-quadratic-C}{gm_subsec-quadratic-D}, we find 
\begin{subequations}
\label{eq:gm_subsec-cubic-E}
\begin{align}
\text{\eq{gm_subsec-cubic-A}}&\sim b^{\frac{3}{2}}\,b^{\frac{3+n_\delta}{2}}\,b^{\frac{3+n_\delta}{2}}\,\,, 
\label{eq:gm_subsec-cubic-E-1} \\
\text{\eq{gm_subsec-cubic-B}}&\sim b^2\,b^{\frac{3}{2}}\,b^{\frac{3+n_\delta}{2}}\,b^{\frac{3+n_\delta}{2}}\,\,, 
\label{eq:gm_subsec-cubic-E-2} \\
\text{\eq{gm_subsec-cubic-C}}&\sim 
b^2\,b^{\frac{3+n_\delta}{2}}\,b^{\frac{3+n_\delta}{2}}\,b^{\frac{3+n_\delta}{2}}\,b^{-n_\delta}\,\,, \label{eq:gm_subsec-cubic-E-3} \\
\text{\eq{gm_subsec-cubic-D}}&\sim b^2\,b^{\frac{3}{2}}\,b^{\frac{3}{2}}\,b^{\frac{3+n_\delta}{2}}\,\,. \label{eq:gm_subsec-cubic-E-4}
\end{align}
\end{subequations}
This tells us that, as expected, the contribution of \eq{gm_subsec-cubic-A} is the most relevant at this order in the fields, 
followed by that of \eq{gm_subsec-cubic-B} and then by those of \eqsII{gm_subsec-cubic-C}{gm_subsec-cubic-D}, 
which are equally important in the infrared.

\subsubsection*{Going beyond the tree-level approximation}

\noindent {Let us conclude this section with a very brief sketch of how 
the calculation at all loops would proceed in presence of 
the cross stochasticity $P_{\eps_g\eps_m}$ (and also of the matter stochasticity $P_{\eps_m}$), 
\ie~of how to extend the calculation of Section~\ref{subsec:P_eps_g} to this case.} 

{The key point of the calculation of Section~\ref{subsec:P_eps_g} is the fact that, 
for zero noise $\eps_m$, the conditional likelihood ${\cal P}[\delta|\delta_{\rm in}]$ 
is a Dirac delta functional. This leads to the joint likelihood factorizing nicely, cf.~\eq{loops_subsubsection_1-G}.} 
{This is no longer true if the evolution of the matter field is noisy. 
However, we see from \eqsII{matter_likelihood-B}{joint_likelihood-B} 
that in the actions for ${\cal P}[\delta]$ and ${\cal P}[\delta_g,\delta]$ 
the field $X$ appears at most quadratically. Hence, it should be possible to compute 
the functional integral over $X$ as a (functional) derivative expansion around zero matter noise, 
i.e.~as an expansion of the Dirac delta functional of \eq{loops_subsubsection_1-E}, 
similarly to how the integral of a very narrow Gaussian (normalized and centered in $x_0$) against a slowly-varying function 
can be approximated by the integral of the function against an infinite sum of derivatives 
of a Dirac delta function $\smash{\delta_{\rm D}^{(1)}(x-x_0)}$, 
the $n$th derivative multiplied by a coefficient proportional to the integral 
$\int_{-\infty}^{+\infty}\dif x\,x^n\exp(-x^2/2\sigma^2)$.} 

{The full computation is beyond the scope of this work, and we leave it for a future publication. 
We mention, however, that the same techniques that we would 
use to compute the conditional likelihood at all orders in loops if $P_{\eps_g\eps_m}$ and $P_{\eps_m}$ are not zero 
can be used to compute the conditional likelihood {${\cal P}[\delta_{g_1},\delta_{g_2},\dots\delta_{g_n}|\delta]$} 
for multiple tracers $\delta_{g_1},\delta_{g_2},\dots\delta_{g_n}$ (always, of course, if we assume Gaussian noise).}

\section{Impact of non-Gaussian stochasticities}
\label{sec:calculation_of_likelihood-nongaussian_stochasticities}

\noindent We now move to the study of higher-order stochasticities. In this section it will become even more clear that the scalings 
of $X_g$ and of the initial matter field $\delta_{\rm in}$, \ie~\eqsI{RG-C}, directly tell us 
the relative importance on large scales of the contributions of these non-Gaussian stochasticities. 
This is because we will be able to match the discussion at the end of 
Section~\ref{subsec:action_and_RG_scalings} to the explicit leading contributions to ${\wp}[\delta_g|\delta]$.

\subsection{\texorpdfstring{Bispectrum of galaxy stochasticity and stochasticity in $b_1$}
{Bispectrum of galaxy stochasticity and stochasticity in b\_1}}
\label{subsec:bispectrum_and_noise_in_LIMD_bias}

{Again, we will focus on the stochasticities for galaxies here, since these are the most important in the large-scale limit.} 
As in the previous section, we work to cubic order {in the fields} and at the lowest order in gradients. 
Paralleling the discussion in Section~\ref{subsec:action_and_RG_scalings} we can gain 
insight on which additional terms we must consider in our action by looking at the bias expansion, 
more precisely by considering the three-point functions $\braket{\delta_g\delta_g\delta_g}$, 
$\braket{\delta_g\delta_g\delta}$ and $\braket{\delta_g\delta \delta}$. Besides the stochasticity in the galaxy bispectrum 
captured by a vertex in $S_{g,{\rm int}}$ with three $X_g$ legs, cf.~\eq{RG-E}, 
we consider the impact of a stochasticity in the bias coefficient $b_1$. 
As we have seen in \eq{recall_eps_delta}, this is captured in the bias 
expansion by the operator $\eps_{g,\delta}$, defined in real space by 
\begin{equation}
\label{eq:bias_expansion_stochasticity-A}
\delta_g(\vec{x}) = b_1\delta(\vec{x}) + \eps_g(\vec{x}) + \eps_{g,\delta}(\vec{x})\delta(\vec{x})\,\,,
\end{equation}
where $\eps_{g,\delta}$ has zero correlation with the long-wavelength matter field. 

Let us start by putting $\eps_{g,\delta}$ to zero. At linear order in perturbations 
(\ie~considering only the linear matter field so that $\braket{\delta\delta\delta} = 0$), 
the only non-zero three point function is $\braket{\delta_g\delta_g\delta_g} = \braket{\eps_g\eps_g\eps_g}$. 
Once we switch $\eps_{g,\delta}$ on, 
we find additional contributions to $\braket{\delta_g\delta_g\delta_g}$ proportional to the linear power spectrum, 
and a non-vanishing $\braket{\delta_g\delta_g\delta}$ correlator. More precisely, we find 
\begin{equation}
\label{eq:bias_expansion_stochasticity-B}
\braket{\delta_g(\vec{x})\delta_g(\vec{y})\delta(\vec{z})} = 
\xi_{\rm L}(\abs{\vec{x}-\vec{z}})\braket{\eps_g(\vec{x})\eps_{g,\delta}(\vec{y})} + (\vec{x}\to\vec{y})\,\,,
\end{equation}
where $\xi_{\rm L}$ is the real-space two-point correlation function, and at leading order in derivatives 
{$\braket{\eps_g(\vec{x})\eps_{g,\delta}(\vec{y})}\propto\delta^{(3)}_{\rm D}(\vec{x}-\vec{y})$}. 
The additional {leading} contribution to the {galaxy-galaxy-galaxy three-point function is} 
\begin{equation}
\label{eq:bias_expansion_stochasticity-C}
\begin{split}
\braket{\delta_g(\vec{x})\delta_g(\vec{y})\delta_g(\vec{z})} = 
b_1\xi_{\rm L}(\abs{\vec{x}-\vec{z}})\braket{\eps_g(\vec{y})\eps_{g,\delta}(\vec{x})} + 
b_1\xi_{\rm L}(\abs{\vec{x}-\vec{z}})\braket{\eps_g(\vec{y})\eps_{g,\delta}(\vec{z})} + \text{$2$ perms.} 
\end{split}
\end{equation}

From Tab.~\ref{tab:summary_Z} we see that the impact of a non-zero $\braket{\eps_g\eps_{g,\delta}}$ is captured by a term 
$S_{g,{\rm int}}\supset X_gX_g\delta_{\rm in}$. Indeed, in Appendix~\ref{app:higher_order_stochasticities} we show that the 
inclusion of the terms in {\eqsII{bias_expansion_stochasticity-B}{bias_expansion_stochasticity-C}} corresponds to 
\begin{equation}
\label{eq:bias_expansion_stochasticity-D}
\begin{split}
S_{g,{\rm int}} \supset \frac{1}{2}\int_{\vec{p}_1,\dots\vec{p}_3}\bigg((2\pi)^3\delta^{(3)}_{\rm D}(\vec{p}_{123})\,
D_1P_{\eps_g\eps_{g,\delta}}(\abs{-\vec{p}_1})\,X_g(\vec{p}_1)X_g(\vec{p}_2)\delta_{\rm in}(\vec{p}_3) 
+ (\vec{p}_1\to\vec{p}_2) \bigg)\,\,,
\end{split}
\end{equation}
where $P_{\eps_g\eps_{g,\delta}}(k)\sim k^0$ (at leading order in derivatives) 
is the Fourier transform of the correlation function $\braket{\eps_g(\vec{x})\eps_{g,\delta}(\vec{y})}$. 

A bispectrum $B_{\eps_g\eps_g\eps_g}$ of the galaxy stochasticity, instead, 
contributes to $S_{g,{\rm int}}$ simply as (see also Tab.~\ref{tab:summary_Z})
\begin{equation}
\label{eq:bias_expansion_stochasticity-F}
S_{g,{\rm int}} \supset {-\frac{\iu}{3!}}\int_{\vec{p}_1,\dots\vec{p}_3}(2\pi)^3\delta^{(3)}_{\rm D}(\vec{p}_{123})\,
B_{\eps_g\eps_g\eps_g}(-\vec{p}_1,-\vec{p}_2,-\vec{p}_3)\,X_g(\vec{p}_1)X_g(\vec{p}_2)X_g(\vec{p}_3)\,\,.
\end{equation}

We are now in position to {study} the contributions of these terms to the conditional 
likelihood up to cubic order in the galaxy and matter fields 
{(the details of the calculation are collected in Appendix~\ref{app:higher_order_stochasticities}).} 
First, we consider the contribution from \eq{bias_expansion_stochasticity-D}, and find 
\begin{equation}
\label{eq:bias_expansion_stochasticity-G}
\begin{split}
\Delta\wp^{(3)}[\delta_g|\delta] = 
\frac{1}{2}\int_{\vec{p}_1,\dots\vec{p}_3}\bigg(&(2\pi)^3\delta^{(3)}_{\rm D}(\vec{p}_{123})\,
P_{\eps_g\eps_{g,\delta}}(\abs{\vec{p}_1})\, 
\frac{\delta_g(\vec{p}_1)-b_1\delta(\vec{p}_1)}{P_{\eps_g}(\vec{p}_1)}\,
\frac{\delta_g(\vec{p}_2)-b_1\delta(\vec{p}_2)}{P_{\eps_g}(\vec{p}_2)}\,\delta(\vec{p}_3) \\
& + (\vec{p}_1\to\vec{p}_2)\bigg)\,\,. 
\end{split}
\end{equation}
At leading order in derivatives all the noise power spectra are constant in $k$: 
we can then rewrite this in real space as\footnote{Notice 
that $\smash{\eps_g}$ and $\smash{\eps_{g,\delta}}$ have the same dimension: this can be seen directly from 
\eq{bias_expansion_stochasticity-A}. Hence the ratio between 
$\smash{P_{\eps_g}}$ and $\smash{P_{\eps_g\eps_{g,\delta}}}$ is dimensionless.} 
\begin{equation}
\label{eq:bias_expansion_stochasticity-H}
\Delta\wp^{(3)}[\delta_g|\delta] = \frac{P^{\{0\}}_{\eps_g\eps_{g,\delta}}}{P^{\{0\}}_{\eps_g}}\int\dif^3x\,
\frac{\big(\delta_g(\vec{x})-b_1\delta(\vec{x})\big)^2\,\delta(\vec{x})}{P^{\{0\}}_{\eps_g}} 
\end{equation}
{(notice the similarity with \eq{gm_subsec-cubic-D}, modulo an additional derivative suppression there).} 

What about \eq{bias_expansion_stochasticity-F}? Again, we stop at leading order in derivatives, 
and we write the noise bispectrum $B_{\eps_g\eps_g\eps_g}$ as (see e.g.~Eq.~(2.86) of \cite{Desjacques:2016bnm}) 
\begin{equation}
\label{eq:definition_of_stochastic_bispectrum}
B^{\{0\}}_{\eps_g\eps_g\eps_g} = \lim_{k_1,k_2\to0}\braket{\eps_g(\vec{k}_1)\eps_g(\vec{k}_2)\eps_g(\vec{k}_3)}' = 
S^{\eps_g\eps_g\eps_g}_{3}\big(P^{\{0\}}_{\eps_g}\big)^2\,\,.
\end{equation}
Hence, with similar manipulations as those that led to \eq{bias_expansion_stochasticity-H}, we arrive at 
\begin{equation}
\label{eq:bias_expansion_stochasticity-I}
\Delta\wp^{(3)}[\delta_g|\delta] = \frac{S^{\eps_g\eps_g\eps_g}_{3}}{3!}
\int\dif^3x\,\frac{\big(\delta_g(\vec{x})-b_1\delta(\vec{x})\big)^3}{P^{\{0\}}_{\eps_g}}\,\,.
\end{equation}

Before proceeding, we can discuss the relative importance of these two terms, and also compare them 
with the cubic terms of Sections~\ref{subsec:P_eps_g} and \ref{subsec:P_eps_g_eps_m}.

\subsection{Relative importance with respect to deterministic evolution}
\label{subsec:NG_scalings}

Let us first compare the two terms of \eqsII{bias_expansion_stochasticity-H}{bias_expansion_stochasticity-I} 
with the contribution at cubic order that we have when 
only the noise $P_{\eps_g}$ is non-vanishing, \ie~\eq{gm_subsec-cubic-A}. 
In this way we can confirm that our predictions for the scalings in the infrared 
of Section~\ref{subsec:action_and_RG_scalings}, \emph{that are derived at the level of the action} 
before computing the actual likelihood, were indeed correct. 

{F}ollowing the same arguments of Section~\ref{subsec:P_eps_g_eps_m} we can see that 
\begin{subequations}
\label{eq:bias_expansion_stochasticity-J}
\begin{align}
\text{\eq{bias_expansion_stochasticity-H}}&\sim 
b^{\frac{3}{2}}\,b^{\frac{3}{2}}\,b^{\frac{3+n_\delta}{2}}\,\,, \label{eq:bias_expansion_stochasticity-J-1} \\
\text{\eq{bias_expansion_stochasticity-I}}&\sim 
b^{\frac{3}{2}}\,b^{\frac{3}{2}}\,b^{\frac{3}{2}}\,\,, \label{eq:bias_expansion_stochasticity-J-2}
\end{align}
\end{subequations}
which we compare to the $b^{9/2+n_\delta}$ scaling, cf.~\eq{gm_subsec-cubic-E-1}, 
for the leading term at cubic order that we have for Gaussian stochasticities. 
We see that these scalings are \emph{exactly} those we derived in \eqsII{RG-F}{RG-G}. 
This confirms that, in general, we can check the relative importance of the various terms in the 
{conditional} likelihood by looking at the (ir)relevance of the different operators 
in the actions $S_{\rm int}$ and $S_{g,{\rm int}}$. 

In Section~\ref{subsec:action_and_RG_scalings} we saw that the term coming from the stochasticity in $b_1$ is 
more relevant on large scales than the one coming from the three-point function of $\eps_g$, and that both 
are less relevant than the one of \eq{gm_subsec-cubic-A}. What about the relative importance with the 
three cubic contributions coming from the cross stochasticity between galaxies and matter? 
If we compare \eqsI{bias_expansion_stochasticity-J} with 
\eqsIII{gm_subsec-cubic-E-2}{gm_subsec-cubic-E-3}{gm_subsec-cubic-E-4}, we see that 
the latter terms are always smaller at large scales, for $n_\delta=\num{-1.7}$: for example, 
we can see \eq{gm_subsec-cubic-D} as a higher-derivative correction to \eq{bias_expansion_stochasticity-H}. 
We also notice that the relative importance of the contribution coming from $B_{\eps_g\eps_g\eps_g}$ 
and the leading one from a mixed galaxy-matter stochasticity, i.e.~\eq{gm_subsec-cubic-B}, scales only as $b^{n_\delta+2} = b^{0.3}$. 

\begin{figure}
\centering
\includegraphics[width=\columnwidth]{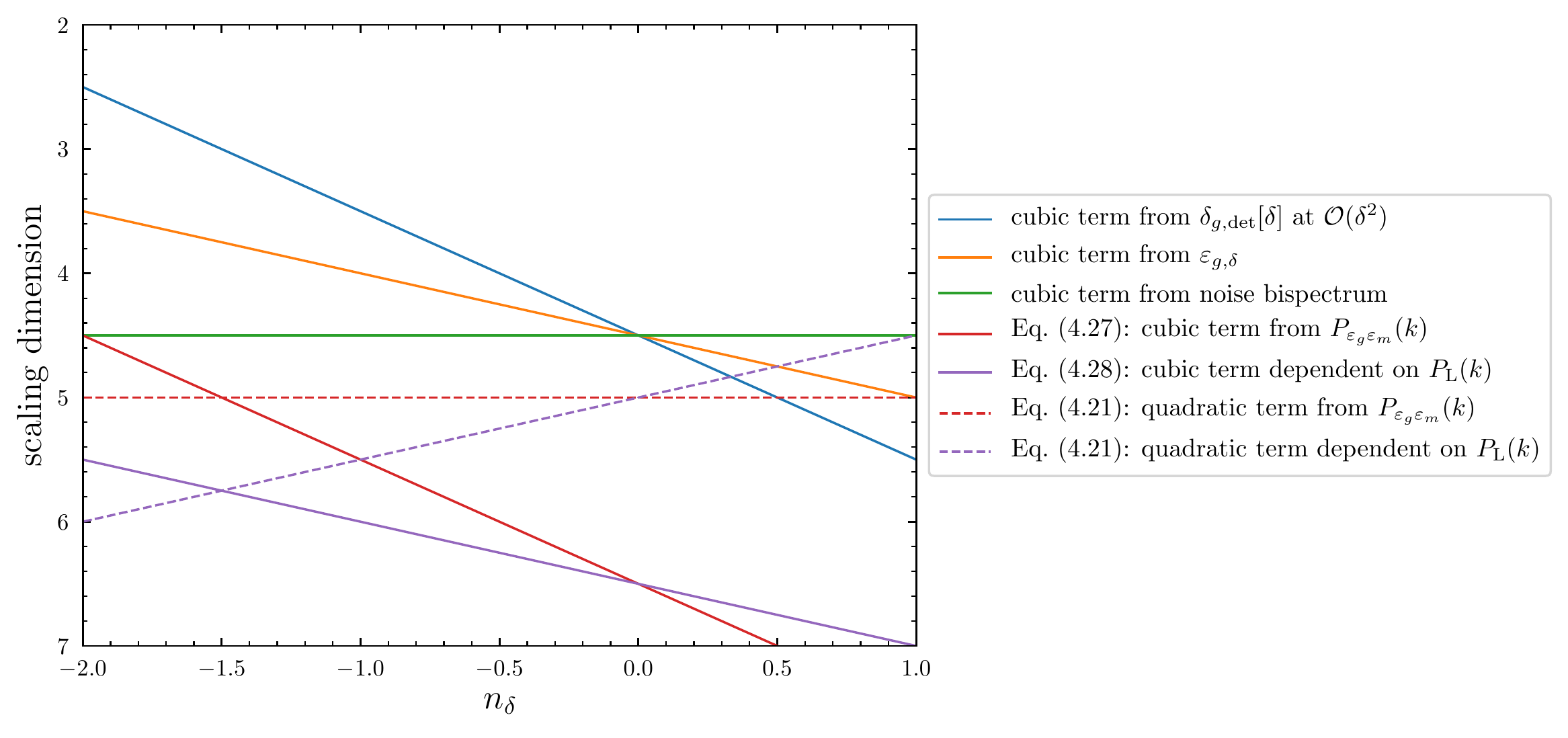} 
\caption{Scaling dimension, i.e.~relative importance, of the various contributions to $\wp[\delta_g|\delta]$ 
discussed in Sections~\ref{subsec:P_eps_g_eps_m} and \ref{subsec:bispectrum_and_noise_in_LIMD_bias}. 
More precisely, we plot their scaling dimension as a function of the 
spectral index $n_\delta$ of the linear power spectrum, ranging from $-2$, the value of $n_\delta$ on very small scales, to $1$, the value of $n_\delta$ on scales much larger
than the equality scale. The contributions considered here are at most at cubic order in 
the fields $\delta_g-b_1\delta$ and $\delta$. At this order we see, for example, that the stochasticity in 
the linear bias is always less relevant than the deterministic evolution at second order and 
always more relevant than the non-Gaussianity of the noise if $n_\delta$ is negative.} 
\label{fig:scalings}
\end{figure}

In the next section we will discuss in more detail these results, which are summarized in Fig.~\ref{fig:scalings}, 
and how all the scalings that we discussed so far reflect the presence of three expansion parameters 
for the logarithm of the conditional likelihood. This section and Section~\ref{subsec:P_eps_g_eps_m} already show very clearly that, 
in addition to the derivative expansion ($n$ derivatives scale as $b^n$) and the expansion 
in the smallness of the matter field on large scales ($n$ powers of the matter field scale as $b^{n(3+n_\delta)/2}$), 
there is the expansion in the galaxy stochasticity $\delta_g(\vec{x})-b_1\delta(\vec{x})$, 
which scales as $b^{{3}/{2}}$. \emph{This, combined with the fact that we can 
anticipate these scalings before actually computing the likelihood \emph{(simply 
by looking at how many powers of the fields $\delta_{\rm in}$ and $X_g$ a given operator in 
the action $S_{g,{\rm int}}$ contains)}, is a key result of this work.} 
We will elaborate on it more throughout the rest of the paper.

\section{Discussion}
\label{sec:discussion}

\noindent Before drawing our conclusions, we stop to discuss in more detail the last point of the 
previous section, \ie~how the non-Gaussianity of the noise introduces a new expansion parameter in 
the game. We do this in Section~\ref{subsec:expansion_parameters} below. 
Section~\ref{subsec:loops} discusses how to go beyond the tree-level approximation 
that we have (mostly) used throughout both Sections~\ref{sec:calculation_of_likelihood-gaussian_stochasticities} 
and \ref{sec:calculation_of_likelihood-nongaussian_stochasticities}.

\subsection{Three expansion parameters}
\label{subsec:expansion_parameters}

\noindent In the previous four sections we have shown in detail how to compute the corrections 
to the conditional likelihood derived in \cite{Schmidt:2018bkr}. 
While the form of these corrections is important {in itself}, as we will argue in Section~\ref{subsec:bayesian_forward_modeling}, 
the most important takeaway point is how to study their relative importance. 
The scalings that we derived from dimensional analysis correspond to the fact that the 
corrections to the ``Gaussian'' likelihood of \cite{Schmidt:2018bkr} are controlled 
by \emph{three} expansion parameters (see also Tab.~\ref{tab:expansion_parameters} for a summary): 
\begin{enumerate}[leftmargin=*]
\item first, we have the derivative expansion. In Fourier space this is an expansion in powers of $k^2$, 
controlled by whatever is the longest nonlocality scale in the process. 
This can be the halo Lagrangian radius $R(M_h)$, if the formation of the tracer is mainly controlled by gravity. 
Pressure forces, radiative-transfer effects and other physical processes affecting the formation 
of galaxies can add new nonlocality scales in the problem;\footnote{Here 
we have in mind the bias expansion of the tracer overdensity in terms of the nonlinear matter overdensity, 
hence the higher-derivative terms controlled by $\vec{\nabla}/k_{\rm NL}$ do not make an appearance.} 
\item then, we have an expansion in the perturbations $\delta(\vec{x})$ of the matter field. On large scales 
their size is controlled by $\sqrt{(k^3/2\pi^2) P_{\rm L}(k)} = \sqrt{(k/k_{\rm NL})^{3+n_\delta}}$ for a power-law power spectrum; 
\item finally, there is the stochasticity of the tracer field. On large scales, 
we have seen that the relative importance of the contributions coming from higher-order stochasticities 
to those coming from the deterministic bias expansion (in which we consider only the scale-independent noise power spectrum $P_{\eps_g}$) 
are controlled by the relative size of $\delta_g(\vec{x})-b_1\delta(\vec{x})$ and $\delta(\vec{x})$, \ie~by the ratio 
$\smash{\sqrt{(k^3/2\pi^2) P_{\eps_g}}/\sqrt{(k^3/2\pi^2) P_{\rm L}(k)}}$. 
For a scale-independent $P_{\eps_g}$ (\ie~at zeroth order in derivatives) 
and a power-law power spectrum, this expansion parameter is 
\begin{equation}
\label{eq:expansion_parameters-A}
\sqrt{\frac{P_{\eps_g}^{\{0\}}}{P_{\rm L}(k)}} = \sqrt{\frac{k^3_{\rm NL}
P_{\eps_g}^{\{0\}}}{2\pi^2}}\,\bigg(\frac{k}{k_{\rm NL}}\bigg)^{-\frac{n_\delta}{2}}\,\,.
\end{equation}
\end{enumerate}

\begin{table}
\myfloatalign
\caption[.]{Expansion parameters for the conditional likelihood ${\cal P}[\delta_g|\delta]$. 
The scale $k_{\eps_g}$, for a power-law power spectrum, is defined by \eq{expansion_parameters-B}. 
Notice that in the contributions of Section~\ref{subsec:P_eps_g_eps_m}, 
which come up when considering the cross stochasticity $P_{\eps_g\eps_m}$, 
the linear power spectrum itself appears, but there is no need to include it as an additional expansion parameter 
{since the second line below already accounts for it}.} 
\label{tab:expansion_parameters}
\centering
\medskip
\begin{tabular}{lll}
\toprule
parameter & Fourier space & power-law power spectrum \\
\midrule
derivatives & $R(M_h) k$ & \\[2ex]
matter perturbations & $\sqrt{\dfrac{k^3P_{\rm L}(k)}{2\pi^2}}$ & $\bigg(\dfrac{k}{k_{\rm NL}}\bigg)^{\frac{3+n_\delta}{2}}$ \\[2ex]
stochasticity & $\sqrt{\dfrac{k^3P^{\{0\}}_{\eps_g}}{2\pi^2}}$ & \\[2ex]
\midrule
$\dfrac{\text{stochasticity}}{\text{matter perturbations}}$ & $\sqrt{\dfrac{P^{\{0\}}_{\eps_g}}{P_{\rm L}(k)}}$ & 
$\bigg(\dfrac{k}{k_{\eps_g}}\bigg)^{{-\frac{n_\delta}{2}}}$ \\[2ex]
\bottomrule
\end{tabular}
\end{table}

What does this tell us? Let us consider, for example, the likelihood of \cite{Schmidt:2018bkr} 
with only $P_{\eps_g}$ that is non-vanishing, \ie~\eq{cubic_conditional_likelihood-E}, 
and focus on the zeroth order in derivatives. If we expand this in perturbations of the matter field, 
and stop at third order in the fields, we get 
\eq{gm_subsec-cubic-A}. In the previous section we have seen that, as long as we restrict to sufficiently large scales 
such that the linear matter power spectrum is larger than that of the noise, 
this term dominates over the corrections from higher-order stochasticities. 
However, once we go to higher order in the fields and consider for example 
the contribution at fourth order in the deterministic bias expansion, 
the resummed likelihood contains, \eg, a term of the form $\sim (\delta_g-b_1\delta)\delta^4$ 
(simply by expanding the square of $\delta_g-\delta_{g,{\rm det}}[\delta]$). 
Even if we focus on $k$ small enough {that} $P_{\rm L}(k)$ is larger than $\smash{P_{\eps_g}^{\{0\}}}$, 
we see that the contribution of \eq{bias_expansion_stochasticity-H} 
from the stochasticity in the $b_1$ bias coefficient 
is expected to be more important than the above-mentioned deterministic term. 
Explicitly, we cutoff our galaxy and matter fields at a scale $\Lambda$, so 
that only modes of $\delta$ and $\delta_g$ with $\abs{\vec{k}} < \Lambda$ are included 
(notice that we cutoff the \emph{final} matter field, 
the one that can be obtained by non-perturbative forward models like N-body simulations\footnote{{The fact 
that it must be the final density field to be cutoff at the scale $\Lambda$ comes out naturally from 
the approach of this paper. To see it recall that, when we cutoff our 
path integrals for ${\cal P}[\delta]$ and ${\cal P}[\delta_g,\delta]$ 
(in order for them to make sense mathematically), what we do is cutoff the fields $X_g$, $X$ and $\delta_{\rm in}$ 
at a scale $\Lambda$, which is taken to be the same for all three fields (for simplicity). 
The fields $\delta_g$ and $\delta$ are nothing but the currents associated with $X_g$ and $X$, to which 
they are coupled \emph{linearly}, cf.~\eqsII{matter_likelihood-A}{joint_likelihood-A}. Hence, 
by imposing a cutoff on $X_g$ and $X$, we ensure that 
the short modes $\abs{\vec{k}}\geq\Lambda$ of $\delta_g$ and $\delta$ are effectively put to zero.}}). 
Then, dropping the universal scaling of the volume element for simplicity, 
the contribution of a term $\sim (\delta_g-b_1\delta)\delta^4$ scales as 
$\Lambda^{3/2}\,\Lambda^{4(3+n_\delta)/2}$ for $\Lambda\to 0^+$, to be compared with 
the scaling $\Lambda^{{3}/{2}}\,\Lambda^{{3}/{2}}\,\Lambda^{({3+n_\delta})/{2}}$ 
of the stochasticity in $b_1$, which is more important for $n_\delta=-1.7$. 

We had seen that this had to be the case already at the end of Section~\ref{subsec:action_and_RG_scalings}, cf.~\eq{RG-G}, 
before doing any calculation of the conditional likelihood. Thus, 
it is not necessarily useful to use the likelihood of 
\eq{cubic_conditional_likelihood-E} at very high orders 
in the deterministic bias expansion, since many of the terms that are 
included in that likelihood are less relevant on large scales than other terms that are already neglected {by it}. 

Let us discuss in more detail the condition ${\sqrt{{P^{\{0\}}_{\eps_g}}/{P_{\rm L}(k)}}} \ll 1$. 
For a power-law power spectrum, this identifies a scale $k_{\eps_g}$ such that 
\begin{equation}
\label{eq:expansion_parameters-B}
\text{$\sqrt{\frac{P^{\{0\}}_{\eps_g}}{P_{\rm L}(k_{\eps_g})}} = 1\,\,,$ \quad i.e.~\quad 
$k_{\eps_g} = k_{\rm NL}\left(\frac{k^3_{\rm NL}P^{\{0\}}_{\eps_g}}{2\pi^2}\right)^{\frac{1}{n_\delta}} $}\,\,.
\end{equation}
On scales shorter than this, the stochasticities become larger than the matter fluctuations, 
and it is not possible to treat terms like those of Section~\ref{subsec:bispectrum_and_noise_in_LIMD_bias} perturbatively. 

Other two important points are the following. First, in order to define this scale $k_{\eps_g}$, 
we have worked under the assumption of a power-law power spectrum with $n_\delta = \num{-1.7}$. 
This tells us that, no matter how large is $\smash{P^{\{0\}}_{\eps_g}}$ with respect to $2\pi^2/k^3_{\rm NL}$, 
there is always a solution to \eq{expansion_parameters-B}, 
as long as we go {to} sufficiently small $k$. However, in the real {U}niverse 
the matter power spectrum is not a power law: we do not have arbitrarily large inhomogeneities on large scales, 
where indeed the matter distribution is homogeneous and isotropic. Then, it is possible that for some tracers 
$\smash{P^{\{0\}}_{\eps_g}}$ is larger than the linear matter power spectrum $P_{\rm L}(k)$ for all $k$. 
In this limit the terms coming from higher-order stochasticities will \emph{always} be more important than the ones contained in the 
``deterministic'' likelihood of Section~\ref{subsec:P_eps_g}, and we cannot treat them perturbatively. 

Second, we also emphasize that these estimates of the importance of the noise hold in an ``average'' sense.\footnote{In 
fact, $\smash{\sqrt{k^3 P_\varphi(k)}}$ gives only the \emph{typical} size of 
a perturbation of momentum $k$ of a field $\varphi(\vec{x})$.} 
On the other hand, if we are close to the point $\delta_g=\delta_{g,{\rm det}}[\delta]$, which is the maximum-likelihood point for the 
``Gaussian'' likelihood of \eq{cubic_conditional_likelihood-E}, 
the contributions from higher-order stochasticities are suppressed (since they are controlled by $\delta_g-b_1\delta$). 
{Contrast this with those coming from the stochasticity in $b_1$: 
they scale as $(\delta_g - b_1 \delta)^2\delta$, and thus are potentially less suppressed.} 
We will have more to say about these points in Section~\ref{subsec:bayesian_forward_modeling}. 

Before proceeding we point out that, as it happens in any effective field theory, 
the estimates of this section and of Section~\ref{subsec:NG_scalings} do not account for possible hierarchies between 
the dimensionless coefficients multiplying the various operators. For example, 
let us compare the contribution of \eq{gm_subsec-cubic-A}, 
assuming $\delta_{g,{\rm det}}^{(2)}[\delta] = b_2\delta^2/2$, 
with that of \eq{bias_expansion_stochasticity-H}, stopping at leading order in derivatives. 
We see that the relative size of the two terms, at a fixed scale, depends on the ratio between $b_2$ 
and $\smash{P^{\{0\}}_{\eps_g\eps_{g,\delta}}/P^{\{0\}}_{\eps_g}}$. 
If we assume that the noise for the galaxy sample under consideration follows closely 
a Poisson distribution, {the ratio of the two spectra is a number of order $1$.} 
Then, it is clear that for a highly biased tracer with $b_2\gg 1$ the importance of the term in \eq{gm_subsec-cubic-A} is enhanced. 
These hierarchies can depend on many things, like the properties of the galaxy sample, 
the redshift, and so on: we will not discuss them further.

\subsection{Regarding loops}
\label{subsec:loops}

\noindent So far {most} of our computations have been done at tree level. What about loops? 
{After our brief encounter with loops of the initial matter field 
$\delta_{\rm in}$ in Section~\ref{subsec:stochastic_terms_and_tadpoles}, 
here we want to ask a different question: what happens to loops of the fields $X_g$ and $X$, 
which arise when we compute the likelihood? This is a question that is 
specific to this paper: a discussion about loops on more generic terms is left to Appendix~\ref{app:loops}.}

\subsubsection*{Gaussian stochasticities}

\noindent Let us first consider the case of only the stochasticity $P_{\eps_g}$ being different from zero. 
Then, looking at the path integral for the joint likelihood of \eqsIII{action_joint-A}{action_joint-B-temp}{action_joint-B}, 
we quickly realize that we cannot have loops with internal lines of $X_g$ since 
we do not have interaction vertices carrying more than one $X_g$ field. At most we can have tree-like diagrams as 
\begin{equation}
\label{eq:loops_subsection_1-A}
\raisebox{-0.0cm}{\includegraphicsbox[scale=0.25,trim={1.5cm 3cm 1.5cm 3cm},clip]{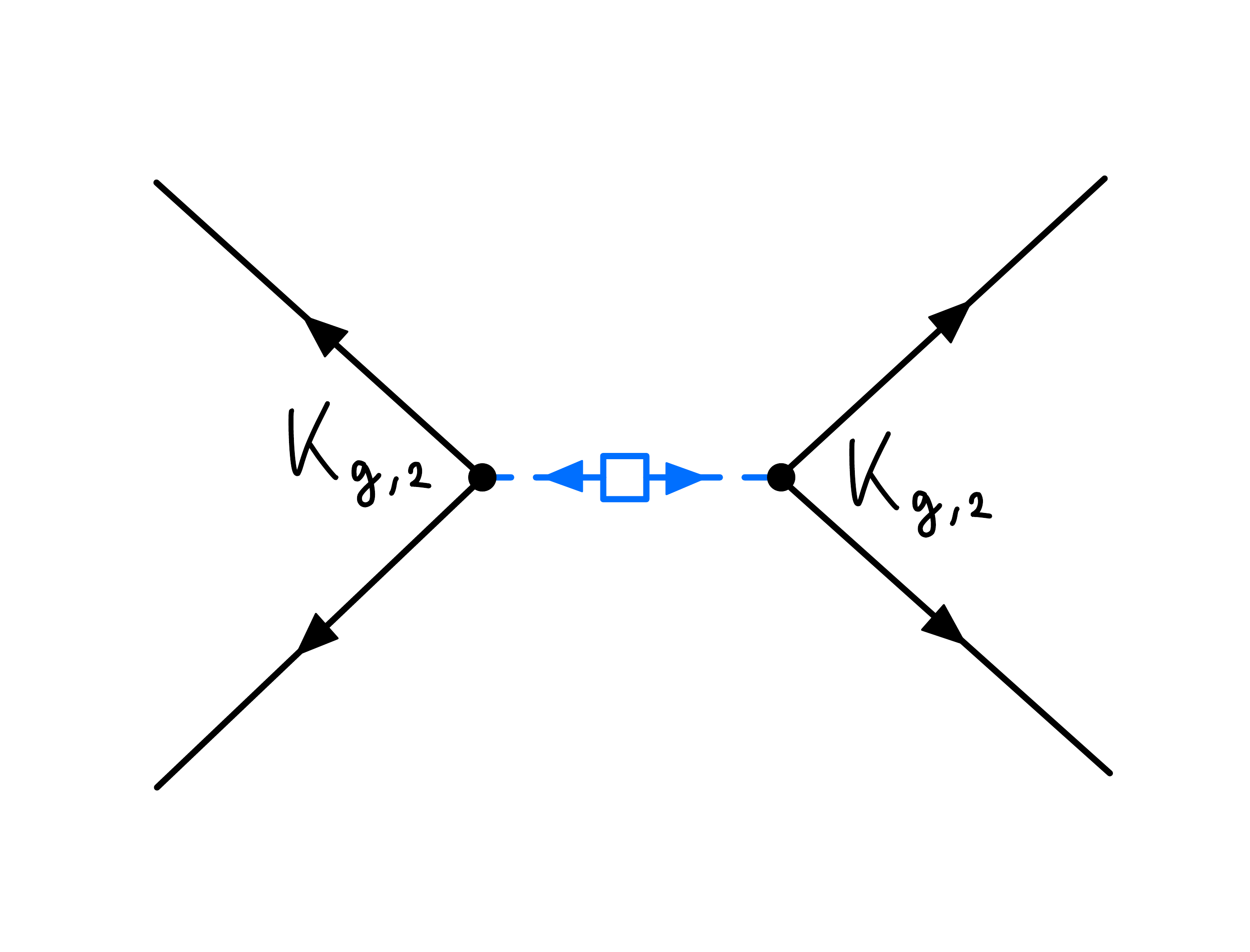}}\,\,, 
\end{equation}
where we use the blue wiggly line to denote the $X_g = \iu J_g$ field {(see Tab.~\ref{tab:feynman})}, 
and we denote its propagator by 
$\braket{X_gX_g}'=\!\!\raisebox{-0.0cm}{\includegraphicsbox[scale=0.1,trim={2cm 8cm 2cm 8cm},clip]{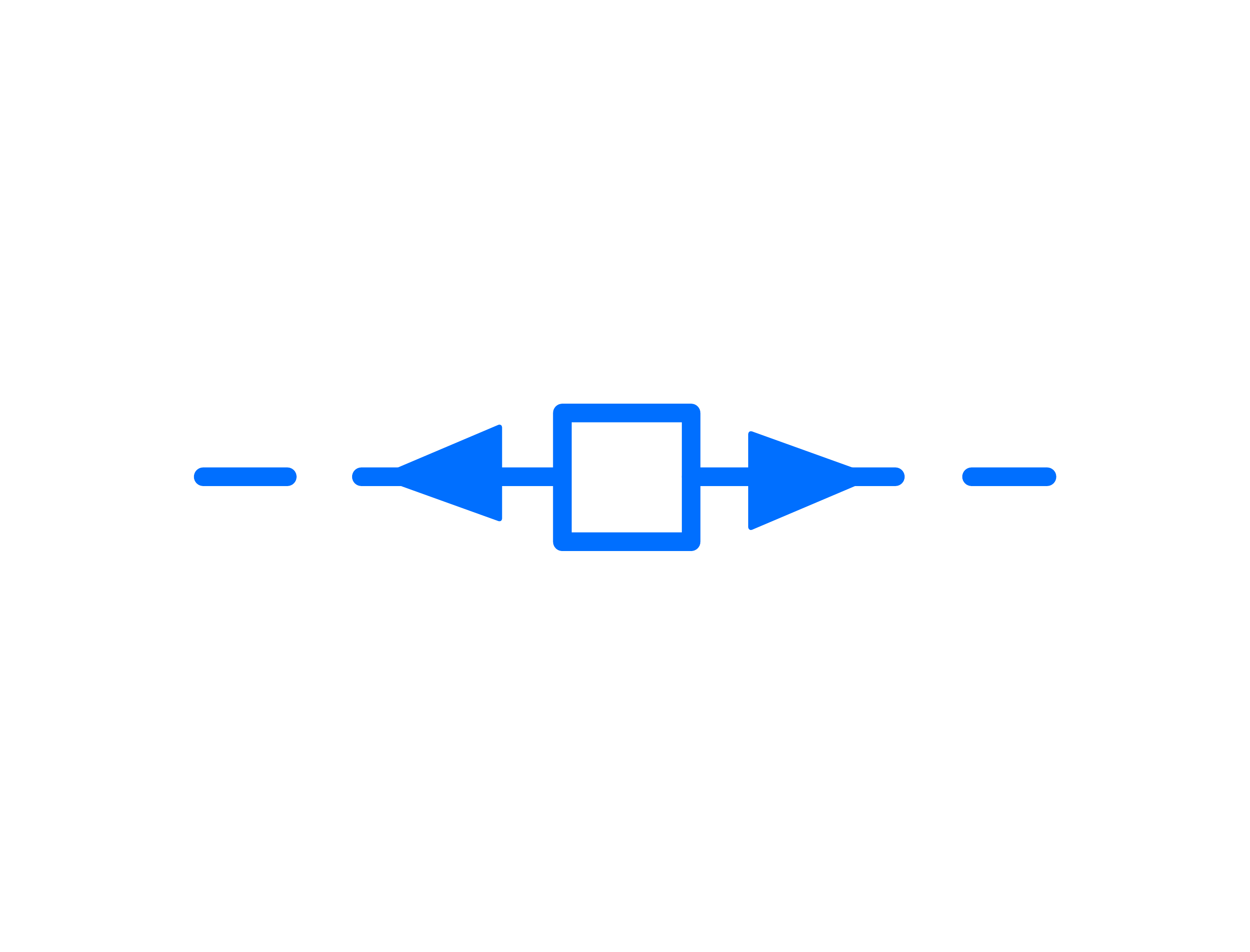}}\!$. 
{Moreover,} in the remainder of {the paper} we will be more schematic with diagrams, since from now on the 
discussion will be mostly qualitative. 
Also, we will not use anymore the shell-by-shell integration of Section~\ref{subsec:stochastic_terms_and_tadpoles}, 
so we will not have thick lines in loops. 

{The important point about diagrams like that of \eq{loops_subsection_1-A} is that they} 
are not UV-sensitive, and already included in the calculation at all orders in loops 
that we carried out {in} Section~\ref{subsec:P_eps_g} (more precisely, the diagram above 
is a fourth-order term in the likelihood, that comes from the last term of \eq{fourth_order_check-B}).

\subsubsection*{Non-Gaussian stochasticities}

\noindent Things change if we include non-Gaussian stochasticities. 
Let us consider, for example, some interaction of the form $X_gX_gX_g$, 
like that of \eq{bias_expansion_stochasticity-F}, or some interaction mixing 
the tracer stochasticity with the matter one, \eg~a $X_gX_gX$ vertex. 
In the joint likelihood these vertices give rise to one-loop diagrams like 
\begin{equation}
\label{eq:loops_subsection_2-A}
\raisebox{-0.0cm}{\includegraphicsbox[scale=0.25,trim={5.5cm 6cm 5.5cm 6cm},clip]{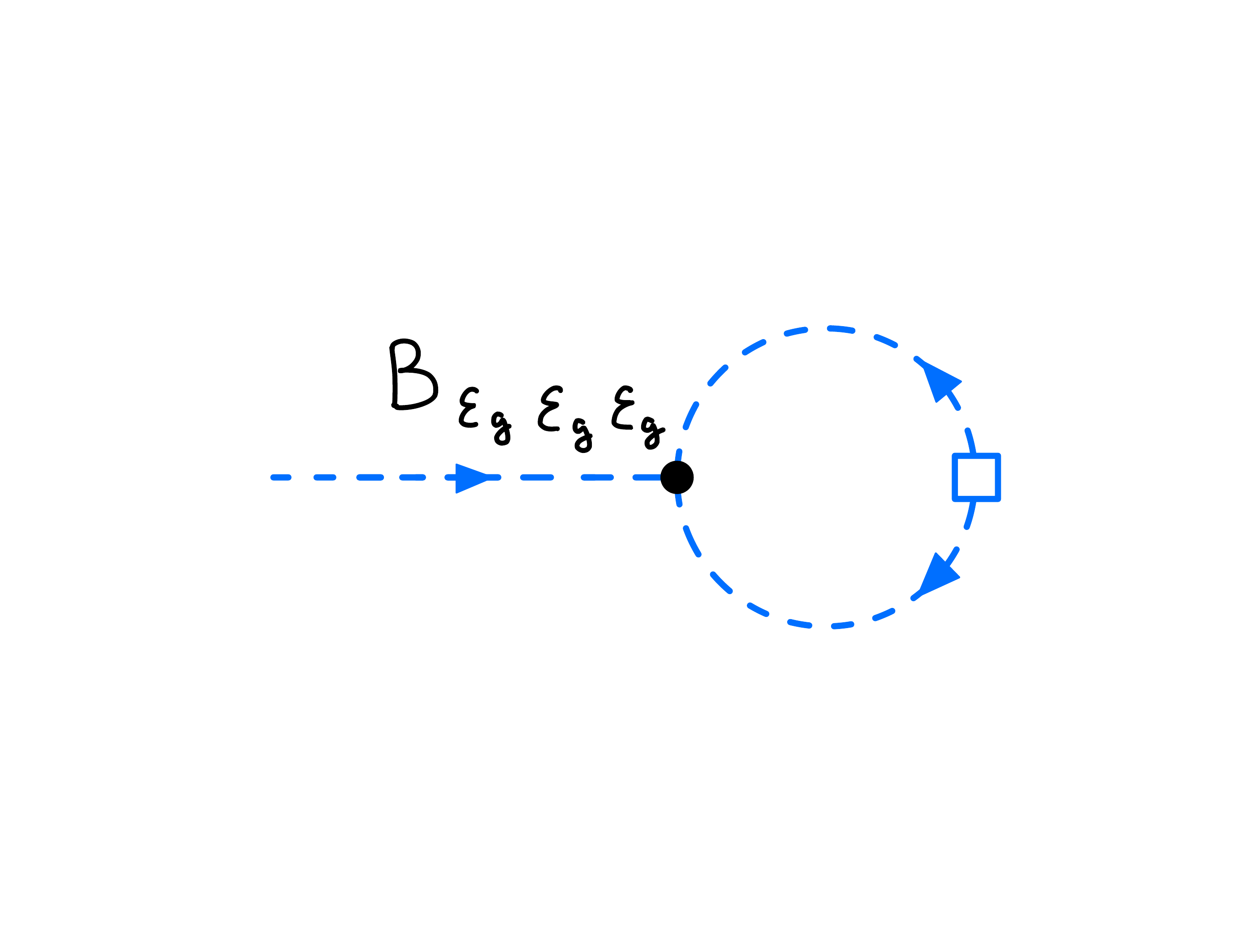}}
\,\,\,\,\,\,,\quad
\raisebox{-0.0cm}{\includegraphicsbox[scale=0.25,trim={5.5cm 6cm 5.5cm 6cm},clip]{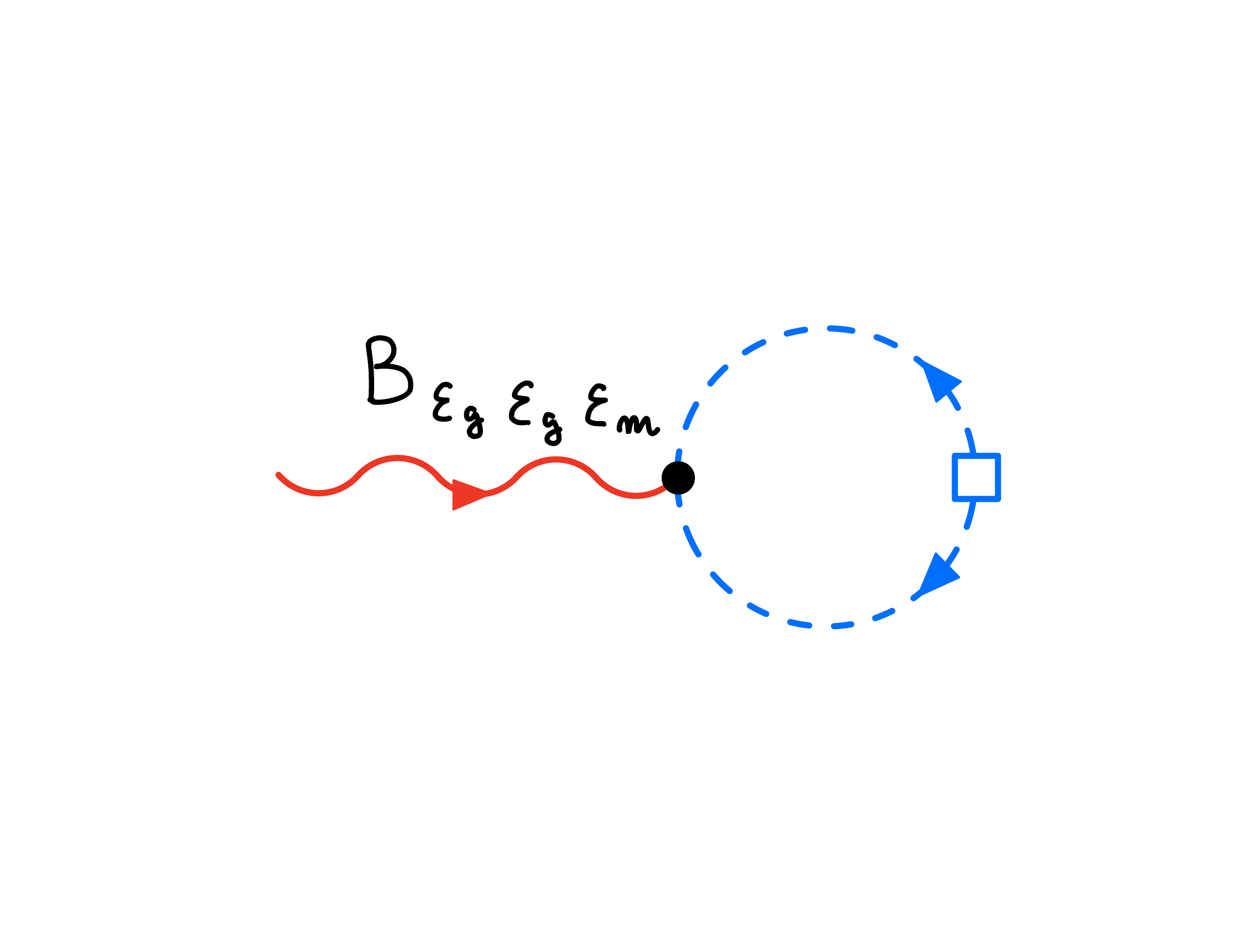}}
\end{equation}
which, similarly to that of \eq{galaxy_tadpole_loop}, generate a tadpole for galaxies and matter, respectively. 
The tadpole for galaxies can be dealt with in the same way discussed at the end of Section~\ref{subsec:stochastic_terms_and_tadpoles}. 
What about the tadpole for matter? The $X_gX_gX$ interaction is zero in the limit that the momentum of $X$ {becomes very soft}, 
again because of matter and momentum conservation: hence, the second diagram of the above equation is actually vanishing. 
This will continue to hold at higher orders: no tadpole for $\delta$ is generated, 
and any term linear in $X_g$ can be reabsorbed by a redefinition of $\delta_g$. 

Another interaction that we discussed was that of \eq{bias_expansion_stochasticity-D}, 
i.e.~the stochastic correction to the linear LIMD bias. 
Two of the diagrams that we get from this vertex are 
\begin{equation}
\label{eq:loops_subsection_2-B}
\raisebox{-0.0cm}{\includegraphicsbox[scale=0.25,trim={1cm 4cm 1cm 4cm},clip]{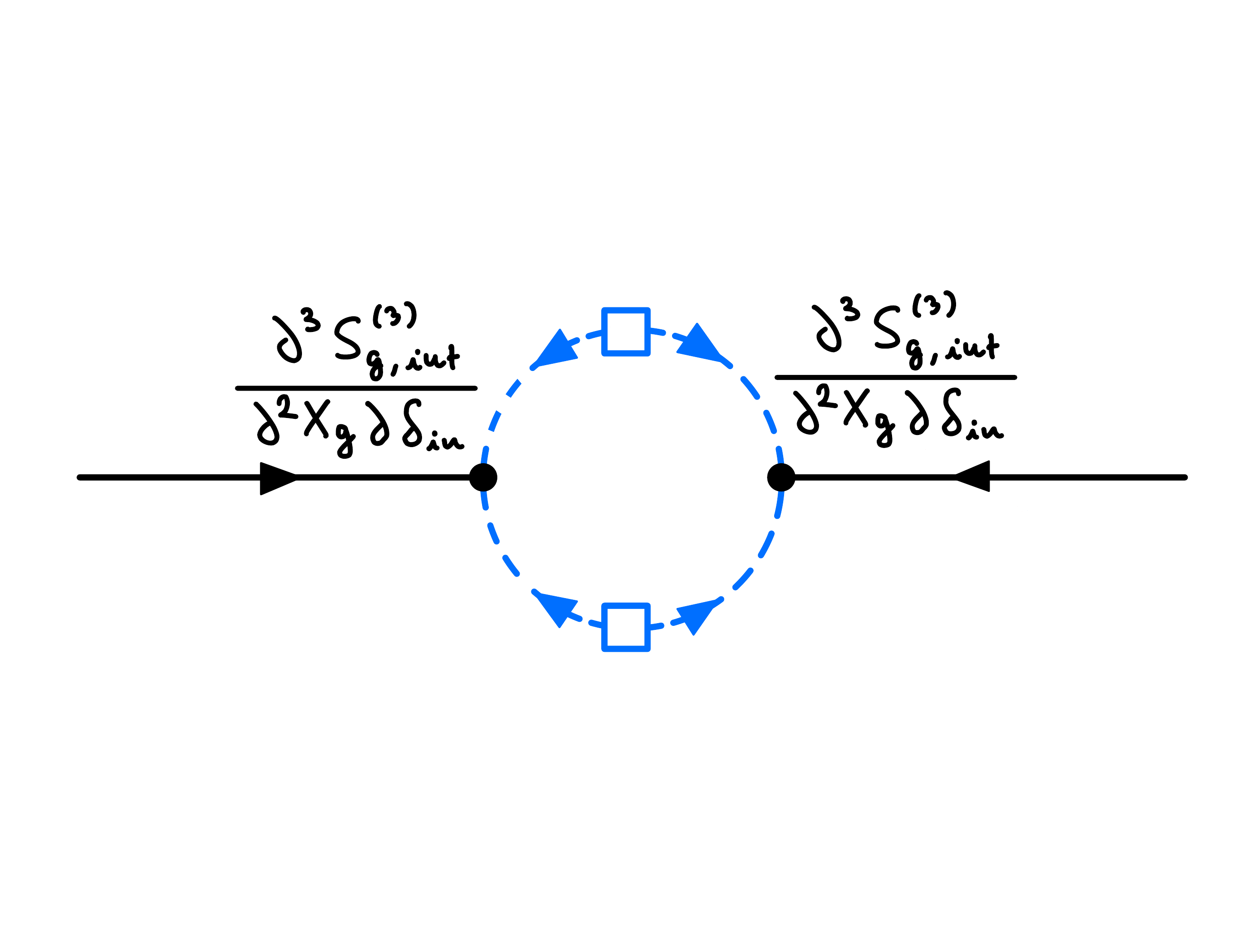}}
\,\,\,\,\,\,,\quad
\raisebox{-0.0cm}{\includegraphicsbox[scale=0.25,trim={1cm 2.5cm 1cm 2.5cm},clip]{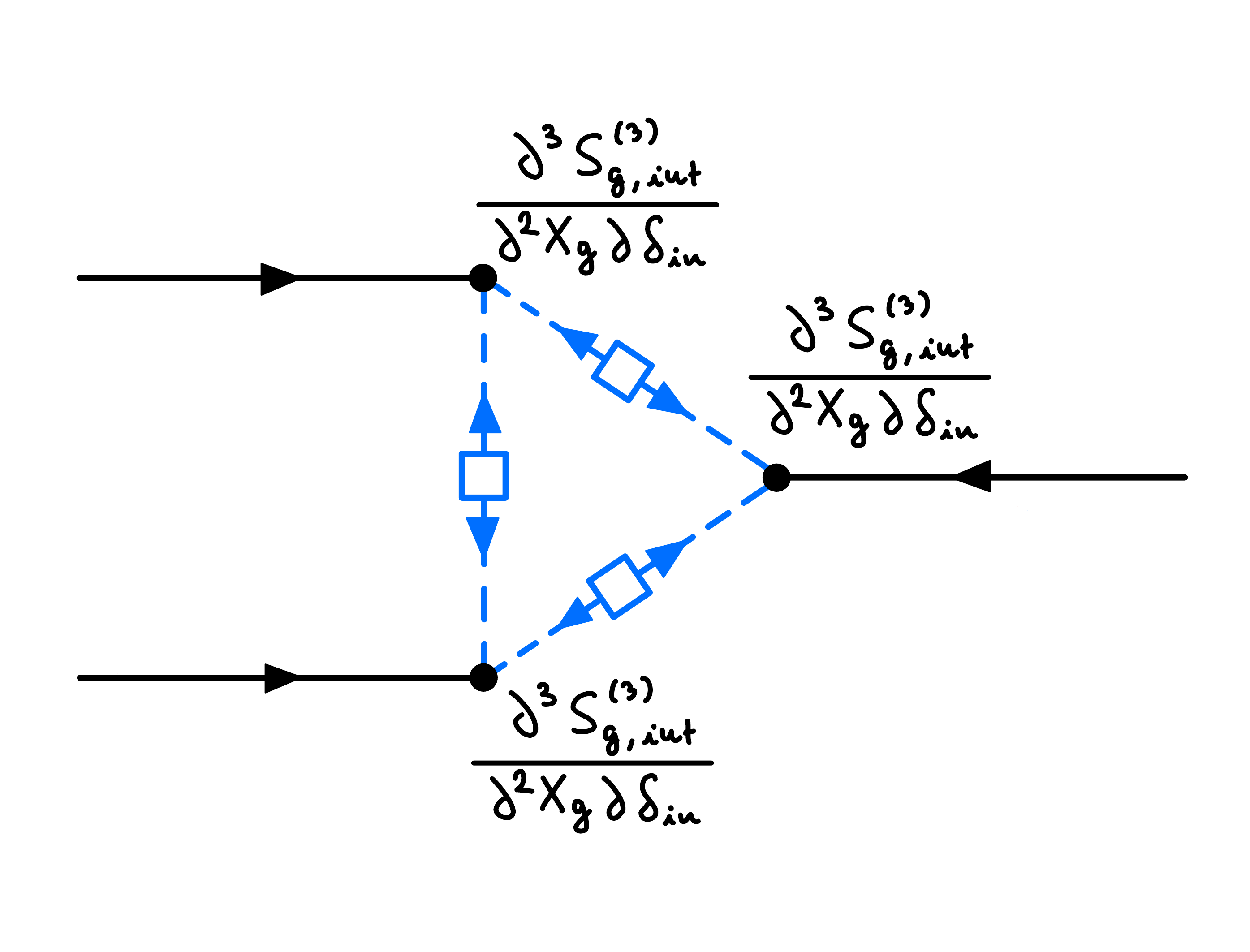}}
\,\,\,\,\,\,.
\end{equation}
At leading order in derivatives, the first of these diagrams renormalizes the inverse of the 
initial power spectrum as $P_{\rm in}^{-1}(k)\to P_{\rm in}^{-1}(k)^{-1}+\text{const.}$, 
while the second {one} generates a cubic local interaction 
for $\delta_{\rm in}$ of the form $\propto\int\dif^3x\,\delta^3_{\rm in}(\vec{x})$. 

What is the impact of these terms on the conditional likelihood? 
{T}he {first} thing {that we have} to notice is that, at this order in derivatives, these terms are generated \emph{only} 
in the action for the joint likelihood: the matter likelihood is left untouched. {More precisely, 
in the EFT for matter we can only have a stochastic correction to the speed of sound, there is no bias parameter 
$b_1$: the equivalent of the first diagram of \eq{loops_subsection_2-B}, with a UV-sensitive 
loop of the field $X$, gives rise to a term of the form 
$\propto\int\dif^3x\,\nabla^2\delta_{\rm in}(\vec{x})\nabla^2\delta_{\rm in}(\vec{x})$, which 
renormalizes the inverse initial power spectrum by a \mbox{term proportional to $k^4$.}} 

{I}n order to assess qualitatively the impact of these terms on 
$\wp[\delta_g|\delta]$, then, we consider only the case where we include them in 
the actions with Gaussian stochasticities, {and} we 
follow the tree-level calculations of {Appendix~\ref{app:tree_level}}. 
{Up} to cubic order in the galaxy and matter fields 
we can just replace $\delta_{\rm in}$ with $\delta/D_1$, see 
\eg~\eqsII{linear_classical_fields_solution-matter}{linear_classical_fields_solution-joint}. 
{T}he quadratic matter likelihood of \eq{quadratic_matter_likelihood}, {\ie~} 
\begin{equation}
\label{eq:quadratic_matter_likelihood-recap}
\wp^{(2)}[\delta] = -\frac{1}{2}\int_{\vec{k}}\frac{\delta(\vec{k})\delta(-\vec{k})}{P_{\rm L}(k)}\,\,,
\end{equation}
{indeed remains unaffected by these new operators}, while the quadratic joint likelihood of \eq{quadratic_joint_likelihood} 
gains a contribution coming from the constant shift of the inverse initial power spectrum. 
{That is, we have} 
\begin{equation}
\label{eq:quadratic_joint_likelihood-recap}
\wp^{(2)}[\delta_g,\delta] = \wp^{(2)}[\delta] 
- \frac{m^2}{2}\int_{\vec{k}}\delta(\vec{k})\delta(-\vec{k}) 
- \frac{1}{2}\int_{\vec{k}}\frac{\abs{\delta_g(\vec{k}) - b_1\delta(\vec{k})}^2}{P_{\eps_g}(k)}\,\,.
\end{equation}
{Here we have called $D^2_1{m^2}$ the constant shift of the inverse initial power spectrum, 
$P^{-1}_{\rm in}(k)\to P^{-1}_{\rm in}(k) + D^2_1{m^2}$:\footnote{
{The $D^2_1$ factor is chosen to make $m^2$ the shift in the inverse linear power spectrum: 
$P^{-1}_{\rm L}(k)\to P^{-1}_{\rm L}(k) + m^2$.}} 
we will see in a moment the reason for this.} 

{\eq{quadratic_joint_likelihood-recap} tells us that} also \eq{gm_subsec-quadratic-A}, 
\ie~the logarithm of the conditional likelihood at quadratic order in the fields, 
gains an additional term $\propto\int\dif^3x\,\delta^2(\vec{x})$. At third order in the fields, the same thing happens: 
we only get an additional contribution to the cubic action $\smash{S^{(3)}_{g,{\rm int}}}$, which turns into a contribution 
$\propto\int\dif^3x\,\delta^3(\vec{x})$ to the logarithm of the conditional likelihood. 

We can identify two main features of these additional terms: 
\begin{itemize}[leftmargin=*]
\item first, in these terms there is no appearance of the kernels $K_{n\geq 2}$ 
for the nonlinear deterministic evolution of the matter field. This 
is similar to what happened {throughout Sections~\ref{sec:calculation_of_likelihood-gaussian_stochasticities} 
and \ref{sec:calculation_of_likelihood-nongaussian_stochasticities}.} 
Indeed, in the conditional likelihood for the galaxy and matter fields 
the only nonlinear deterministic evolution we should care about is the one of $\delta_g$ 
with respect to the \emph{nonlinear} matter field $\delta$. {We say more about this in Appendix~\ref{app:loops}}; 
\item the second (and maybe most important) feature of these terms is that they involve the matter field only. 
So, according to our scalings and the discussions of Section~\ref{subsec:expansion_parameters}, they can dominate at large scales. 
Of course, this is especially true for the quadratic term $\propto\int\dif^3x\,\delta^2(\vec{x})$. 
What is the expected size of the coefficient in front of this operator? 
Recalling that this term comes from the first diagram of \eq{loops_subsection_2-B}, which is UV-sensitive, 
we cannot predict the exact value of this Wilson coefficient. However, 
we know that there is no loss of generality in assuming such coefficient to be positive, 
since the loop corrections coming from \eq{loops_subsection_2-B} 
are manifestly positive (this can be seen directly from the form of the interaction of \eq{bias_expansion_stochasticity-D}, 
and by the fact that $\braket{X_gX_g}'$ and $P_{\eps_g\eps_{g,\delta}}$ are real numbers). 
In other words, the inverse initial power spectrum $\smash{P_{\rm in}^{-1}}$ gets renormalized by a positive-definite constant, 
similarly to mass renormalization in quantum field theory ({this is why we have called} 
this constant $D^2_1{m^2}$, even if it has dimensions of a mass cubed). 
Then, since in the diagrams of \eq{loops_subsection_2-B} it is $X_g$ 
that is running in the loops, we can argue that the size of 
${m^2}$ is controlled by $\widebar{n}_g$ in the 
case that the galaxy stochasticity follows the Poisson distribution. 
\end{itemize}
We will discuss 
a bit more about the importance of these terms in Section~\ref{subsec:bayesian_forward_modeling} below. 
For now, we conclude {this short section} by noticing 
that it is not possible to reabsorb these new terms in the result of \eq{cubic_conditional_likelihood-E}. 
Let us stop for simplicity at quadratic order in the fields. 
In principle, at this order we have a term $\propto\int\dif^3x\,b_1^2\delta^2(\vec{x})$ from expanding the square, 
and we might think that a redefinition of $b_1\to\widebar{b}_1$ 
could reabsorb the new term coming from the diagram of \eq{loops_subsection_2-B}. 
However, this cannot work because we are lacking the term $\sim\int\dif^3x\,\delta_g(\vec{x})\delta(\vec{x})$ 
that is needed to make up the square $(\delta_g-\widebar{b}_1\delta)^2$.

\section{Conclusions and future directions}
\label{sec:conclusions}

\noindent In this work we have outlined how to compute the conditional likelihood ${\cal P}[\delta_g|\delta]$ 
to observe a tracer field $\delta_g$ given an underlying matter field $\delta$. 
In the limit that we consider only Gaussian noise for the tracer $\delta_g$, and at leading order in derivatives in the noise fields 
(but to arbitrary order in the deterministic bias expansion), we recover the results of \cite{Schmidt:2018bkr}. 
Corrections to this result arise when we consider the galaxy-matter cross stochasticity, 
the non-Gaussianity of the galaxy noise field, and the noise in the bias coefficients. 

We find that these corrections are controlled by three expansion parameters. 
The first two are the smallness of the matter density field on large scales, 
and the fact that higher-derivative terms are suppressed for momenta much smaller 
than the typical nonlocality scale of tracer formation. 
The third expansion parameter is the amplitude of the galaxy noise field with respect to the matter field on large scales: 
if the noise power spectrum is much larger than the matter power spectrum, 
we expect the corrections coming from the non-Gaussianity of the noise 
and the stochasticity in the bias coefficients to be important. 

Let us first summarize the implications of our results for applications of Bayesian forward modeling, 
and then briefly sketch possible future developments. Finally, we will conclude the paper by summarizing the key equations and 
what are the new terms in the conditional likelihood that we find with respect to \cite{Schmidt:2018bkr}.

\subsection{Impact on applications of Bayesian forward modeling}
\label{subsec:bayesian_forward_modeling}

\noindent We start from the impact of the corrections coming from the non-Gaussianity of the stochasticity 
and the noise in the bias coefficients. 
As discussed in Section~\ref{subsec:expansion_parameters}, 
the size of these terms is controlled by 
the square root of $P_{\eps_g}^{\{0\}}/P_{\rm L}(k)$, which, assuming Poisson noise, is $\sqrt{1/\widebar{n}_g P_{\rm L}(k)}$. 
If this parameter approaches unity, then all higher-order stochastic contributions in principle become 
comparable and the perturbative control of the likelihood breaks down. 

This discussion ties in with the question of how much information 
we can extract from galaxy surveys. If the tracer under consideration is very noisy, 
do applications of forward modeling based 
on the ``minimal'' likelihood of \eq{cubic_conditional_likelihood-E} fail? 
Is it possible to still extract some (or any) cosmological information from measurements of galaxy clustering? 

A way in which we could try to answer this is looking in more detail at this ``Poisson limit'', 
$\smash{{P_{\rm L}(k)}/{P^{\{0\}}_{\eps_g}} \ll 1}$. 
In this limit the loop corrections to the stochasticities are suppressed, 
since their overall size is controlled exactly by the matter power spectrum, 
as we see for example from the diagrams of \eqsII{matter_stochasticity_loop-A}{galaxy_stochasticity_loop}. 
Hence, if we assume that the galaxy noise follows a Poisson distribution, 
this assumption is protected against loop corrections. It makes sense, then, to try and do a resummation of all terms coming 
from higher-order correlation functions of the noise {(both 
from non-Gaussianity of $\eps_g$ and noise in the bias coefficients, like $\eps_{g,\delta}$)}, 
whose amplitude is now fixed by the Poisson assumption. 

Finally, another important difference with \cite{Schmidt:2018bkr} are the terms discussed in Section~\ref{subsec:loops}. 
These are made up of the matter field $\delta$ only, and the most relevant on 
large scales is the one coming from the first diagram of \eq{loops_subsection_2-B}, proportional to $\int\dif^3x\,\delta^2(\vec{x})$. 
Since $\delta_g$ never appears by itself, these terms do not change the position of the maximum{-}likelihood point. 
They also do not affect any Fisher analysis carried out with this likelihood, since 
they vanish once we take two derivatives of the logarithm of the likelihood with respect to $\delta_g$. 
Further study is required to see what is their impact when we want to extract errors 
on cosmological parameters with a full Monte Carlo sampling of the likelihood. 
{However, the fact that they do not couple to the data $\delta_g$, 
together with the fact that their free EFT coefficients are independent of the bias coefficients (which instead couple to $\delta_g$), 
suggest that also the full inference will not be affected.} 

This leads to another question. Do we expect all the corrections that we discussed in this work 
to lead to a shift in the maximum{-}likelihood point? 
For the likelihood of \eq{cubic_conditional_likelihood-E}, \ie~that 
of \cite{Schmidt:2018bkr}, we have that the maximum occurs at $\delta_g=\delta_{g,{\rm det}}[\delta]$. 
In this regard, the solutions for the classical fields of \eq{linear_classical_fields_solution-joint} 
{(that are the fundamental ingredients for the tree-level calculation of the likelihood)} suggest that 
this property still holds, at least on large scales where $\delta_{g,{\rm det}}[\delta]\approx b_1\delta$. 
However, it is not guaranteed that all the corrections we found in this work 
resum to functions of $\delta_g-\delta_{g,{\rm det}}[\delta]$ 
to some power (even if the {fact that we recover} 
a Dirac delta functional of $\delta_g-\delta_{g,{\rm det}}[\delta]$ 
in the limit of zero {Gaussian} noise{, as discussed in 
Section~\ref{subsec:P_eps_g}}, seems to suggest {this).} 
For this reason, it is hard to answer the 
question of whether or not there is a shift in the maximum{-}likelihood point, 
and consequently a possible bias on the estimated cosmological parameters. 
A resummation in the Poisson limit would be useful to shed light on this issue, as well.

\subsection{Further developments}
\label{subsec:further_developments}

\noindent An important extension of this work is that of including ``photon propagation effects'', 
\ie~move from the tracers' rest frame to redshift space. 

What are the differences that we expect in the computation of the likelihood? First, when we go to redshift space 
the galaxy density changes as the zeroth component of a four-vector. The coordinate change depends on the 
matter field $\delta$, so it will add many new terms to the action 
$S_g$. However, since we know how it acts non-perturbatively on the 
galaxy field, we expect that we can account for it with functional techniques 
similar to those used in theories with local (gauged) spacetime symmetries, like gravity. 
This is not all: redshift-space distortions involve the velocity field of galaxies that, like $\delta_g$, comes with its noise. 
While there cannot be a white-noise term $\smash{P_{\eps_v}^{\{0\}}}$ in the power 
spectrum of the stochastic relative velocity between galaxies and matter, because of the equivalence principle, 
this stochastic velocity bias plays an important role for galaxy statistics in redshift space 
(see \cite{Perko:2016puo,Ding:2017gad,delaBella:2018fdb,Desjacques:2018pfv}, for example). 
It would then be interesting to see how this new field can be included in the ``action-based'' formalism of this paper. 

Another obvious way to extend our result is that of including primordial 
non-Gaussianity in the computation of the conditional likelihood. 
This would allow to extend Bayesian forward modeling techniques to put constraints also 
on the parameters describing the primordial bispectrum.\footnote{Of course, if we 
include non-Gaussianity from inflationary dynamics 
we should also consider the non-Gaussianity induced by nonlinear dynamics at recombination, 
see \eg~\cite{Senatore:2008vi,Senatore:2008wk}.} 
We can already have an idea of how the implementation of PNG would work. 
Let us consider for example the case of Gaussian stochasticities, and 
consider only the noise in the galaxy power spectrum. 
PNG would then affect the computation of the conditional likelihood in two ways: 
\begin{enumerate}[leftmargin=*]
\item it would modify the likelihood for the initial matter field, ${\cal P}[\delta_{\rm in}]$. 
For example, a primordial bispectrum would add a $\delta_{\rm in}\delta_{\rm in}\delta_{\rm in}$ 
term to the simple quadratic term of \eq{initial_likelihood} (this modification of the prior must be 
taken into account also when we want to carry out the actual inference, not only in the computation 
of ${\cal P}[\delta_g{|}\delta]$); 
\item it would modify the deterministic bias expansion, 
for example by requiring a $\nabla^{-2}\delta$-like counterterm (``scale-dependent bias''). 
This is already well-studied in the literature (for a review, see Section 7 of \cite{Desjacques:2016bnm}). 
\end{enumerate}

If we now trace back the steps that led us to the conditional likelihood at all orders, discussed in Section~\ref{subsec:P_eps_g}, 
we see that the modification of the likelihood for the initial matter field does not affect ${\cal P}[\delta_g|\delta]$ in any way. 
Indeed, it would only change the expression of the likelihood for the nonlinear matter field $\delta$, 
which always simplifies when we divide ${\cal P}[\delta_g,\delta]$ by 
${\cal P}[\delta]$ to get the conditional likelihood, cf.~\eq{loops_subsubsection_1-G}. 
Therefore the only impact of PNG on ${\cal P}[\delta_g|\delta]$ is the 
augmentation of the deterministic bias expansion 
$\delta_{g,{\rm det}}[\delta]$ by the PNG-induced counterterms: the form of \eq{cubic_conditional_likelihood-E} is left unchanged. 
This makes sense since, as argued by \cite{Schmidt:2018bkr}, this result is really just a consequence 
of assuming that $\eps_g$ is a Gaussian-distributed field, 
and does not make any reference to the likelihood for the initial conditions. 
All of this of course is under the assumption of a weak non-Gaussianity 
of the initial conditions, in the sense of a small skewness of $\delta_{\rm in}$, which corresponds to $f_{\rm NL}\ll\num{d4}$. 
A more detailed analysis, taking into account the {additional terms} that we have dropped in {the calculation 
of Section~\ref{subsec:P_eps_g}}, is left for future work.

\subsection{Comparison with \texorpdfstring{Schmidt et al., 2018 \cite{Schmidt:2018bkr}}{Schmidt et al., 2018}}
\label{subsec:comparison_with_fabian}

\noindent We conclude this work by highlighting the key equations and summarizing what are the new terms 
that we find in the conditional likelihood with respect to the result of \cite{Schmidt:2018bkr}. 
\begin{itemize}[leftmargin=*]
\item The main result is \eq{cubic_conditional_likelihood-E}, i.e.~the conditional likelihood under the assumption 
that only the noise power spectrum $P_{\eps_g}$ is non-vanishing, resummed at all orders in the deterministic bias expansion. 
This matches the result of \cite{Schmidt:2018bkr} under the same assumptions. 
\item Then, further results are those of Sections~\ref{subsec:P_eps_g_eps_m} and \ref{subsec:bispectrum_and_noise_in_LIMD_bias}. 
There we have studied how the higher-derivative noise spectra, the non-Gaussianity of the noise and the stochasticities in 
the bias coefficients modify the above result. 
\item The presence of a cross stochasticity $\smash{P_{\eps_g\eps_m}}$ modifies the covariance of the Gaussian likelihood 
of \eq{cubic_conditional_likelihood-E}. This is shown in \eq{gm_subsec-A}, and it reproduces the findings of \cite{Schmidt:2018bkr}. 
However, we find additional new terms that were missed by \cite{Schmidt:2018bkr}. These are the terms of 
\eqsIII{gm_subsec-B}{gm_subsec-C}{gm_subsec-D}. Interestingly, the first two terms show that once we include 
the cross stochasticity $\smash{P_{\eps_g\eps_m}}$ there will be contributions to the conditional likelihood that know about 
the initial matter power spectrum (although these are relatively suppressed, see Fig.~\ref{fig:scalings}). 
\item The stochasticity in the linear bias and the bispectrum of the noise give rise, respectively, to the terms of 
\eq{bias_expansion_stochasticity-H} and of \eq{bias_expansion_stochasticity-I}. Comparing \eq{bias_expansion_stochasticity-H} 
with \eq{gm_subsec-D} we see that the latter is effectively a higher-derivative correction to the contribution 
coming from $\eps_{g,\delta}$. 
\end{itemize}

\section*{Acknowledgements}

\noindent It is a pleasure to thank Simeon Bird, Linda Blot, Lorenzo Bordin, Jonathan Braden, 
Marco Celoria, Pier Stefano Corasaniti, Paolo Creminelli, Takeshi Kobayashi, Kaloian Lozanov, 
Mehrdad Mirbabayi, Enrico Pajer and Giovanni Tambalo for useful discussions. 
We acknowledge support from the Starting Grant (ERC-2015-STG 678652) ``GrInflaGal'' from the European Research Council.

\appendix

\section{Regarding the scalings under the renormalization group}
\label{app:scalings}

\noindent In this brief appendix we elaborate on how we derived the scalings of Section~\ref{subsec:action_and_RG_scalings}. 
First, let us recall how this is done in textbook examples, such as the Standard Model (without loss of generality, 
we will consider only the Euclidean case and real fields). 

In these theories the action for some field multiplet $\phi^a$ is given by 
a quadratic part plus some cubic and higher-order interactions. The quadratic part is the sum of a kinetic term, 
which is typically given by $\delta^{ab}\delta^{ij}\partial_i\phi^a\partial_j\phi^b$, 
and a (hermitian) mass matrix, not necessarily diagonal. 
In order to find the mass eigenstates, then, one diagonalizes the mass matrix via a rotation of the field multiplet. The diagonalization 
of the mass matrix does not affect the kinetic matrix, since $\delta^{ab}$ is left invariant by a rotation. 
The final step is looking at the scaling dimensions. In order to do this, one has to choose the 
free-theory fixed point: under the rescaling $\vec{k} = b\vec{k}'$ that 
we introduced in Section~\ref{subsec:action_and_RG_scalings}, what part of the quadratic action is left invariant? 
This choice determines the scalings of the fields $\phi^a$. What is typically done is choose the fixed point \emph{by requiring that 
the kinetic term is left invariant.} Combined with the assumption of the kinetic term being 
$\delta^{ab}\delta^{ij}\partial_i\phi^a\partial_j\phi^b$ 
this results in the scaling dimensions that we are used to: 
for example, for a scalar field in $D$ dimensions we have a scaling $b^{D/2-1}$, and so on. 

We can then see what are the differences with our case (we focus on the joint likelihood for simplicity, since 
everything we will say carries over to the matter likelihood). 
First, in our case the would-be mass matrix in \eq{action_joint-B-2} is not a hermitian matrix: 
to see this it is enough to look at the form of ${\cal M}^{ab}_g$ in \eq{action_joint-C} at leading 
order in derivatives (so that $K_{g,1}$ is simply $b_1$ and $K_1$ is equal to $1$). 
Therefore it would not be diagonalized by a rotation. 
Even if we disregard this fact, however, we see that the most important difference is that 
our kinetic matrix is \emph{not} given by the identity (it is not even a diagonal matrix). 
Indeed, we are treating the kinetic term of the $X_g$ field, $\smash{P_{\eps_g}(k)\supset P_{\eps_g}^{\{2\}} k^2}$, 
as an interaction, so there is no $X_gX_g$ term in the kinetic matrix. 

This is just the fact that we are expanding around a different free-theory fixed point than 
the ones we are used to: the field $X_g$ is like a very massive field, 
and the constant term $\smash{P_{\eps_g}^{\{0\}}}$ dominates in its propagator 
at the scales that we can probe. 

What about the field $X$? The stochasticity for matter does not contain any term at order $k^0$ or $k^2$: we 
have $P_{\eps_m}(k)\sim k^4$ on large scales. There is, then, neither a ``canonical'' kinetic term nor a mass term that 
we can use to fix the scaling of $X$, and the propagator of $X$ (which is the \emph{inverse} of $P_{\eps_m}(k)$: see 
\eq{action_joint-C} and compare the ${\cal M}_g^{22}$ and ${\cal M}_g^{33}$ entries) diverges very fast in the infrared, as $k^{-4}$. 

This seems to be a problem at first glance. Indeed, if we now fix the scaling of $X$ by choosing the fixed point where 
such ``non-canonical'' kinetic term is left invariant (so that in addition to leaving 
$\smash{S_g^{(2)}\supset X_gX_g}$ and $\smash{S_g^{(2)}\supset \delta_{\rm in}\delta_{\rm in}}$ fixed under $\vec{k} = b\vec{k}'$, 
also $\smash{S_g^{(2)}\supset XX}$ does not change), we have that the scaling of $X$ is $b^{-7/2}$ in Fourier space, 
which becomes $b^{-1/2}$ in real space. 

This suggests that adding more and more powers of $X$ makes a given operator more and more relevant, 
so that the effective-field-theory expansion does not make sense. 
This clearly cannot be true, so where is the mistake? 
The point is that any interaction involving more than one power of $X$ is a higher-derivative one. 
Roughly speaking, for each $X$ power in the operators describing the noise in the SPT counterterms 
(fourth line of Tab.~\ref{tab:summary_Z}), in those describing the mixed higher-order stochasticities 
(sixth line of Tab.~\ref{tab:summary_Z}), and in those describing the non-Gaussianity of the matter stochasticity 
(last line of Tab.~\ref{tab:summary_Z}), one has to add a $\nabla^2$ factor: this factor ``cures'' 
the scaling of the operator in the infrared.\footnote{This argument seems to suggest that the ``effective'' scaling 
of the field $X$ is $b^{3/2}$, from $\nabla^2 X\sim b^2\, b^{-1/2}$, making it scale as the field $X_g$. 
We did not check this in enough detail, so we refrain from making more precise statements. Moreover, 
see the second and third points at the end of this appendix.} 

This is a bit less trivial than it might look like. Indeed, 
a fundamental requirement for this argument to work is that loops \emph{never} generate operators 
involving higher powers of $X$ without derivative suppression. 
In our case mass and momentum conservation guarantees that this does not happen, as we 
have seen many times throughout Sections \ref{sec:setting_up_the_integrals} and \ref{sec:discussion}.\footnote{Notice 
that for a (Euclidean) field theory of a scalar $\phi$ one might try to use a shift symmetry 
to expand around the $S^{(2)}=(-1/2)\int{\rm d}^3x\,\phi\nabla^4\phi$ free-theory fixed point. Loops are forbidden 
from generating non-shift-symmetric interactions, and the resulting effective field theory might make sense. 
Indeed, operators without derivatives ($S\supset\phi^n$) and operators that carry too few derivatives 
to be irrelevant in the infrared (like $S\supset\vec{\nabla}\phi^2\cdot\vec{\nabla}\phi^2$) are not shift-symmetric. 
If the shift symmetry is broken and such operators can be generated by loops 
it does not seem possible to have a stable EFT expansion around this fixed point. 
This argument can be generalized to generic $S^{(2)}=(-1/2)\int{\rm d}^3x\,\phi\nabla^{2n}\phi$ fixed points 
($n>2$) if the shift symmetry is augmented to a ``Galilean-like'' symmetry $\phi\to\phi + c + \vec{d}\cdot\vec{x}+\cdots$.} 

We conclude this appendix with a few comments:
\begin{itemize}[leftmargin=*]
\item first, we notice that fixing the scaling of $X$ by requiring $\smash{S_g^{(2)}\supset XX}$ to be left invariant 
under $\vec{k} = b\vec{k}'$ leaves also the mixing between $X_g$ and $X$ invariant, as it can be seen straightforwardly by looking at 
the ${\cal M}_g^{12}$ entry of \eq{action_joint-C} and using $X_g\sim b^{3/2}$, $X\sim b^{-1/2}$ in real space. 
This is a nice confirmation that our choice of the free-theory fixed point is the {physically} correct one; 
\item so far we have not discussed operators with only one power of $X$ and no powers of $X_g$, i.e.~the 
operators describing the deterministic evolution for the matter field. In these operators, $X$ does \emph{not} come 
with an additional derivative suppression. Even so, we see that such operators are irrelevant: indeed, 
the scaling of the operator for the evolution at order $n$ in perturbations 
(including the $b^{-3}$ from the volume term $\dif^3x$) is $b^{-3}\,b^{-1/2}\,b^{n(3+n_\delta)/2}$, 
where we have used the fact that we are stopping at leading order in derivatives 
and that the SPT kernels are invariant under rescaling of the momenta. 
Hence we see that for higher and higher $n$ the operator becomes less and less important in the infrared, 
and the EFT expansion makes sense. {At a fixed $n$}, though, 
we see that such operators are more important on large scales than those 
that describe the deterministic evolution of the galaxy field: this is the reason for the ordering of the first 
two lines of Tab.~\ref{tab:summary_Z}; 
\item the fact that operators of the form $X\delta_{\rm in}\delta_{\rm in}\cdots$ 
are more relevant than those of the form $X_g\delta_{\rm in}\delta_{\rm in}\cdots$ does not 
mean that, when we consider the deterministic galaxy evolution at a given order in perturbations, 
the contributions coming from the nonlinear evolution of matter are more important than the 
nonlinear terms in the bias expansion. 
Indeed, both these nonlinear contributions are included in the kernels $K_{g,n}$, which 
are entering in the action $S_{g,{\rm int}}$ as the vertices 
only of operators of the form $X_g\delta_{\rm in}\delta_{\rm in}\cdots$. 
For example, for $\delta_g = b_1\delta + b_2\delta^2/2$ coupled with SPT evolution for matter, 
at second order these would be the contributions $K_{g,2}\supset b_1F_2(\vec{p}_1,\vec{p}_2)$ 
and $K_{g,2}\supset b_2/2$, c.f.~\eq{K_2_example-A}. Since $F_2(\vec{p}_1,\vec{p}_2)$ is invariant 
under rescaling of $\vec{p}_1,\vec{p}_2$, we see that these two contributions are equally relevant in the infrared. 
This is actually important when we compute the likelihood perturbatively (see the next section). 
The reason is that if we consider the deterministic evolution of galaxies up to, e.g., second order, and we then include operators 
$S_{g,{\rm int}}\supset X_g\delta_{\rm in}\delta_{\rm in}$, we are forced to include also operators 
$S_{g,{\rm int}}\supset X\delta_{\rm in}\delta_{\rm in}$ (recall that $S_g$ is the action for the \emph{joint} 
likelihood), since these are more important in the infrared. These operators will be fundamental 
to remove the dependence of ${\cal P}[\delta_g|\delta]$ on the kernels $K_n$ describing 
the nonlinear evolution of matter, what in the calculation at all loops of Section~\ref{subsec:P_eps_g} is 
achieved by the Dirac delta functional of \eq{loops_subsubsection_1-E}; 
\item finally, the discussions in Section~\ref{subsec:action_and_RG_scalings} and in this appendix 
{point very strongly towards the full identification of} the fields $X_g$ and $X$ with the fields $\eps_g$ and $\eps_m$ that 
are routinely used in the bias expansion. We leave a more detailed analysis to a future work. 
\end{itemize}

\section{Tree-level calculation with \texorpdfstring{$P_{\eps_g}$}{P\_\{\textbackslash epsilon\_g\}} only}
\label{app:tree_level}

\noindent The tree-level expression for the (field-dependent part) of the logarithms 
of the two likelihoods is given by (see \eg~\cite{ZinnJustin:2002ru})
\begin{subequations}
\label{eq:tree_level_formula}
\begin{alignat}{3}
\wp[\delta] &\equiv\ln {\cal P}[\delta] &&= -S\big[\vec{\phi}_{\rm cl}[\delta]\big] 
+ \int_{\vec{k}}\vec{\mathcal{J}}\cdot\vec{\phi}_{\rm cl}[\delta]\,\,, \label{eq:tree_level_formula-1} \\
\wp[\delta_g,\delta] &\equiv\ln {\cal P}[\delta_g,\delta] &&= -S_g\big[\vec{\phi}_{g,{\rm cl}}[\delta_g,\delta]\big] 
+ \int_{\vec{k}}\vec{\mathcal{J}}_g\cdot\vec{\phi}_{g,{\rm cl}}[\delta_g,\delta]\,\,, \label{eq:tree_level_formula-2}
\end{alignat}
\end{subequations}
where, as we started to do in Section~\ref{subsec:P_eps_g_eps_m}, 
we use the letter $\wp$ to denote the logarithm of the likelihoods, and the two ``classical fields'' 
$\vec{\phi}_{\rm cl}$, $\vec{\phi}_{g,{\rm cl}}$ are obtained by solving the equations for the two saddle points, i.e.~
\begin{equation}
\label{eq:saddle_point_equations}
\frac{\partial S[\vec{\phi}]}{\partial\vec{\phi}}\bigg|_{\vec{\phi}=\vec{\phi}_{\rm cl}} = \vec{\mathcal{J}}\,\,, \quad
\frac{\partial S_g[\vec{\phi}_g]}{\partial\vec{\phi}_g}\bigg|_{\vec{\phi}=\vec{\phi}_{g,{\rm cl}}} = \vec{\mathcal{J}}_g\,\,. 
\end{equation}

In absence of non-Gaussian stochasticities, the interaction parts of the actions $S$ and $S_g$ are simply given by 
\begin{subequations}
\label{eq:deterministic_interaction_actions}
\begin{align}
S_{\rm int} &= \iu\int_{\vec{k}}X(\vec{k})\delta_{{\rm fwd}}[\delta_{\rm in}](-\vec{k})
\,\,, \label{eq:deterministic_interaction_actions-1} \\
S_{g,{\rm int}} &= S_{\rm int} 
+ \iu\int_{\vec{k}}X_g(\vec{k})\delta_{g,{\rm fwd}}[\delta_{\rm in}](-\vec{k})\,\,. \label{eq:deterministic_interaction_actions-2}
\end{align}
\end{subequations} 
Let us then start by considering only terms up to second order in perturbations 
in the deterministic expansions of \eqsI{G_and_M_kernels}, 
so that the actions $S_{\rm int}$, $S_{g,{\rm int}}$ contain only cubic terms. 
To find the matter and joint likelihoods we need to use \eqsII{tree_level_formula}{saddle_point_equations}. 
We show in Appendix~\ref{subapp:third_order_check} that, up to cubic order in the galaxy and matter fields, 
the formulas for $\wp[\delta]$ and $\wp[\delta_g,\delta]$ reduce to 
\begin{subequations}
\label{eq:W_to_cubic_order}
\begin{align}
\wp[\delta] &= \wp^{(2)}[\delta]+\wp^{(3)}[\delta]+\cdots \nonumber \\
&= \frac{1}{2}\int_{\vec{k},\vec{k}'}\mathcal{J}^a(\vec{k})({\cal M}^{-1})^{ab}(\vec{k},\vec{k}')\mathcal{J}^b(\vec{k}') - 
S^{(3)}_{\rm int}\big[\vec{\phi}_{\rm cl}^{(1)}[\delta]\big] + {\cal O}(\vec{\mathcal{J}}^4)\,\,, \label{eq:W_to_cubic_order-1} \\
\wp[\delta_g,\delta] &= \wp^{(2)}[\delta_g,\delta]+\wp^{(3)}[\delta_g,\delta]+\cdots \nonumber \\
&= \frac{1}{2}\int_{\vec{k},\vec{k}'}\mathcal{J}^a_g(\vec{k})({\cal M}^{-1})^{ab}_g(\vec{k},\vec{k}')\mathcal{J}^b_g(\vec{k}') - 
S^{(3)}_{g,{\rm int}}\big[\vec{\phi}_{g,{\rm cl}}^{(1)}[\delta_g,\delta]\big] 
+ {\cal O}(\vec{\mathcal{J}}_g^4)\,\,, \label{eq:W_to_cubic_order-2}
\end{align}
\end{subequations}
where the classical fields $\vec{\phi}_{\rm cl}$, $\vec{\phi}_{g,{\rm cl}}$ are evaluated at first order 
in an expansion in powers of the ``currents'' $\vec{\mathcal{J}}$, $\vec{\mathcal{J}}_g$ 
{(we do not use a superscript ``$(1)$'' to avoid cluttering the notation too much: from now 
on $\vec{\phi}_{\rm cl}$, $\vec{\phi}_{g,{\rm cl}}$ are always intended to be 
first-order quantities in $\vec{\mathcal{J}}$ and $\vec{\mathcal{J}}_g$)}. 
That is, they are obtained by solving the ``free-theory'' equations for the saddle points. 

We now have to actually solve 
these two equations {for the classical fields} $\vec{\phi}_{\rm cl}$, $\vec{\phi}_{g,{\rm cl}}$. They are simply given by 
\begin{equation}
\label{eq:saddle_point_equations-quadratic}
\frac{\partial S^{(2)}[\vec{\phi}]}{\partial\vec{\phi}}\bigg|_{\vec{\phi}=\vec{\phi}_{\rm cl}} = \vec{\mathcal{J}}\,\,, \quad
\frac{\partial S^{(2)}_g[\vec{\phi}_g]}{\partial\vec{\phi}_g}\bigg|_{\vec{\phi}=\vec{\phi}_{g,{\rm cl}}} = \vec{\mathcal{J}}_g\,\,, 
\end{equation}
where the quadratic actions are those of \eqsII{action_matter-C-2}{action_joint-B-2}. 
Solving these equations requires only applying the inverse of the matrices ${\cal M}$ and ${\cal M}_g$ 
{of \eqsII{kinetic_matrix_for_matter}{action_joint-C}} to the currents of \eqsII{action_matter-B}{action_joint-B-temp-2}. 
Thanks to ${\cal M}$ and ${\cal M}_g$ being proportional to Dirac delta functions, 
it is enough to work at a fixed $\vec{k}$ (we refer to Appendix~\ref{subapp:third_order_check} for more details). 
For simplicity we drop all higher-derivative corrections to the growth factor and the linear bias, \ie~we take 
$K_{g,1}=b_1$ and $K_1=1$: we discuss below how to include these corrections. 
Conversely, we allow for a generic scale dependence of the galaxy noise power spectrum 
(we do not stop at order $k^0$ in its derivative expansion). 
Dropping the $\vec{k}$ argument for simplicity of notation, the solution for the linear classical fields is then given by 
\begin{equation}
\label{eq:linear_classical_fields_solution-matter}
\vec{\phi}_{\rm cl} = 
\begin{pmatrix}
X_{\rm cl} \\
\delta_{{\rm in},{\rm cl}}
\end{pmatrix}
=
\begin{pmatrix}
\dfrac{\iu\delta}{D^2_1P_{\rm in}} \\[0.9em] 
\dfrac{\delta}{D_1}
\end{pmatrix}
\end{equation}
for the matter likelihood, and by 
\begin{equation}
\label{eq:linear_classical_fields_solution-joint}
\vec{\phi}_{g,{\rm cl}} = 
\begin{pmatrix}
X_{g,{\rm cl}} \\
X_{\rm cl} \\
\delta_{{\rm in},{\rm cl}}
\end{pmatrix}
=
\begin{pmatrix}
\dfrac{\iu(\delta_g-b_1\delta)}{P_{\eps_g}} \\[0.9em] 
\dfrac{\iu\delta}{D^2_1P_{\rm in}} - \dfrac{\iu b_1(\delta_g-b_1\delta)}{P_{\eps_g}} \\[0.9em] 
\dfrac{\delta}{D_1}
\end{pmatrix}
=
\begin{pmatrix}
\dfrac{\iu(\delta_g-b_1\delta)}{P_{\eps_g}} \\[0.9em] 
\dfrac{\iu\delta}{D^2_1P_{\rm in}} - b_1X_{g,{\rm cl}} \\[0.9em] 
\dfrac{\delta}{D_1}
\end{pmatrix}
\end{equation}
for the joint likelihood. 

We can now compute $\wp[\delta]$ and $\wp[\delta_g,\delta]$. 
We start from the matter likelihood. At quadratic order in the fields, we get 
\begin{equation}
\label{eq:quadratic_matter_likelihood}
\wp^{(2)}[\delta] = -\frac{1}{2}\int_{\vec{k}}\frac{\delta(\vec{k})\delta(-\vec{k})}{P_{\rm L}(k)}\,\,.
\end{equation}
It is straightforward to see that, instead, the quadratic joint likelihood is given by 
(we use the fact that all the fields we are considering are real, 
\ie~$\varphi(-\vec{k}) = \varphi^\ast(\vec{k})$ for any field $\varphi$) 
\begin{equation}
\label{eq:quadratic_joint_likelihood}
\wp^{(2)}[\delta_g,\delta] = \wp^{(2)}[\delta] 
- \frac{1}{2}\int_{\vec{k}}\frac{\abs{\delta_g(\vec{k}) - b_1\delta(\vec{k})}^2}{P_{\eps_g}(k)}\,\,.
\end{equation}
This tells us that, at this order, the logarithm $\wp[\delta_g|\delta]$ of the conditional likelihood is given by 
\begin{equation}
\label{eq:quadratic_conditional_likelihood}
\wp^{(2)}[\delta_g|\delta] = -\frac{1}{2}\int_{\vec{k}}\frac{\abs{\delta_g(\vec{k}) - b_1\delta(\vec{k})}^2}{P_{\eps_g}(k)}\,\,.
\end{equation}

We can then move to cubic order. First, comparing the first-order classical fields $X_{\rm cl}$ and $\delta_{{\rm in},{\rm cl}}$ 
in \eq{linear_classical_fields_solution-matter} and in \eq{linear_classical_fields_solution-joint}, 
and using the fact that the field $X$ enters linearly in 
both $S_{\rm int}$ and $S_{g,{\rm int}}$, we see that subtracting the logarithm of the matter-only 
likelihood from that of the joint likelihood 
exactly cancels the term in $\wp[\delta_g,\delta]$ that comes from $S_{\rm int}$ 
evaluated at $X_{\rm cl}$ equal to ${\iu}\delta/D^2_1P_{\rm in}$. 
Then, the cubic-order contribution to the logarithm of the conditional likelihood is simply given by 
\begin{equation}
\label{eq:cubic_conditional_likelihood-A}
\wp^{(3)}[\delta_g|\delta] = \bigg({-\iu}\int_{\vec{k}}X_g(\vec{k})\delta^{(2)}_{g,{\rm fwd}}[\delta_{\rm in}](-\vec{k}) 
- \iu\int_{\vec{k}}X(\vec{k})\delta^{(2)}_{\rm fwd}[\delta_{\rm in}](-\vec{k})
\bigg)\bigg|_{\vec{\phi}_g=\vec{\phi}_{g,{\rm cl}}}\,\,, 
\end{equation}
with $\vec{\phi}_{g,{\rm cl}}$ now given by
\begin{equation}
\label{eq:cubic_conditional_likelihood-B}
\vec{\phi}_{g,{\rm cl}}=\begin{pmatrix}
X_{g,{\rm cl}} \\
X_{\rm cl} \\
\delta_{{\rm in},{\rm cl}}
\end{pmatrix}
=
\begin{pmatrix}
\dfrac{\iu(\delta_g-b_1\delta)}{P_{\eps_g}} \\[0.9em] 
-\dfrac{\iu b_1(\delta_g-b_1\delta)}{P_{\eps_g}} \\[0.9em] 
\dfrac{\delta}{D_1}
\end{pmatrix}
=
\begin{pmatrix}
\dfrac{\iu(\delta_g-b_1\delta)}{P_{\eps_g}} \\[0.9em] 
{-b_1}X_{g,{\rm cl}} \\[0.9em] 
\dfrac{\delta}{D_1}
\end{pmatrix}
\,\,.
\end{equation}
Using the expressions for $\delta_{\rm fwd}$ and $\delta_{g,{\rm fwd}}$ of \eqsI{G_and_M_kernels} at second order in perturbations, 
we see that \eq{cubic_conditional_likelihood-A} becomes 
\begin{equation}
\label{eq:cubic_conditional_likelihood-C}
\begin{split}
\wp^{(3)}[\delta_g|\delta] = \int_{\vec{k}}\frac{\delta_g(\vec{k})-b_1\delta(\vec{k})}{P_{\eps_g}(k)}
\int_{\vec{p}_1,\vec{p}_2}&(2\pi)^{3}\delta^{(3)}(-\vec{k}-\vec{p}_{12}) \\ 
&\times\big(K_{g,2}(-\vec{k};\vec{p}_1,\vec{p}_2)-b_1K_{2}(-\vec{k};\vec{p}_1,\vec{p}_2)\big)\,\delta(\vec{p}_1)\delta(\vec{p}_2)\,\,,
\end{split}
\end{equation}
where we recognize in $K_{g,2}(-\vec{k};\vec{p}_1,\vec{p}_2)-b_1K_{2}(-\vec{k};\vec{p}_1,\vec{p}_2)$ 
the kernel for the \emph{deterministic} bias expansion \emph{up to quadratic order} in the \emph{nonlinear} matter field 
(see \eqsII{K_2_example-A}{deterministic_bias_expansion-B}, for example). That is, we find 
\begin{equation}
\label{eq:cubic_conditional_likelihood-D}
\begin{split}
\wp^{(3)}[\delta_g|\delta] &= \int_{\vec{k}}\frac{\big(\delta_g(\vec{k})-b_1\delta(\vec{k})\big)\,
\delta^{(2)}_{g,{\rm det}}[\delta](-\vec{k})}{P_{\eps_g}(k)} = 
\int_{\vec{k}}\frac{\big(\delta_g(\vec{k})-\delta^{(1)}_{g,{\rm det}}[\delta](\vec{k})\big)\,
\delta^{(2)}_{g,{\rm det}}[\delta](-\vec{k})}{P_{\eps_g}(k)}\,\,,
\end{split}
\end{equation}
where 
\begin{equation}
\label{eq:cubic_conditional_likelihood-D-help-appendix}
\delta_{g,{\rm det}}[\delta] = \delta^{(1)}_{g,{\rm det}}[\delta] + \delta^{(2)}_{g,{\rm det}}[\delta] + \cdots 
\end{equation}
is the deterministic bias expansion of $\delta_g$, defined by \eq{deterministic_bias_expansion-A}. 
Summing this to \eq{quadratic_conditional_likelihood}, and keeping track of signs and factors of $1/2$, 
we see that {up to cubic order in the fields} the logarithm of the conditional likelihood is given by 
{\eq{cubic_conditional_likelihood-E}, \ie~} 
\begin{equation}
\label{eq:cubic_conditional_likelihood-E-appendix}
\wp[\delta_g|\delta] = 
{-\frac{1}{2}}\int_{\vec{k}}\frac{\abs{\delta_g(\vec{k})-\delta_{g,{\rm det}}[\delta](\vec{k})}^2}{P_{\eps_g}(k)}\,\,.
\end{equation}

Clearly, it is possible to extend \eqsII{quadratic_conditional_likelihood}{cubic_conditional_likelihood-D} 
at higher orders in the fields. 
More precisely, we explicitly checked that \eq{cubic_conditional_likelihood-E} 
holds also at cubic order in the deterministic bias expansion 
(quartic order in the interactions), by using the extension of the tree-level formulas of \eqsI{W_to_cubic_order} 
at fourth order in the fields (collected in Appendix~\ref{subapp:fourth_order_check}). 
It is actually possible to resum the contributions from the deterministic evolution at all orders in the fields, 
and reproduce exactly the result of \cite{Schmidt:2018bkr} in the limit of zero matter and galaxy-matter stochasticities. 
This is shown in Section~\ref{subsec:P_eps_g}. 

{We conclude by briefly discussing how to add to the tree-level 
calculation of this appendix the higher-derivative terms in the deterministic bias expansion 
for the galaxy field and in the deterministic evolution of the matter field. 
We see that the higher-derivative contributions at second (and higher) order in perturbations are automatically included: 
indeed, in our derivation we have left the kernels $K_{g,n\geq 2}$ and $K_{n\geq 2}$ completely general. 
The higher-derivative linear biases contained in $b(k)$ and the scale-dependent corrections to the growth factor given by 
$K_1(k)$ are also included straightforwardly. The first enters wherever 
the deterministic bias expansion of $\delta_g$ in terms of the nonlinear matter field $\delta$ appears, 
while the second whenever $D_1$ appears. 
In all the relations from \eq{linear_classical_fields_solution-matter} to 
\eq{cubic_conditional_likelihood-D} we can just substitute $b_1$ with $b(k)$ 
and $D_1$ with $K_1(k)D_1$.}

\section{Adding higher-derivative quadratic stochasticities}
\label{app:higher_derivative_stochasticities}

\noindent Here we sketch how to compute, at tree level, the higher-derivative contributions discussed in 
Section~\ref{subsec:P_eps_g_eps_m}, coming from a non-vanishing cross stochasticity between galaxies and matter. 

Then, let us put $P_{\eps_m}$ to zero in {\eqsII{kinetic_matrix_for_matter}{action_joint-C}}. 
As a consequence, it is straightforward to {see} that we can forget about the matter-only likelihood, 
since its expression (and then its impact on the conditional likelihood) 
will be the same as that of the tree-level calculation of Appendix~\ref{app:tree_level}. 
What we need, then, is the correction to the {(linear in the galaxy and matter fields $\delta_g$ and $\delta$)} 
classical solution $\vec{\phi}_{g,{\rm cl}}$ at leading order in $P_{\eps_g\eps_m}$
It is given by (again suppressing the arguments to streamline the notation) 
\begin{equation}
\label{eq:linear_classical_fields_solution-joint-correction}
\vec{\phi}_{g,{\rm cl}} = \text{\eq{linear_classical_fields_solution-joint}} + 
\begin{pmatrix}
\dfrac{P_{\eps_g\eps_m}}{P_{\eps_g}}\bigg(2b_1X_{g,{\rm cl}} - \dfrac{\iu\delta}{D^2_1P_{\rm in}}\bigg) \\[0.9em] 
-\dfrac{b_1P_{\eps_g\eps_m}}{P_{\eps_g}}\bigg(2b_1X_{g,{\rm cl}} - \dfrac{\iu\delta}{D^2_1P_{\rm in}}\bigg) - \dfrac{P_{\eps_g\eps_m}X_{g,{\rm cl}}}{D^2_1P_{\rm in}} \\[0.9em] 
\dfrac{\iu P_{\eps_g\eps_m}X_{g,{\rm cl}}}{D_1}
\end{pmatrix}
\,\,,
\end{equation}
where, as in \eq{linear_classical_fields_solution-joint}, we have 
\begin{equation}
\label{eq:classical_field_reminder}
X_{g,{\rm cl}} = \frac{\iu(\delta_g-b_1\delta)}{P_{\eps_g}}\,\,.
\end{equation}
Also, notice that we have dropped all higher-derivative corrections to the growth factor and the linear bias, \ie~we have taken 
$K_{g,1}=b_1$ and $K_1=1$ as we did in {Appendix~\ref{app:tree_level}}. 

With the classical field of \eq{linear_classical_fields_solution-joint-correction} 
we can compute the corrections coming from $P_{\eps_g\eps_m}$ 
using, for example, the formulas of \eqsII{W_to_cubic_order-2}{cubic_conditional_likelihood-A}. 
At quadratic order these corrections are given by 
\begin{equation}
\label{eq:appendix_higher_derivative_stochasticities-A}
{-{}}\int_{\vec{k}}\frac{b_1P_{\eps_g\eps_m}(k)\abs{\delta_g(\vec{k})-b_1\delta(\vec{k})}^2}{P^2_{\eps_g}(k)}
\end{equation}
and 
\begin{equation}
\label{eq:appendix_higher_derivative_stochasticities-B}
\int_{\vec{k}}\frac{P_{\eps_g\eps_m}(k)\big(\delta_g(\vec{k})-b_1\delta(\vec{k})\big)\delta(-\vec{k})}{P_{\eps_g}(k)P_{\rm L}(k)}\,\,.
\end{equation}
At cubic order, instead, we first have 
\begin{equation}
\label{eq:appendix_higher_derivative_stochasticities-C}
2\int_{\vec{k}}\frac{b_1P_{\eps_g\eps_m}(k)}{P_{\eps_g}(k)} 
\frac{\big(\delta_g(\vec{k})-b_1\delta(\vec{k})\big)\,
\delta^{(2)}_{g,{\rm det}}[\delta](-\vec{k})}{P_{\eps_g}(k)}\,\,.
\end{equation}
Then we have 
\begin{equation}
\label{eq:appendix_higher_derivative_stochasticities-D}
\begin{split}
{-{}}\int_{\vec{k}}\frac{P_{\eps_g\eps_m}(k)}{P_{\eps_g}(k)}\frac{\delta(\vec{k})}{P_{\rm L}(k)}
\int_{\vec{p}_1,\vec{p}_2}(2\pi)^3\delta^{(3)}_{\rm D}(-\vec{k}-\vec{p}_{12})\,K_{g,{\rm det},2}(-\vec{k};\vec{p}_1,\vec{p}_2)\, 
\delta(\vec{p}_1)\delta(\vec{p}_2)\,\,,
\end{split}
\end{equation}
which can be rewritten as 
\begin{equation}
\label{eq:appendix_higher_derivative_stochasticities-E}
{-{}}\int_{\vec{k}}\frac{P_{\eps_g\eps_m}(k)}{P_{\eps_g}(k)}
\frac{\delta(\vec{k})\delta^{(2)}_{g,{\rm det}}[\delta](-\vec{k})}{P_{\rm L}(k)}\,\,.
\end{equation}
Finally we have 
\begin{equation}
\label{eq:appendix_higher_derivative_stochasticities-F}
\begin{split}
{-{}}\int_{\vec{k}}\frac{\delta_g(\vec{k})-b_1\delta(\vec{k})}{P_{\eps_g}(k)} 
\int_{\vec{p}_1,\vec{p}_2}\bigg[&(2\pi)^3\delta^{(3)}_{\rm D}(-\vec{k}-\vec{p}_{12})\,K_{g,{\rm det},2}(-\vec{k};\vec{p}_1,\vec{p}_2) \\
&\times\bigg(\frac{P_{\eps_g\eps_m}(p_2)}{P_{\eps_g}(p_2)}
\delta(\vec{p}_1)\big(\delta_g(\vec{p}_2) - b_1\delta(\vec{p}_2)\big)\bigg) \\
&+ (\vec{p}_1\to\vec{p}_2)\bigg]\,\,.
\end{split}
\end{equation}

As it was for the tree-level calculations of {Appendix~\ref{app:tree_level}}, 
these results do not depend on the assumptions of $K_{g,1}=b_1$ and $K_1=1$. 
This can be easily seen also from the formulas of Appendix~\ref{app:saddle_point_formulas}: 
basically, there ${\cal M}^{ij}$ is allowed to be a generic (symmetric) matrix. 
Then, re-introducing $b(k)$ and the scale dependence of $D_1$ (so that $P_{\rm L}(k)=D^2_1K^2_1(k)P_{\rm in}(k)$, as well), 
and resumming \eq{appendix_higher_derivative_stochasticities-A} with \eq{appendix_higher_derivative_stochasticities-C} in 
perturbations and derivatives,\footnote{{Notice that even if we resum the impact of 
the cross stochasticity $P_{\eps_g\eps_m}(k)$ as $\big(P_{\eps_g}(k) - 2b_1P_{\eps_g\eps_m}(k)\big)^{-1}$ to all orders in $k^2$, 
cf.~the first term of \eq{gm_subsec-A}, such functional form cannot be trusted for arbitrary high $k$. 
Momenta cannot be pushed beyond the longest ``unresolved'' nonlocality scale present in the theory, 
which in the case of galaxies is usually assumed to be at least of order of the halo Lagrangian radius $R(M_h)$.}} 
we get \eqsIV{gm_subsec-A}{gm_subsec-B}{gm_subsec-C}{gm_subsec-D} of Section~\ref{subsec:P_eps_g_eps_m}.

\section{Adding non-Gaussian stochasticities}
\label{app:higher_order_stochasticities}

\noindent In this appendix we collect some of the details for the calculations of 
Section~\ref{sec:calculation_of_likelihood-nongaussian_stochasticities}. 
First, we discuss how to include the three-point function $\braket{\eps_g\eps_g\eps_g}$ of the galaxy stochasticity. Then, we move to 
the contributions from the stochasticity $\eps_{g,\delta}$ in the linear LIMD bias coefficient $b_1$.

\subsection{\texorpdfstring{$\braket{\eps_g\eps_g\eps_g}$ bispectrum}
{<\textbackslash epsilon\_g\textbackslash epsilon\_g\textbackslash epsilon\_g> bispectrum}}
\label{subapp:epsilon_g_bispectrum}

\noindent As discussed in Section~\ref{subsec:joint}, we have a contribution from the bispectrum of $\eps_g$ 
to the galaxy-galaxy-galaxy three-point function. Calling $B_{\eps_g\eps_g\eps_g}$ this bispectrum, we can include its contribution 
to the generating functional $Z[J_g,J]$ as 
\begin{equation}
\label{eq:epsilon_g_bispectrum-A}
Z[J_g,J] = \text{\eq{joint_likelihood-B}}\times\eu^{\frac{1}{3!}\int_{\vec{p}_1,\dots\vec{p}_3}(2\pi)^3\delta^{(3)}_{\rm D}
(\vec{p}_{123})
B_{\eps_g\eps_g\eps_g}(\vec{p}_1,\vec{p}_2,\vec{p}_3)J_g(\vec{p}_1)J_g(\vec{p}_2)J_g(\vec{p}_3)}\,\,.
\end{equation}
Moving to the action for the joint likelihood, we see that $S_{g,{\rm int}}$ gets a contribution at cubic order of the form 
\begin{equation}
\label{eq:epsilon_g_bispectrum-B}
\begin{split}
S_{g,{\rm int}} &= \text{\eq{deterministic_interaction_actions-2}} \\
&\;\;\;\; - \frac{\iu}{3!}\int_{\vec{p}_1,\dots\vec{p}_3}(2\pi)^3\delta^{(3)}_{\rm D}(\vec{p}_{123})\,
B_{\eps_g\eps_g\eps_g}(-\vec{p}_1,-\vec{p}_2,-\vec{p}_3)\,X_g(\vec{p}_1)X_g(\vec{p}_2)X_g(\vec{p}_3)\,\,.
\end{split}
\end{equation}

\subsection{\texorpdfstring{Contributions from $\eps_{g,\delta}$}
{Contributions from \textbackslash epsilon\_\{g,\textbackslash delta\}}}
\label{subapp:contributions_from_epsilon_delta}

\noindent The contribution to $\braket{\delta_g\delta_g\delta}$ in real space is given by 
\eq{bias_expansion_stochasticity-B}, \ie\footnote{{This is the leading contribution from $\eps_{g,\delta}$ on large scales: 
even if we disregard higher-order terms in the deterministic evolution for the matter field and in 
the deterministic bias expansion, the galaxy-galaxy-galaxy three-point function contains also three additional terms controlled 
by the $\braket{\eps_g(\vec{x})\eps_{g,\delta}(\vec{y})\eps_{g,\delta}(\vec{z})}$ correlator, 
{proportional to $\delta^{(3)}(\vec{x}-\vec{y})\delta^{(3)}(\vec{x}-\vec{z})$ on large scales}. 
Their (ir)relevance on large scales can be studied with the same methods of Section~\ref{subsec:action_and_RG_scalings}.}} 
\begin{equation}
\label{eq:contributions_from_epsilon_delta-A}
\braket{\delta_g(\vec{x})\delta_g(\vec{y})\delta(\vec{z})} = 
\xi_{\rm L}(\abs{\vec{x}-\vec{z}})\braket{\eps_g(\vec{x})\eps_{g,\delta}(\vec{y})} + (\vec{x}\to\vec{y})\,\,.
\end{equation}
Taking the Fourier transform, we get 
\begin{equation}
\label{eq:contributions_from_epsilon_delta-B}
\bigg(\int_{\vec{x}}\int_{\vec{y}}\int_{\vec{z}}
\eu^{-\iu\vec{k}_1\cdot\vec{x}-\iu\vec{k}_2\cdot\vec{y}-\iu\vec{k}_3\cdot\vec{z}}
\braket{\delta_g(\vec{x})\delta_g(\vec{y})\delta(\vec{z})}\bigg)' = 
P_{\eps_g\eps_{g,\delta}}(k_1)P_{\rm L}(k_3)+(\vec{k}_1\to\vec{k}_2)\,\,,
\end{equation}
where $P_{\eps_g\eps_{g,\delta}}(k)$ (which is $\sim k^0$ at leading order in derivatives) 
is the Fourier transform of the correlation function $\braket{\eps_g(\vec{x})\eps_{g,\delta}(\vec{y})}$ 
{with respect to $\vec{x}-\vec{y}$}. 

We can now see how to implement this term in the action. In order to do this, 
we study its impact on the joint generating functional for galaxies and matter, $Z[J_g,J]$. 
We take the generating functional to be 
\begin{equation}
\label{eq:contributions_from_epsilon_delta-C}
\begin{split}
Z[J_g,J] = \int{\cal D}\delta_{\rm in}\,&\text{integrand of \eq{joint_likelihood-B}} \\
&\times\eu^{\frac{1}{2}\int_{\vec{p}_1,\dots\vec{p}_3}(2\pi)^3\delta^{(3)}_{\rm D}(\vec{p}_{123}) D_1 
s_{J_gJ_g\delta_{\rm in}}(-\vec{p}_1,-\vec{p}_2,-\vec{p}_3)J_g(\vec{p}_1)J_g(\vec{p}_2)\delta_{\rm in}(\vec{p}_3)}\,\,,
\end{split}
\end{equation}
where we assume $s_{J_gJ_g\delta_{\rm in}}$ to be symmetric in $\vec{p}_1\leftrightarrow\vec{p}_2$ and, 
from now on, we consider \emph{only} the leading order in derivatives in the deterministic evolution of the galaxy and matter fields. 
That is, we take $K_{g,1} = b_1$ and $K_1 = 1$ since, as we pointed out also in Appendix~\ref{app:higher_derivative_stochasticities}, 
none of the following results depend on them being $k$-independent: 
including their scale dependence is straightforward, following the discussion {at the end of} Section~\ref{subsec:P_eps_g}. 
{P}utting to zero the nonlinear terms in \eq{joint_likelihood-B} 
that come from the deterministic evolution for 
galaxies and matter, the contribution of this additional term in the generating functional 
can be computed by rewriting its logarithm as 
\begin{equation}
\label{eq:contributions_from_epsilon_delta-D}
\int_{\vec{k}}b_1D_1\tilde{s}(\vec{k})\delta_{\rm in}(-\vec{k})\,\,,
\end{equation}
where we define $\tilde{s}$ as 
\begin{equation}
\label{eq:contributions_from_epsilon_delta-E}
\tilde{s}(\vec{k}) = \frac{1}{2b_1}\int_{\vec{p}_1,\vec{p}_2}(2\pi)^3\delta^{(3)}_{\rm D}
(\vec{p}_{12}-\vec{k})\,s_{J_gJ_g\delta_{\rm in}}(-\vec{p}_1,-\vec{p}_2;\vec{k})\,
J_g(\vec{p}_1)J_g(\vec{p}_2)\,\,,
\end{equation}
and (in this appendix only) we use a semicolon to distinguish the momentum that is associated with the initial matter field. 
If we recall that in the exponent of the integrand of \eq{joint_likelihood-B} the term that couples linearly $J_g$ 
with $\delta_{\rm in}$ is of the form $\int_{\vec{k}}J_g(\vec{k})b_1D_1\delta_{\rm in}(-\vec{k})$, 
we can see that this new term basically corresponds to a stochasticity in $b_1$ 
of the form $\smash{b_1\to b_1 + P_{\eps_g\eps_{g,\delta}}^{\{0\}}J(\vec{x})}$ 
(in real space and at leading order in derivatives), since in a moment we will indeed identify $s_{J_gJ_g\delta_{\rm in}}$ 
with the cross spectrum between $\eps_g$ and $\eps_{g,\delta}$. 

Let us then consider the quadratic, noise-free theory 
(since the terms we are considering here will simply add to the logarithm of $Z[J_g,J]$ there is no loss of generality in doing so). 
As far as the integral over $\delta_{\rm in}$ is concerned, its effect is just redefining $J_g\to J_g + \tilde{s}$. 
The logarithm of $Z[J_g,J]$, then, is simply given by 
\begin{equation}
\label{eq:contributions_from_epsilon_delta-G}
\ln\bigg(\frac{Z[J_g,J]}{Z[0,0]}\bigg) = W[J_g,J] = 
\frac{1}{2}\int_{\vec{k}}P_{\rm in}(k)\big(D_1J(\vec{k}) + b_1D_1J_g(\vec{k}) + b_1D_1\tilde{s}(\vec{k})\big)
\times(\vec{k}\to-\vec{k})\,\,.
\end{equation}
Taking the cubic order of the expansion in the currents $J_g$ and $J$, we simply have 
\begin{equation}
\label{eq:contributions_from_epsilon_delta-H}
W^{(3)}[J_g,J] = \int_{\vec{k}}b_1 D_1 P_{\rm in}(k)\big(D_1J(\vec{k}) + b_1D_1J_g(\vec{k})\big)\tilde{s}(-\vec{k})\,\,.
\end{equation}

Then, to make contact with \eq{contributions_from_epsilon_delta-B}, 
we study the functional derivatives of this term with respect to $J_g$ and $J$. 
First, we notice that taking one derivative with respect to $J_g$ and two with respect to $J$ gives zero: 
this is consistent with the fact that $\eps_{g,\delta}$ does not contribute to 
the galaxy-matter-matter three-point function at leading order in the deterministic expansion. 
Then, we compute separately the derivatives with respect to two powers of $J_g$ 
and one of $J$, and with respect to three powers of $J_g$.

\subsubsection*{\texorpdfstring{$J_g J_g J$ derivative}{J\_g J\_g J derivative}}

\noindent First, the derivative of $W^{(3)}[J_g,J]$ with respect to $J(\vec{k})$ is given by 
\begin{equation}
\label{eq:contributions_from_epsilon_delta-sub_1-A}
\frac{\partial W^{(3)}[J_g,J]}{\partial J(\vec{k})} = b_1 P_{\rm L}(k)\tilde{s}(\vec{k})\,\,.
\end{equation}
Then, using the symmetry of $s_{J_gJ_g\delta_{\rm in}}$ for $\vec{p}_1\leftrightarrow\vec{p}_2$, we find 
\begin{equation}
\label{eq:contributions_from_epsilon_delta-sub_1-B}
\frac{\partial\tilde{s}(-\vec{k})}{J_g(\vec{k}')} = \frac{1}{b_1}\int_{\vec{p}}(2\pi)^3\delta^{(3)}_{\rm D}(\vec{p}-\vec{k}'+\vec{k})\,
s_{J_gJ_g\delta_{\rm in}}(\vec{k}',-\vec{p};-\vec{k})\,J_g(\vec{p})\,\,.
\end{equation}
Applying this formula to \eq{contributions_from_epsilon_delta-sub_1-A} we get 
\begin{equation}
\label{eq:contributions_from_epsilon_delta-sub_1-C}
\begin{split}
\frac{\partial^2 W^{(3)}[J_g,J]}{\partial J(\vec{k})\partial J_g(\vec{l})} &= 
b_1 P_{\rm L}(k)\frac{\partial\tilde{s}(\vec{k})}{\partial J_g(\vec{l})} \\
&= P_{\rm L}(k)\int_{\vec{p}}(2\pi)^3\delta^{(3)}_{\rm D}(\vec{p}-\vec{l}-\vec{k})\,
s_{J_gJ_g\delta_{\rm in}}(\vec{l},-\vec{p};\vec{k})\,J_g(\vec{p})\,\,.
\end{split}
\end{equation}
Then, an additional derivative with respect to $J_g(\vec{m})$ gives 
\begin{equation}
\label{eq:contributions_from_epsilon_delta-sub_1-D}
\frac{\partial^3 W^{(3)}[J_g,J]}{\partial J(\vec{k})\partial J_g(\vec{l})\partial J_g(\vec{m})} = 
(2\pi)^3\delta^{(3)}_{\rm D}(\vec{k}+\vec{l}+\vec{m})\,P_{\rm L}(k)\,s_{J_gJ_g\delta_{\rm in}}(\vec{l},\vec{m};\vec{k})\,\,.
\end{equation}
Hence, comparing with \eq{contributions_from_epsilon_delta-B} 
with $\vec{k}_3=\vec{k}$, $\vec{k}_1 = \vec{l}$ and $\vec{k}_2=\vec{m}$, we find 
\begin{equation}
\label{eq:contributions_from_epsilon_delta-sub_1-E}
s_{J_gJ_g\delta_{\rm in}}(\vec{l},\vec{m};\vec{k}) = P_{\eps_g\eps_{g,\delta}}(l) + P_{\eps_g\eps_{g,\delta}}(m)\,\,.
\end{equation}

\subsubsection*{\texorpdfstring{$J_g J_g J_g$ derivative}{J\_g J\_g J\_g derivative}}

\noindent We can also check what is the derivative of \eq{contributions_from_epsilon_delta-H} with respect to three powers of $J_g$. 
This gives us the contribution to the galaxy-galaxy-galaxy bispectrum. 
The derivative of $W^{(3)}[J_g,J]$ with respect to $J_g(\vec{k})$ is
\begin{equation}
\label{eq:contributions_from_epsilon_delta-sub_2-A}
\frac{\partial W^{(3)}[J_g,J]}{\partial J_g(\vec{k})} = b^2_1 P_{\rm L}(k)\tilde{s}(\vec{k}) + \int_{\vec{k}'}b_1D_1P_{\rm in}(k')
\big(D_1J(\vec{k}') + b_1D_1J_g(\vec{k}')\big)\frac{\partial\tilde{s}(-\vec{k}')}{\partial J_g(\vec{k})}\,\,.
\end{equation}
Then, using \eq{contributions_from_epsilon_delta-sub_1-B}, we find 
\begin{equation}
\label{eq:contributions_from_epsilon_delta-sub_2-B}
\frac{\partial^2\tilde{s}(-\vec{k}')}{J_g(\vec{k})\partial J_g(\vec{l})} = 
\frac{1}{b_1}\,(2\pi)^3\delta^{(3)}_{\rm D}(-\vec{l}-\vec{k}+\vec{k}')\,
s_{J_gJ_g\delta_{\rm in}}(\vec{k},\vec{l};-\vec{k}')\,\,.
\end{equation}
This can be used in the equation for the second derivative of $W^{(3)}[J_g,J]$, \ie~
\begin{equation}
\label{eq:contributions_from_epsilon_delta-sub_2-C}
\begin{split}
\frac{\partial^2 W^{(3)}[J_g,J]}{\partial J_g(\vec{k})\partial J_g(\vec{l})} &= 
b^2_1P_{\rm L}(k)\frac{\partial\tilde{s}(\vec{k})}{\partial J_g(\vec{l})} 
+ b^2_1P_{\rm L}(l)\frac{\partial\tilde{s}(\vec{l})}{\partial J_g(\vec{k})} \\
&\;\;\;\; + \int_{\vec{k}'}b_1D_1P_{\rm in}(k')\big(D_1J(\vec{k}') + b_1D_1J_g(\vec{k}')\big)
\frac{\partial^2\tilde{s}(-\vec{k}')}{\partial J_g(\vec{k})\partial J_g(\vec{l})}\,\,.
\end{split}
\end{equation}
More precisely, we find 
\begin{equation}
\label{eq:contributions_from_epsilon_delta-sub_2-D}
\begin{split}
\frac{\partial^3 W^{(3)}[J_g,J]}{\partial J_g(\vec{k})\partial J_g(\vec{l})\partial J_g(\vec{m})} 
&= b^2_1P_{\rm L}(k)\frac{\partial^2\tilde{s}(\vec{k})}{\partial J_g(\vec{l})\partial J_g(\vec{m})} 
+ b^2_1P_{\rm L}(l)\frac{\partial^2\tilde{s}(\vec{l})}{\partial J_g(\vec{k})\partial J_g(\vec{m})} \\
&\;\;\;\; + b^2_1P_{\rm L}(m)\frac{\partial^2\tilde{s}(\vec{m})}{\partial J_g(\vec{k})\partial J_g(\vec{l})}\,\,,
\end{split}
\end{equation}
which is equal to 
\begin{equation}
\label{eq:contributions_from_epsilon_delta-sub_2-E}
\frac{\partial^3 W^{(3)}[J_g,J]}{\partial J_g(\vec{k})\partial J_g(\vec{l})\partial J_g(\vec{m})} = 
(2\pi)^3\delta^{(3)}_{\rm D}(\vec{k}+\vec{l}+\vec{m})\,
\big(b_1P_{\rm L}(k)s_{J_gJ_g\delta_{\rm in}}(\vec{l},\vec{m};\vec{k}) + \text{2 perms.}\big)\,\,.
\end{equation}
It is then straightforward to confirm that such result corresponds to the Fourier transform of \eq{bias_expansion_stochasticity-C} 
once we use \eq{contributions_from_epsilon_delta-sub_1-E}. 
Hence, comparing with \eq{contributions_from_epsilon_delta-C}, 
we find that the {additional} contribution to the action of \eq{deterministic_interaction_actions-2} is 
\begin{equation}
\label{eq:contributions_from_epsilon_delta-sub_2-F}
\begin{split}
S_{g,{\rm int}} &= \text{\eq{deterministic_interaction_actions-2}} \\
&\;\;\;\; + \frac{1}{2}\int_{\vec{p}_1,\dots\vec{p}_3}(2\pi)^3\delta^{(3)}_{\rm D}(\vec{p}_{123})\,
D_1s_{J_gJ_g\delta_{\rm in}}(-\vec{p}_1,-\vec{p}_2,-\vec{p}_3)\,X_g(\vec{p}_1)X_g(\vec{p}_2)\delta_{\rm in}(\vec{p}_3)\,\,.
\end{split}
\end{equation}

\subsection{Impact on conditional likelihood}
\label{subapp:effect_on_likelihood}

{\noindent Once we have \eqsII{epsilon_g_bispectrum-B}{contributions_from_epsilon_delta-sub_2-F} it 
is straightforward to compute their impact on the conditional likelihood that 
we studied in Section~\ref{sec:calculation_of_likelihood-nongaussian_stochasticities}.} 

{Indeed, since these two contributions to $S_g$ modify only its interaction part, we can use the solutions of 
\eqsII{linear_classical_fields_solution-matter}{linear_classical_fields_solution-joint} 
(this also implies that the quadratic part of the likelihood is left untouched). 
Moreover, since the action $S$ for the matter likelihood is left unmodified, 
it is enough to check what happens to the joint likelihood. 
Stopping at leading order in derivatives in the deterministic evolution, and dropping the matter stochasticity and the galaxy-matter 
stochasticity in the solution for the classical field $\vec{\phi}_g$, 
we arrive at \eq{bias_expansion_stochasticity-G} (i.e.~the contribution from the noise in $b_1$) 
and \eq{bias_expansion_stochasticity-I} (the contribution from the three-point function of $\eps_g$).}

\section{Saddle-point approximation}
\label{app:saddle_point_formulas}

\noindent In this appendix we expand the tree-level solution{s} for the likelihoods, 
\ie~\eqsII{tree_level_formula}{saddle_point_equations}, at third and fourth order in the fields. 
We will derive those equations for a generic action $S$, functional of some fields $\vec{\phi}(\vec{k})$. 
Moreover, we will condense the indices $(a,\vec{k})$ into a single index $i$, and use Einstein's summation convention on $i$ 
(without distinction between upper and lower indices). 
Therefore our fields $\phi^a(\vec{k})$ become simply $\phi^i$, and the same happens for the matrices from 
which we construct the quadratic action. 

The action $S$ is written as a quadratic part plus an interaction part, \ie~
\begin{equation}
\label{eq:saddle_point_formulas-A}
S[\phi] = \frac{1}{2}\phi^i{\cal M}^{ij}\phi^j + S_{\rm int}[\phi]\,\,,
\end{equation}
and at tree level the (field-dependent) part of the generating functional 
for connected diagrams (which we call $W$) is given by 
\begin{equation}
\label{eq:saddle_point_formulas-B}
W[{\cal J}] = -S\big[\phi_{\rm cl}[{\cal J}]\big] + {\cal J}^i\phi_{\rm cl}^i[{\cal J}]\,\,,
\end{equation}
where $\phi_{\rm cl}$ solves the equation 
\begin{equation}
\label{eq:saddle_point_formulas-C}
\frac{\partial S[\phi]}{\partial\phi^i}\bigg|_{\phi=\phi_{\rm cl}} = {\cal J}^i\,\,.
\end{equation}
Our goal here is to solve these two equations at cubic and quartic order in the currents. 

Before proceeding, we notice that we solve \eq{saddle_point_formulas-C} taking the homogeneous solution to be 
$\phi_{\rm cl}$ equal to zero, given that we want homogeneity and isotropy and there are no tadpoles in the classical action.\footnote{Of 
course this relationship must be enforced order-by-order in loops by renormalizing 
the tadpoles, as we discussed in Section~\ref{subsec:stochastic_terms_and_tadpoles}.}

\subsection{At third order in the fields}
\label{subapp:third_order_check}

\noindent Let us first find the solution to \eq{saddle_point_formulas-C}. Rewriting it as 
\begin{equation}
\label{eq:third_order_check-A}
{\cal M}^{ij}\phi_{\rm cl}^j + \frac{\partial S_{\rm int}[\phi]}{\partial\phi^i}\bigg|_{\phi=\phi_{\rm cl}} = {\cal J}^i 
\end{equation}
we see that we can solve it order-by-order in $\cal J$. Up to second order in the fields (cubic order in the interactions), we find 
\begin{equation}
\label{eq:third_order_check-B}
\begin{split}
\phi_{\rm cl}^i &= \big(\phi_{\rm cl}^{(1)}\big)^i + \big(\phi_{\rm cl}^{(2)}\big)^i 
= {\cal M}^{-1}_{ij}{\cal J}^j 
- {\cal M}^{-1}_{ij}\,\frac{\partial S^{(3)}_{\rm int}[\phi]}{\partial\phi^j}\bigg|_{\phi=\phi^{(1)}_{\rm cl}}\equiv 
{\cal M}^{-1}_{ij}{\cal J}^j - {\cal M}^{-1}_{ij}V^j\,\,.
\end{split}
\end{equation}
We then plug this solution into \eq{saddle_point_formulas-B}. 
After expanding up to cubic order in the currents and using the symmetry of ${\cal M}^{-1}$, 
\ie~${\cal M}^{-1}_{ij} = {\cal M}^{-1}_{ji}$, straightforward manipulations give us 
\begin{equation}
\label{eq:third_order_check-C}
\begin{split}
W &= -\frac{1}{2}{\cal M}^{-1}_{ij}{\cal J}^j\,{\cal M}_{ik}\,{\cal M}^{-1}_{kl}{\cal J}^l 
+ \frac{1}{2}{\cal M}^{-1}_{ij}V^j\,{\cal M}_{ik}\,{\cal M}^{-1}_{kl}{\cal J}^l 
+ \frac{1}{2}{\cal M}^{-1}_{ij}{\cal J}^j\,{\cal M}_{ik}\,{\cal M}^{-1}_{kl}V^l \\
&\;\;\;\; - S^{(3)}_{\rm int}\big[\phi^{(1)}_{\rm cl}\big] + {\cal J}^i{\cal M}^{-1}_{ij}{\cal J}^j - {\cal J}^i{\cal M}^{-1}_{ij}V^j \\
&= \frac{1}{2}{\cal J}^i{\cal M}^{-1}_{ij}{\cal J}^j + {\cal J}^i{\cal M}^{-1}_{ij}V^j 
- S^{(3)}_{\rm int}\big[\phi^{(1)}_{\rm cl}\big] - {\cal J}^i{\cal M}^{-1}_{ij}V^j \\ 
&= \frac{1}{2}{\cal J}^i{\cal M}^{-1}_{ij}{\cal J}^j - S^{(3)}_{\rm int}\big[\phi^{(1)}_{\rm cl}\big]\,\,.
\end{split}
\end{equation}
We see that the dependence on $V^i$ has dropped. This means that we do not need the solution 
for the classical fields at second order in the currents to get $W$ at cubic order. 

Before proceeding to fourth order in the currents, we also comment on the fact that 
the matrices ${\cal M}$ and ${\cal M}^{-1}$ are diagonal in $\vec{k},\vec{k}'$: 
both are proportional to $\delta^{(3)}_{\rm D}(\vec{k}+\vec{k}')$. Hence, 
using the fact that the kernels for the deterministic bias expansion and matter evolution 
are invariant under reflection of the momenta, we are justified in dropping 
all the momentum dependencies and work at fixed $\vec{k}$ when solving for the classical 
fields at linear order (and for the likelihoods at cubic order) in the currents, 
as we did throughout Appendices~\ref{app:tree_level} and \ref{app:higher_derivative_stochasticities}.

\subsection{At fourth order in the fields}
\label{subapp:fourth_order_check}

\noindent First, let us find the solution for the classical fields at cubic order in $\cal J$. 
Expanding \eq{third_order_check-A} at cubic order, we find 
\begin{equation}
\label{eq:fourth_order_check-A}
\begin{split}
\phi_{\rm cl}^i &= \big(\phi_{\rm cl}^{(1)}\big)^i + \big(\phi_{\rm cl}^{(2)}\big)^i + \big(\phi_{\rm cl}^{(3)}\big)^i \\
&= {\cal M}^{-1}_{ij}{\cal J}^j - {\cal M}^{-1}_{ij}\,
\frac{\partial S^{(3)}_{\rm int}[\phi]}{\partial\phi^j}\bigg|_{\phi=\phi^{(1)}_{\rm cl}} 
- {\cal M}^{-1}_{ij}\,\frac{\partial S^{(4)}_{\rm int}[\phi]}{\partial\phi^j}\bigg|_{\phi=\phi^{(1)}_{\rm cl}} 
- {\cal M}^{-1}_{ij}\,\frac{\partial^2S_{\rm int}^{(3)}}{\partial\phi^j\partial\phi^k}\bigg|_{\phi=\phi^{(1)}_{\rm cl}}\,
\big(\phi^{(2)}_{\rm cl}\big)^k \\
&\equiv {\cal M}^{-1}_{ij}{\cal J}^j - {\cal M}^{-1}_{ij}V^j - {\cal M}^{-1}_{ij}T^j 
+ {\cal M}^{-1}_{ij}\,E_{jk}\,{\cal M}^{-1}_{kl}V^l\,\,. 
\end{split}
\end{equation}
Plugging this solution in \eq{saddle_point_formulas-B}, after many cancellations we get 
\begin{equation}
\label{eq:fourth_order_check-B}
W = \frac{1}{2}{\cal J}^i{\cal M}^{-1}_{ij}{\cal J}^j - S^{(3)}_{\rm int}\big[\phi^{(1)}_{\rm cl}\big] 
- S^{(4)}_{\rm int}\big[\phi^{(1)}_{\rm cl}\big] + \frac{1}{2}V^i{\cal M}^{-1}_{ij}V^j\,\,.
\end{equation}
Again, here we see the that we only need the solution for the classical fields at order $n-2$ 
if we want the logarithm of the tree-level generating functional at order $n$ in the currents. 

{U}sing \eq{fourth_order_check-B} we have checked that the results of {Appendix~\ref{app:tree_level} hold beyond cubic order}: 
more precisely we have checked that \eq{cubic_conditional_likelihood-E} holds also at fourth order in the galaxy and matter fields. 
The calculation is straightforward but involved, and we do not report it here. 
The resummation at all orders is discussed in Section~\ref{subsec:P_eps_g}.

\section{More about loops}
\label{app:loops}

\noindent While in the case of Gaussian noise, with zero matter and galaxy-matter stochasticities, 
we have been able to carry out the path integral exactly (see Section~\ref{sec:calculation_of_likelihood-gaussian_stochasticities}), 
the remaining calculations of this paper were done at tree level. 
What happens if we go beyond this approximation? This question can be made more precise: 
do we expect loops to drastically change the tree-level result? 

To understand this, let us think about the chief example of an effective quantum field theory, the Fermi theory of weak interactions. 
Such a theory is non-renormalizable: once we start computing loops we see that we need 
an infinite number of counterterms to absorb the divergences. 
In other words, since all the operators compatible with the symmetries are generated, 
we have to include all of them in the classical action to begin with, and the role of the counterterms 
is just that of reabsorbing the UV-sensitivity of the coefficients of these operators. 

From this point of view, loops do not seem really important. However, 
the fundamental point is that the counterterms required to absorb the divergences are \emph{local}: 
in principle loops give also {finite} \emph{non-analytic} corrections to 
the correlation functions, that are nothing but the logarithmic runnings of the coefficients of the various operators. 
These runnings are the {definite} prediction of loops, and in order 
to compute them we need to go beyond the leading order in the saddle-point 
formulas for the computation of the generating functional. 
We already caught a glimpse of what happens in our case in Section~\ref{subsec:stochastic_terms_and_tadpoles}: 
\begin{itemize}[leftmargin=*]
\item loops over the initial matter field introduce stochasticities 
but also renormalize the coefficients of the deterministic bias expansion. 
For example, let us consider following two diagrams 
\begin{equation}
\label{eq:loops_intro}
\raisebox{-0.0cm}{\includegraphicsbox[scale=0.25,trim={5cm 3cm 5cm 3cm},clip]{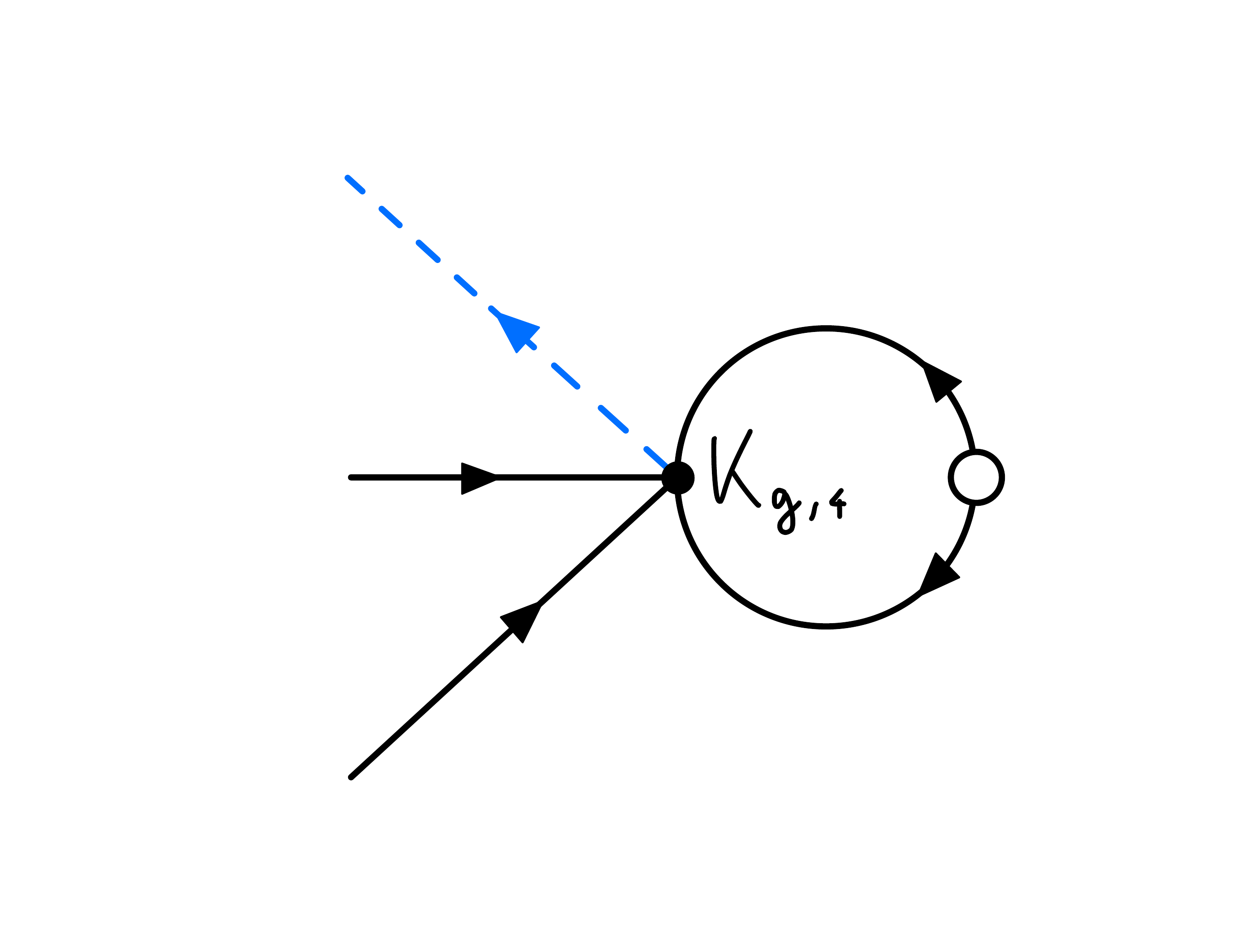}}
\,\,\,\,\,\,,\quad
\raisebox{-0.0cm}{\includegraphicsbox[scale=0.25,trim={2cm 3cm 2cm 3cm},clip]{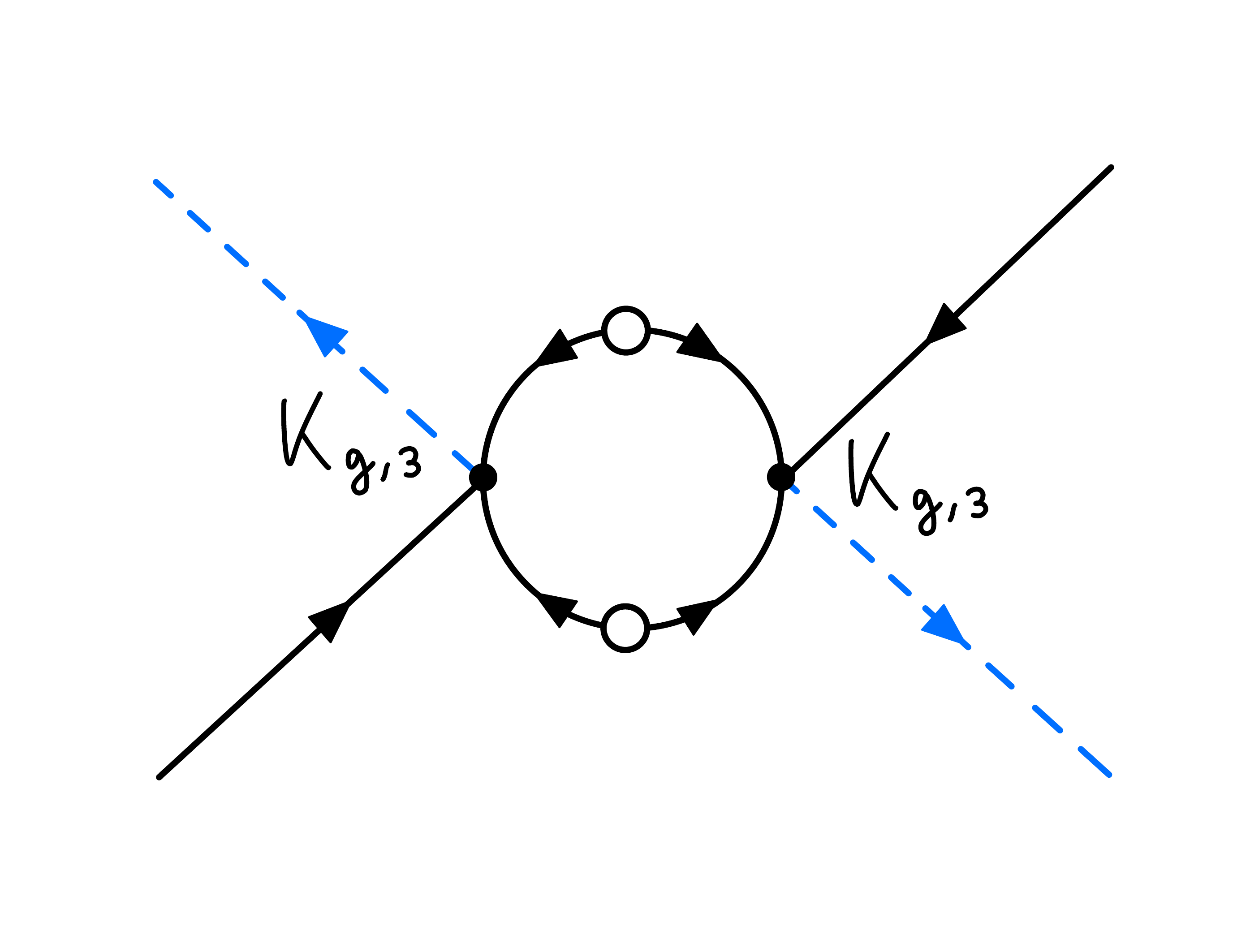}}
\,\,\,\,\,\,.
\end{equation}
The first will renormalize the second-order deterministic galaxy evolution, 
i.e.~the $X_g\delta_{\rm in}\delta_{\rm in}$ vertex, 
while the second one will add a stochasticity to the bias coefficients of the second-order bias expansion; 
\item let us now focus on the stochastic terms only (\ie~diagrams with only $X_g$ or 
$X$ as external legs), for simplicity. 
What happens is that the UV dependence of any loop over $\delta_{\rm in}$ 
in the path integrals for $Z[J]$ and $Z[J_g,J]$ of \eqsII{matter_likelihood-B}{joint_likelihood-B} 
renormalizes the coefficients of these terms. However, following the discussion below \eq{linear_matter_power_spectrum}, 
they must be included anyway 
also if the initial conditions were deterministic since they represent the effect 
of integrating out short-scale modes to arrive at a hydrodynamical description of the matter 
and galaxy fields on large scales. 
\end{itemize}
This seems to suggest that from these loops we do not get any non-analytic 
running of the coefficients of the stochastic terms (or of any other bias coefficient), 
so that we can renormalize any divergence at the renormalization scale $k=0$ 
and end up with an expansion in powers of $k^2$ controlled by the scales $k_{\rm NL}$ and $R(M_h)$. 

Upon further inspection, however, this argument seems a bit circular. 
Let us review in more detail the approach of Section~\ref{subsec:stochastic_terms_and_tadpoles}. 
If we focus on the diagram of \eq{galaxy_stochasticity_loop}, for example, we found 
that loops of $\delta_{\rm in}$ give rise to a UV-sensitive contribution $\sim k^0$ to 
the $J_gJ_g$ term in the generating functional $Z[J_g,J]$, which is indeed reabsorbed by the stochasticity in the galaxy power spectrum. 
However, this result was based on us taking the $k/p\to 0$ limit, in which we can expand 
the argument of the power spectrum $P_{\rm in}(\abs{\vec{p}-\vec{k}/2})$ in the loop. 
This is enough only to confirm that loops of $\delta_{\rm in}$ do not generate any new counterterm 
that is not already included in the full bias expansion. Instead, when we consider a full loop instead of integrating shell by shell, 
$p$ runs down to $0$ and this expansion is not justified.
Physically we do not expect very important effects from large scales, 
since for small momenta the power spectrum {actually} goes to zero instead of growing indefinitely 
{as in \eq{RG-A}}. 
However, Ref.~\cite{Pajer:2013jj} carried out a detailed analysis of loops in the EFT of LSS 
for the total matter field in the case of power-law power spectra, and showed that for some values of 
$n_\delta$ the loop integrals give rise to logarithmic runnings $\sim\ln k/k_{\rm NL}$. 
It would be interesting to rigorously study what happens in the case of biased tracers and 
for a generic (i.e.~not a power law) linear matter power spectrum.\footnote{To reiterate, 
this discussion is not about whether we require some counterterm different 
from the ones considered so far, i.e.~stochasticities with analytic power spectra, bispectra and so on. 
What we want to confirm is that loop corrections do not generate non-analytic runnings 
of the bias coefficients or of the coefficients of these stochasti{c terms}. 
To make a parallel with the Fermi theory, this is the difference between a loop divergence of the form $s\ln (s/\Lambda^2)$ ($s$ being 
the center-of-mass energy squared, $s=E^2_{\rm CM}$), 
and the \emph{local} counterterm $s\ln(\Lambda^2/s_0)$ that reabsorbs it, leading to a running 
coupling $\sim\ln (s/s_0)$. \mbox{See \eg~Section 22.2 of \cite{Schwartz:2013pla}.}} 

There is another point concerning loop effects that is interesting to address. 
In all of our computations so far we have seen that the kernels $K_{n\geq 2}$ for the 
deterministic evolution of the matter field (these kernels being the usual $F_n$ kernels of SPT at leading order in derivatives) 
always drop out from the conditional likelihood {(there is a dependence of ${\cal P}[\delta_g|\delta]$ 
on $K_1$ once we allow for a non-vanishing galaxy-matter stochasticity, 
as we discussed in Section~\ref{subsec:P_eps_g_eps_m}, but 
this is degenerate with other higher-derivative contributions anyway).} 
This makes sense, since our field $\delta$ is already the nonlinear matter field. 

However, these computations hardly count has a full proof. 
In order to prove that the $K_{n}$ kernels drop out at all orders we should first have 
a way to systematically find all the operators generated by loops, and then study their impact on the likelihood. 
A quick way to do this would be to use the real-space Polchinski equation, see \eg~\cite{SkinnerLectures}. 
This is basically the generalization of what we did in Section~\ref{subsec:stochastic_terms_and_tadpoles}, where 
we computed loop diagrams integrating one shell of momentum at a time. By progressively lowering the cutoff $b$ 
we can see what terms are generated in the action in order to keep the 
full path integral (which is independent of $b$) invariant, effectively obtaining all the counterterms. 
Why work in real space? This would have the advantage that we can keep 
{the nonlocal contributions in the bias expansion that come from the displacement terms} under better control, 
{using the ``convective SPT'' approach of \cite{Mirbabayi:2014zca}, 
reviewed in Section~B.5 of \cite{Desjacques:2016bnm} (see Eq.~(B.47) there, for example).} 

This analysis is well beyond the scope of this work. Here we just discuss quickly what can 
happen if we include the stochasticity in the quadratic LIMD bias coefficient $b_2$. 
Adding the stochasticity $\eps_{{{g}},\delta^2}$ leads to a term of the form 
$X_gX_g\delta_{\rm in}\delta_{\rm in}$ in the action $S_{g,{\rm int}}$ (see Tab.~\ref{tab:summary_Z}). 
Differently from the stochasticity in $b_1$, the vertex for this operator will now contain the kernel $K_2$. 
{We can see this because this vertex comes from a shift (schematically)} 
\begin{equation}
\label{eq:loops_subsection_3-A}
b_2\to b_2+P_{\eps_g\eps_{\smash{{{g}},\delta^2}}}^{\{0\}}J_g(\vec{x})\,\,.
\end{equation}
Then, the interaction vertex that couples $X_g$ with two powers of $\delta_{\rm in}$ 
in the action $S_{g,{\rm int}}$ of \eq{deterministic_interaction_actions-2} 
is given by $K_{g,2}$, which is in turn equal to the sum of $b(k)K_2$ and the kernel for the second-order deterministic 
bias expansion $K_{g,{\rm det},2}$, cf.~\eqsII{K_2_example-A}{deterministic_bias_expansion-B}. 
For example, if we consider only LIMD operators and evolve matter via SPT we have 
\begin{equation}
\label{eq:K_2_example-A-recall}
K_{g,2}(\vec{k};\vec{p}_1,\vec{p}_2) = \frac{b_2}{2} + b_1F_2(\vec{p}_1,\vec{p}_2)\,\,.
\end{equation}
As a consequence, the diagram 
\begin{equation}
\label{eq:loops_subsection_3-B}
\raisebox{-0.0cm}{\includegraphicsbox[scale=0.25,trim={1.5cm 3cm 1.5cm 3cm},clip]{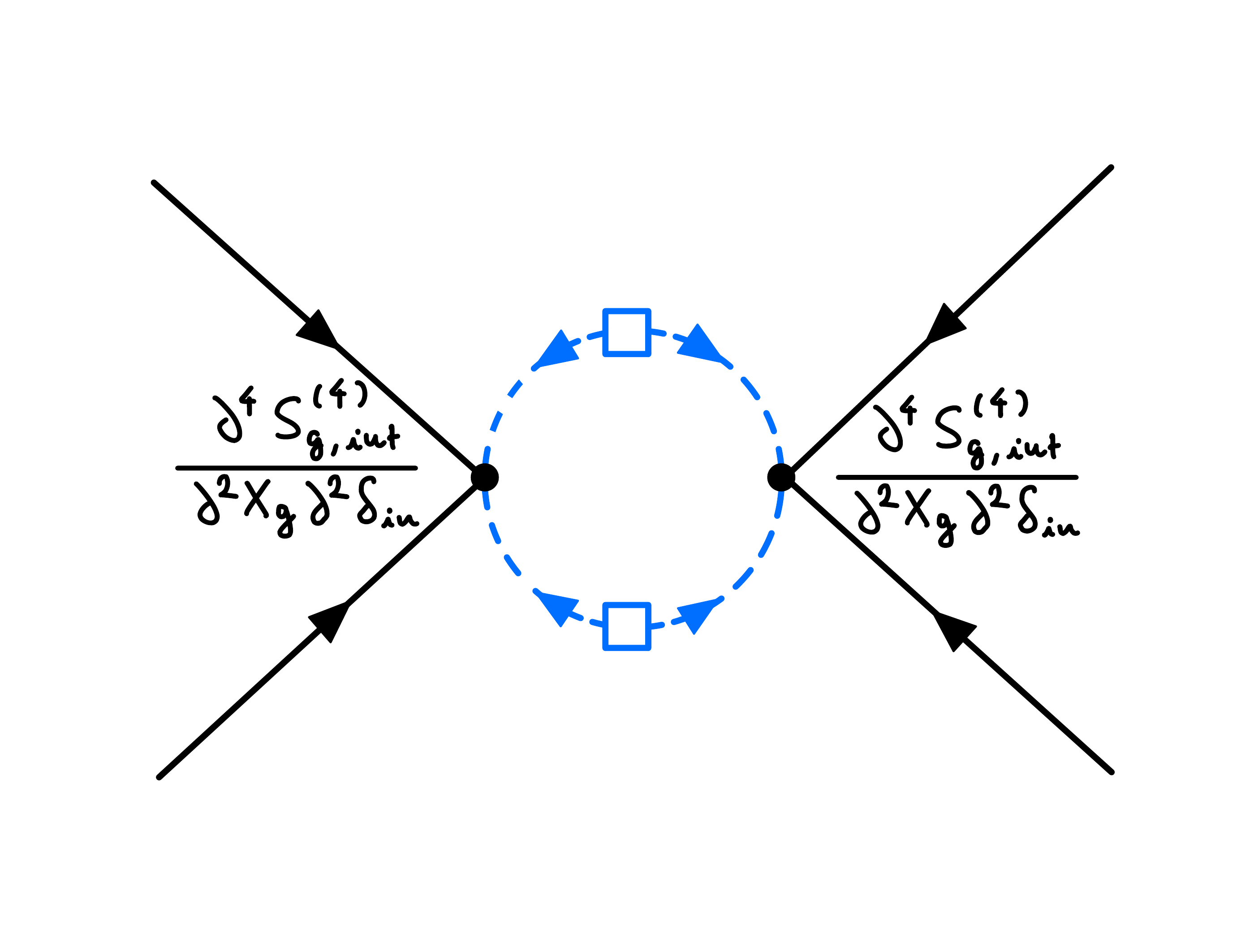}}
\end{equation}
requires a $\delta_{\rm in}\delta_{\rm in}\delta_{\rm in}\delta_{\rm in}$ counterterm that 
apparently is also depending on the $K_2$ kernel. However, 
the important point here is that the SPT kernels are untouched by any loop corrections at zeroth order in derivatives. Therefore, 
when we compute the impact of this term using the tree-level formulas of 
Appendices~\ref{app:tree_level}, \ref{app:higher_derivative_stochasticities} 
and \ref{app:higher_order_stochasticities} (more precisely, their extension 
at fourth order in perturbations), we expect a cancellation between the terms coming from 
the action at fourth order evaluated on the linear solution for the classical fields 
$\smash{\vec{\phi}_{g,{\rm cl}}}$, and those coming from the solution for the classical fields at 
second order (i.e.~the third and last term in \eq{fourth_order_check-B} of Appendix~\ref{subapp:fourth_order_check}).



\clearpage

\bibliographystyle{utphys}
\bibliography{refs}

\providecommand{\href}[2]{#2}\begingroup\raggedright\begin{thebibliography}{10}

\bibitem{Schmidt:2018bkr}
F.~Schmidt, F.~Elsner, J.~Jasche, N.~M. Nguyen, and G.~Lavaux, ``{A rigorous
  EFT-based forward model for large-scale structure},''
  \href{http://dx.doi.org/10.1088/1475-7516/2019/01/042}{{\em JCAP} {\bfseries
  1901} (2019) 042},
\href{http://arxiv.org/abs/1808.02002}{{\ttfamily arXiv:1808.02002
  [astro-ph.CO]}}.

\bibitem{Desjacques:2016bnm}
V.~Desjacques, D.~Jeong, and F.~Schmidt, ``{Large-Scale Galaxy Bias},''
  \href{http://dx.doi.org/10.1016/j.physrep.2017.12.002}{{\em Phys. Rept.}
  {\bfseries 733} (2018) 1--193},
\href{http://arxiv.org/abs/1611.09787}{{\ttfamily arXiv:1611.09787
  [astro-ph.CO]}}.

\bibitem{1995MNRAS.272..885F}
K.~B. {Fisher}, O.~{Lahav}, Y.~{Hoffman}, D.~{Lynden-Bell}, and S.~{Zaroubi},
  ``{Wiener reconstruction of density, velocity and potential fields from
  all-sky galaxy redshift surveys},''
  \href{http://dx.doi.org/10.1093/mnras/272.4.885}{{\em \mnras} {\bfseries 272}
  (Feb., 1995) 885--908},
  \href{http://arxiv.org/abs/astro-ph/9406009}{{\ttfamily astro-ph/9406009}}.
  \url{http://adsabs.harvard.edu/abs/1995MNRAS.272..885F}.

\bibitem{2010MNRAS.406...60J}
J.~{Jasche}, F.~S. {Kitaura}, B.~D. {Wandelt}, and T.~A. {En{\ss}lin},
  ``{Bayesian power-spectrum inference for large-scale structure data},''
  \href{http://dx.doi.org/10.1111/j.1365-2966.2010.16610.x}{{\em \mnras}
  {\bfseries 406} (July, 2010) 60--85},
  \href{http://arxiv.org/abs/0911.2493}{{\ttfamily arXiv:0911.2493}}.
  \url{http://adsabs.harvard.edu/abs/2010MNRAS.406...60J}.

\bibitem{2010MNRAS.409..355J}
J.~{Jasche}, F.~S. {Kitaura}, C.~{Li}, and T.~A. {En{\ss}lin}, ``{Bayesian
  non-linear large-scale structure inference of the Sloan Digital Sky Survey
  Data Release 7},''
  \href{http://dx.doi.org/10.1111/j.1365-2966.2010.17313.x}{{\em \mnras}
  {\bfseries 409} (Nov., 2010) 355--370},
  \href{http://arxiv.org/abs/0911.2498}{{\ttfamily arXiv:0911.2498}}.
  \url{http://adsabs.harvard.edu/abs/2010MNRAS.409..355J}.

\bibitem{2013MNRAS.432..894J}
J.~{Jasche} and B.~D. {Wandelt}, ``{Bayesian physical reconstruction of initial
  conditions from large-scale structure surveys},''
  \href{http://dx.doi.org/10.1093/mnras/stt449}{{\em \mnras} {\bfseries 432}
  (June, 2013) 894--913}, \href{http://arxiv.org/abs/1203.3639}{{\ttfamily
  arXiv:1203.3639}}. \url{http://adsabs.harvard.edu/abs/2013MNRAS.432..894J}.

\bibitem{Wang:2014hia}
H.~Wang, H.~J. Mo, X.~Yang, Y.~P. Jing, and W.~P. Lin, ``{ELUCID - Exploring
  the Local Universe with reConstructed Initial Density field I: Hamiltonian
  Markov Chain Monte Carlo Method with Particle Mesh Dynamics},''
  \href{http://dx.doi.org/10.1088/0004-637X/794/1/94}{{\em Astrophys. J.}
  {\bfseries 794} no.~1, (2014) 94},
\href{http://arxiv.org/abs/1407.3451}{{\ttfamily arXiv:1407.3451
  [astro-ph.CO]}}.

\bibitem{2015MNRAS.446.4250A}
M.~{Ata}, F.-S. {Kitaura}, and V.~{M{\"u}ller}, ``{Bayesian inference of cosmic
  density fields from non-linear, scale-dependent, and stochastic biased
  tracers},'' \href{http://dx.doi.org/10.1093/mnras/stu2347}{{\em \mnras}
  {\bfseries 446} (Feb., 2015) 4250--4259},
  \href{http://arxiv.org/abs/1408.2566}{{\ttfamily arXiv:1408.2566}}.
  \url{http://adsabs.harvard.edu/abs/2015MNRAS.446.4250A}.

\bibitem{1989ApJ...336L...5B}
E.~{Bertschinger} and A.~{Dekel}, ``{Recovering the full velocity and density
  fields from large-scale redshift-distance samples},''
  \href{http://dx.doi.org/10.1086/185348}{{\em \apjl} {\bfseries 336} (Jan.,
  1989) L5--L8}. \url{http://adsabs.harvard.edu/abs/1989ApJ...336L...5B}.

\bibitem{Schmittfull:2017uhh}
M.~Schmittfull, T.~Baldauf, and M.~Zaldarriaga, ``{Iterative initial condition
  reconstruction},'' \href{http://dx.doi.org/10.1103/PhysRevD.96.023505}{{\em
  Phys. Rev.} {\bfseries D96} no.~2, (2017) 023505},
\href{http://arxiv.org/abs/1704.06634}{{\ttfamily arXiv:1704.06634
  [astro-ph.CO]}}.

\bibitem{Seljak:2017rmr}
U.~Seljak, G.~Aslanyan, Y.~Feng, and C.~Modi, ``{Towards optimal extraction of
  cosmological information from nonlinear data},''
  \href{http://dx.doi.org/10.1088/1475-7516/2017/12/009}{{\em JCAP} {\bfseries
  1712} (2017) 009},
\href{http://arxiv.org/abs/1706.06645}{{\ttfamily arXiv:1706.06645
  [astro-ph.CO]}}.

\bibitem{Modi:2019hnu}
C.~Modi, M.~White, A.~Slosar, and E.~Castorina, ``{Reconstructing large-scale
  structure with neutral hydrogen surveys},''
\href{http://arxiv.org/abs/1907.02330}{{\ttfamily arXiv:1907.02330
  [astro-ph.CO]}}.

\bibitem{Carroll:2013oxa}
S.~M. Carroll, S.~Leichenauer, and J.~Pollack, ``{Consistent effective theory
  of long-wavelength cosmological perturbations},''
  \href{http://dx.doi.org/10.1103/PhysRevD.90.023518}{{\em Phys. Rev.}
  {\bfseries D90} no.~2, (2014) 023518},
\href{http://arxiv.org/abs/1310.2920}{{\ttfamily arXiv:1310.2920 [hep-th]}}.

\bibitem{Blas:2015qsi}
D.~Blas, M.~Garny, M.~M. Ivanov, and S.~Sibiryakov, ``{Time-Sliced Perturbation
  Theory for Large Scale Structure I: General Formalism},''
  \href{http://dx.doi.org/10.1088/1475-7516/2016/07/052}{{\em JCAP} {\bfseries
  1607} no.~07, (2016) 052},
\href{http://arxiv.org/abs/1512.05807}{{\ttfamily arXiv:1512.05807
  [astro-ph.CO]}}.

\bibitem{Blas:2016sfa}
D.~Blas, M.~Garny, M.~M. Ivanov, and S.~Sibiryakov, ``{Time-Sliced Perturbation
  Theory II: Baryon Acoustic Oscillations and Infrared Resummation},''
  \href{http://dx.doi.org/10.1088/1475-7516/2016/07/028}{{\em JCAP} {\bfseries
  1607} no.~07, (2016) 028},
\href{http://arxiv.org/abs/1605.02149}{{\ttfamily arXiv:1605.02149
  [astro-ph.CO]}}.

\bibitem{McDonald:2017ths}
P.~McDonald and Z.~Vlah, ``{Large-scale structure perturbation theory without
  losing stream crossing},''
  \href{http://dx.doi.org/10.1103/PhysRevD.97.023508}{{\em Phys. Rev.}
  {\bfseries D97} no.~2, (2018) 023508},
\href{http://arxiv.org/abs/1709.02834}{{\ttfamily arXiv:1709.02834
  [astro-ph.CO]}}.

\bibitem{Elsner:2019rql}
F.~Elsner, F.~Schmidt, J.~Jasche, G.~Lavaux, and N.-M. Nguyen, ``{Cosmology
  Inference from Biased Tracers using the EFT-based Likelihood},''
\href{http://arxiv.org/abs/1906.07143}{{\ttfamily arXiv:1906.07143
  [astro-ph.CO]}}.

\bibitem{Scoccimarro:2003wn}
R.~Scoccimarro, E.~Sefusatti, and M.~Zaldarriaga, ``{Probing Primordial
  non-Gaussianity with Large-Scale Structure},''
  \href{http://dx.doi.org/10.1103/PhysRevD.69.103513}{{\em Phys. Rev.}
  {\bfseries D69} (2004) 103513},
\href{http://arxiv.org/abs/astro-ph/0312286}{{\ttfamily arXiv:astro-ph/0312286
  [astro-ph]}}.

\bibitem{2011JCAP...04..006B}
T.~{Baldauf}, U.~{Seljak}, and L.~{Senatore}, ``{Primordial non-Gaussianity in
  the bispectrum of the halo density field},''
  \href{http://dx.doi.org/10.1088/1475-7516/2011/04/006}{{\em \jcap} {\bfseries
  2011} no.~4, (Apr., 2011) 006},
  \href{http://arxiv.org/abs/1011.1513}{{\ttfamily arXiv:1011.1513
  [astro-ph.CO]}}.

\bibitem{Assassi:2015fma}
V.~Assassi, D.~Baumann, and F.~Schmidt, ``{Galaxy Bias and Primordial
  Non-Gaussianity},''
  \href{http://dx.doi.org/10.1088/1475-7516/2015/12/043}{{\em JCAP} {\bfseries
  1512} no.~12, (2015) 043},
\href{http://arxiv.org/abs/1510.03723}{{\ttfamily arXiv:1510.03723
  [astro-ph.CO]}}.

\bibitem{Mirbabayi:2014zca}
M.~Mirbabayi, F.~Schmidt, and M.~Zaldarriaga, ``{Biased Tracers and Time
  Evolution},'' \href{http://dx.doi.org/10.1088/1475-7516/2015/07/030}{{\em
  JCAP} {\bfseries 1507} no.~07, (2015) 030},
\href{http://arxiv.org/abs/1412.5169}{{\ttfamily arXiv:1412.5169
  [astro-ph.CO]}}.

\bibitem{Senatore:2014eva}
L.~Senatore, ``{Bias in the Effective Field Theory of Large Scale
  Structures},'' \href{http://dx.doi.org/10.1088/1475-7516/2015/11/007}{{\em
  JCAP} {\bfseries 1511} no.~11, (2015) 007},
\href{http://arxiv.org/abs/1406.7843}{{\ttfamily arXiv:1406.7843
  [astro-ph.CO]}}.

\bibitem{ZinnJustin:2002ru}
J.~Zinn-Justin, ``{Quantum field theory and critical phenomena},''
{\em Int. Ser. Monogr. Phys.} {\bfseries 113} (2002) 1--1054.

\bibitem{Peskin:1995ev}
M.~E. Peskin and D.~V. Schroeder, {\em {An Introduction to quantum field
  theory}}.
\newblock Addison-Wesley, Reading, USA, 1995.
\newblock
\url{http://www.slac.stanford.edu/~mpeskin/QFT.html}.
\newblock

\bibitem{SkinnerLectures}
{D. Skinner}, ``{Quantum Field Theory II},''.
  \url{http://www.damtp.cam.ac.uk/user/dbs26/AQFT.html}.

\bibitem{Baumann:2010tm}
D.~Baumann, A.~Nicolis, L.~Senatore, and M.~Zaldarriaga, ``{Cosmological
  Non-Linearities as an Effective Fluid},''
  \href{http://dx.doi.org/10.1088/1475-7516/2012/07/051}{{\em JCAP} {\bfseries
  1207} (2012) 051},
\href{http://arxiv.org/abs/1004.2488}{{\ttfamily arXiv:1004.2488
  [astro-ph.CO]}}.

\bibitem{McDonald:2006mx}
P.~McDonald, ``{Clustering of dark matter tracers: Renormalizing the bias
  parameters},'' \href{http://dx.doi.org/10.1103/PhysRevD.74.103512,
  10.1103/PhysRevD.74.129901}{{\em Phys. Rev.} {\bfseries D74} (2006) 103512},
  \href{http://arxiv.org/abs/astro-ph/0609413}{{\ttfamily
  arXiv:astro-ph/0609413 [astro-ph]}}.
[Erratum: Phys. Rev.D74,129901(2006)].

\bibitem{2014JCAP...08..056A}
V.~{Assassi}, D.~{Baumann}, D.~{Green}, and M.~{Zaldarriaga}, ``{Renormalized
  halo bias},'' \href{http://dx.doi.org/10.1088/1475-7516/2014/08/056}{{\em
  \jcap} {\bfseries 8} (Aug., 2014) 056},
  \href{http://arxiv.org/abs/1402.5916}{{\ttfamily 1402.5916}}.
  \url{http://adsabs.harvard.edu/abs/2014JCAP...08..056A}.

\bibitem{Abolhasani:2015mra}
A.~A. Abolhasani, M.~Mirbabayi, and E.~Pajer, ``{Systematic Renormalization of
  the Effective Theory of Large Scale Structure},''
  \href{http://dx.doi.org/10.1088/1475-7516/2016/05/063}{{\em JCAP} {\bfseries
  1605} no.~05, (2016) 063},
\href{http://arxiv.org/abs/1509.07886}{{\ttfamily arXiv:1509.07886 [hep-th]}}.

\bibitem{Perko:2016puo}
A.~Perko, L.~Senatore, E.~Jennings, and R.~H. Wechsler, ``{Biased Tracers in
  Redshift Space in the EFT of Large-Scale Structure},''
\href{http://arxiv.org/abs/1610.09321}{{\ttfamily arXiv:1610.09321
  [astro-ph.CO]}}.

\bibitem{Ding:2017gad}
Z.~Ding, H.-J. Seo, Z.~Vlah, Y.~Feng, M.~Schmittfull, and F.~Beutler,
  ``{Theoretical Systematics of Future Baryon Acoustic Oscillation Surveys},''
  \href{http://dx.doi.org/10.1093/mnras/sty1413}{{\em Mon. Not. Roy. Astron.
  Soc.} {\bfseries 479} no.~1, (2018) 1021--1054},
\href{http://arxiv.org/abs/1708.01297}{{\ttfamily arXiv:1708.01297
  [astro-ph.CO]}}.

\bibitem{delaBella:2018fdb}
L.~F. de~la Bella, D.~Regan, D.~Seery, and D.~Parkinson, ``{Impact of bias and
  redshift-space modelling for the halo power spectrum: Testing the effective
  field theory of large-scale structure},''
\href{http://arxiv.org/abs/1805.12394}{{\ttfamily arXiv:1805.12394
  [astro-ph.CO]}}.

\bibitem{Desjacques:2018pfv}
V.~Desjacques, D.~Jeong, and F.~Schmidt, ``{The Galaxy Power Spectrum and
  Bispectrum in Redshift Space},''
  \href{http://dx.doi.org/10.1088/1475-7516/2018/12/035}{{\em JCAP} {\bfseries
  1812} no.~12, (2018) 035},
\href{http://arxiv.org/abs/1806.04015}{{\ttfamily arXiv:1806.04015
  [astro-ph.CO]}}.

\bibitem{Senatore:2008vi}
L.~Senatore, S.~Tassev, and M.~Zaldarriaga, ``{Cosmological Perturbations at
  Second Order and Recombination Perturbed},''
  \href{http://dx.doi.org/10.1088/1475-7516/2009/08/031}{{\em JCAP} {\bfseries
  0908} (2009) 031},
\href{http://arxiv.org/abs/0812.3652}{{\ttfamily arXiv:0812.3652 [astro-ph]}}.

\bibitem{Senatore:2008wk}
L.~Senatore, S.~Tassev, and M.~Zaldarriaga, ``{Non-Gaussianities from
  Perturbing Recombination},''
  \href{http://dx.doi.org/10.1088/1475-7516/2009/09/038}{{\em JCAP} {\bfseries
  0909} (2009) 038},
\href{http://arxiv.org/abs/0812.3658}{{\ttfamily arXiv:0812.3658 [astro-ph]}}.

\bibitem{Pajer:2013jj}
E.~Pajer and M.~Zaldarriaga, ``{On the Renormalization of the Effective Field
  Theory of Large Scale Structures},''
  \href{http://dx.doi.org/10.1088/1475-7516/2013/08/037}{{\em JCAP} {\bfseries
  1308} (2013) 037},
\href{http://arxiv.org/abs/1301.7182}{{\ttfamily arXiv:1301.7182
  [astro-ph.CO]}}.

\bibitem{Schwartz:2013pla}
M.~D. Schwartz, {\em {Quantum Field Theory and the Standard Model}}.
\newblock Cambridge University Press, 2014.
\newblock
\url{http://www.cambridge.org/us/academic/subjects/physics/theoretical-physics-and-mathematical-physics/quantum-field-theory-and-standard-model}.
\newblock

\end{thebibliography}\endgroup


\end{document}